\documentclass[11pt]{article}
\usepackage[backend=biber, style=numeric, sorting=none]{biblatex}
\usepackage{geometry}                
\usepackage{graphicx}
\graphicspath{figures/}  

\usepackage{hyperref}    
\usepackage[all]{hypcap} 

\usepackage{multirow}
\usepackage{amssymb}
\usepackage{epstopdf}
\usepackage{amssymb}
\usepackage{siunitx}
\usepackage[inline]{enumitem} 
\setlist[itemize]{noitemsep}
\setlist[enumerate]{noitemsep}
\usepackage{siunitx}
\usepackage{ booktabs} 
\usepackage{rotating}
\usepackage{lscape}
\usepackage{lineno}
\usepackage{subcaption}
\usepackage{tabularx}
\usepackage{bm}
\usepackage{adjustbox}
\usepackage{wrapfig}

\usepackage {tikz-imagelabels}
\usepackage {circuitikz}
\usepackage {venndiagram}
\usepackage{ascmac}

\usepackage{titlesec}
\titleformat{\paragraph}
{\normalfont\normalsize\bfseries}{\theparagraph}{1em}{}
\titlespacing*{\paragraph}
{0pt}{3.25ex plus 1ex minus .2ex}{1.5ex plus .2ex}
\nolinenumbers

\newcommand{\editor}[1]{\par {\em #1} \par }
\usepackage[official]{eurosym}

\newcommand\snowmass{\begin{center}\rule[-0.2in]{\textwidth}{0.01in}\\\rule{\textwidth}{0.01in}\\
\vskip 0.1in Submitted to the  Proceedings of the US Community Study\\
on the Future of Particle Physics (Snowmass 2021)\\
\rule{\textwidth}{0.01in}\\\rule[+0.2in]{\textwidth}{0.01in} \end{center}}

\title{Snowmass Whitepaper: \\ The Belle II Detector Upgrade Program}
\author{Belle II Collaboration}
\date{\today} 

\begin{document}

\snowmass
\begingroup
\let\newpage\relax
\maketitle
\endgroup
\begin{abstract}
   We describe the planned near-term and potential longer-term upgrades of the Belle II detector at the 
SuperKEKB electron-positron collider in Tsukuba, Japan. These upgrades will allow increasingly sensitive 
searches for possible new physics beyond the Standard Model in flavor, tau, electroweak and dark sector 
physics that are both complementary to and competitive with the LHC and other experiments. We encourage the instrumentation-frontier community to contribute and study upgrade ideas as part of the Snowmass process.

\end{abstract}

\vspace{0.2in}

\noindent {\large \bf Corresponding author:}
\begin{verbatim}
Francesco Forti <francesco.forti@pi.infn.it>
\end{verbatim}
\noindent {\large \bf Editors:}
\begin{verbatim}
Sven Vahsen <sevahsen@hawaii.edu>
Peter Krizan <peter.krizan@ijs.si>
Phillip Urquijo <phillip.urquijo@unimelb.edu.au>
Laci Andricek <laci.andricek@hll.mpg.de>
Katsuro Nakamura <katsuro@post.kek.jp>
Carlos Marinas <cmarinas@ific.uv.es>
Jerome Baudot <jerome.baudot@iphc.cnrs.fr>
Akimasa Ishikawa <akimasa.ishikawa@kek.jp>
Nanae Taniguchi <nanae@post.kek.jp>
Ezio Torassa <ezio.torassa@pd.infn.it>
Rok Pestotnik <Rok.Pestotnik@ijs.si>
Claudia Cecchi <claudia.cecchi@pg.infn.it>
Leo Piilonen <piilonen@vt.edu>
XiaoLong Wang <xiaolong@fudan.edu.cn>
Gary S. Varner <varner@phys.hawaii.edu>
Taichiro Koga <taichiro@post.kek.jp>
Satoru YAMADA <satoru.yamada@kek.jp>
Oskar Hartbrich <ohartbri@hawaii.edu>
Umberto Tamponi <tamponi@to.infn.it>
Peter M. Lewis <plewis@uni-bonn.de>
\end{verbatim}
\vspace{0.2in}

\noindent {\large \bf Thematic Areas: Instrumentation Frontier}






 \noindent $\blacksquare$ (IF02) Photon Detectors
 
 \noindent $\blacksquare$ (IF03) Solid State Detectors \& Tracking
 
 \noindent $\blacksquare$ (IF04) Trigger \& DAQ
 
 \noindent $\blacksquare$ (IF05) Micro Pattern Gas Detectors

 \noindent $\blacksquare$ (IF06) Calorimetry
 
 \noindent $\blacksquare$ (IF07) Electronics/ASICs


\tableofcontents

\clearpage


%

\section{Introduction and Overview}
\begin{refsection}
\label{sec:Introduction}
%
%
%
%
%
%


Belle II\cite{Abe:2010gxa} is an international collaboration of 1000 members at more than 100 institutions in
26 countries/regions. U.S. Belle II accounts for 120 members at 18 U.S. universities and national labs.
Primary U.S. Belle II responsibilities include the imaging Time Of Propagation subdetector 
(iTOP) 
used for charged particle identification, the K-Long Muon subdetector (KLM), beam background mitigation, 
computing operations, and data production.

SuperKEKB/Belle II was commissioned with colliding beams in 2018. At the end of 2018 the Vertex Detector VXD (composed of the pixel detector PXD and strip detector SVD) was installed, but with an incomplete PXD. 
In spring 2020, SuperKEKB 
surpassed the highest recorded instantaneous luminosities of the B factories and LHC, obtaining a peak instantaneous luminosity of $\mathcal{L}_{\rm peak}$ = \SI{3.81 e 34}{cm^{-2}s^{-1}} in December 2021. 
SuperKEKB is expected to be able to reach 
$\mathcal{L}_{\rm peak}$ of \SI{2 e 35}{cm^{-2}s^{-1}} 
with the existing accelerator complex. 
In order to reach \SI{6.5 e 35}{cm^{-2}s^{-1}}, however, an upgrade of the interaction region 
and the QCS superconducting final focus will probably be required. An International Task Force has been setup to provide advice to SuperKEKB on the possible technical solutions.  
The current program foresees the accumulation of  $\mathcal{L}_{\rm int}$ = \SI{50}{ab^{-1}} by the first years of the next decade. 
Belle II detector upgrades are developed on three time scales:

\begin{enumerate}
\item {\bf short term}:
year 2022. Long Shutdown 1 (LS1) is planned for approximately 15 months starting in July 2022, to install a complete pixel detector (PXD). This upgrade is well defined and only minimal additions are possible. 
\item {\bf medium term}:
approximately year 2026-27. Long Shutdown 2 (LS2) will probably be needed for the upgrade of the interaction region to reach $\mathcal{L}_{\rm peak}$ = \SI{6.5 e 35}{cm^{-2}s^{-1}}. A new Vertex Detector might be required to accommodate the new IR design, and other sub-detector upgrades are possible, with various options being evaluated. 
\item {\bf long term}: 
years $> 2032$. Studies have started to explore upgrades beyond the currently planned program, such as beam polarization and ultra-high luminosity, possibly $\mathcal{L}_{\rm peak}$ in excess of \SI{1 e 36}{cm^{-2}s^{-1}}. While the beam polarization has a concrete proposal, for ultra-high luminosity studies have just started. 
\end{enumerate}

Belle II is designed to operate efficiently even if the backgrounds extrapolated to target luminosity, but safety margins are not large. The expected evolution of these backgrounds with luminosity is described in a separate Snowmass White Paper titled ”Beam background expectations for Belle II at SuperKEKB". In the case of a redesign of the interaction region large uncertainties in the background extrapolations are unavoidable.
The upgrade program is therefore motivated by a number of considerations:
\begin{itemize}
    \item Improve detector robustness against backgrounds
    \item  Provide larger safety factors for running at higher luminosity 
    \item Increase longer term subdetector radiation resistance
    \item Develop the technology to cope with different future paths, for instance if a major IR redesign is required to reach the target luminosity
    \item Improve overall physics performance
\end{itemize}

Table \ref{tab:upgrades} shows the current ideas for Belle II detector upgrades.
The collaboration is planning to start preparing a Conceptual Design Report in summer 2022 when upgrade options for the accelerator should become clearer. A more focused and well defined set of options will be described in the CDR.

The U.S. groups plan participation in iTOP, KLM, and readout/DAQ upgrades, and have proposed the ``STOPGAP'' system to increase the geometric TOP acceptance. There is also U.S. interest in studying the feasibility of replacing the drift chamber with a time projection chamber (TPC) with charge readout via Micropattern Gasous Detectors (MPGDs). 

Upgrading SuperKEKB to have electron beams with left and right longitudinal polarization of approximately 70\% at the Belle~II interaction point creates
 a unique and versatile facility for probing new physics with precision electroweak measurements that no other experiments, current or planned, 
 can achieve. This program will provide unprecedented precision in neutral current universality measurements involving b-quarks, c-quarks, $\tau$-leptons, muons and electrons, as well as a measurement of $\sin^2\theta_W$ with a precision equivalent to that of the current world average from measurements at the Z-pole~\cite{ALEPH:2005ab}, but at 10~GeV. This provides unique sensitivity to new physics revealed through deviations from the SM running of $\sin^2\theta_W$, such as would be expected from  a dark sector analogue of the Z-boson.
 It also enables a set of other precision measurements, including the $\tau$ anomalous magnetic moment form-factor, $F_2(10~GeV)$~\cite{Bernabeu:2007rr}, with unrivaled precision, thereby providing a measurement analogous to that of the $(g-2)$ of the muon, but in the third generation.
 
 The upgrade involves three hardware projects: 1) a low-emittance polarized electron source in which  electron beams are produced via a polarized laser illuminating a ``strained lattice'' GaAs photocathode as was done for SLD\cite{ALEPH:2005ab};
 2) a pair of spin-rotators, one positioned before and the other after the interaction region, to rotate the spin to longitudinal
 prior to collisions and back to transverse following collisions; and
 3) a Compton polarimeter that measures the beam polarization before the beam enters the interaction region in real-time. One scenario has these upgrades commencing during LS2 and completed during the following summer shutdowns with  initial collisions with polarized beams beginning shortly after. The physics case and R\&D status is described in the Belle~II Snowmass White Paper entitled  "Upgrading SuperKEKB with a Polarized Electron Beam: Discovery Potential and Proposed Implementation".

The medium and long term upgrades are highly relevant to the 2021 Snowmass process. We encourage the wider instrumentation community participating in Snowmass to contribute and study existing or new upgrade ideas for Belle II.

In the following sections a more detailed description of the various options is provided. Some approximate cost estimates are indicated. Given the level of precision of these estimates one can assume 1~M\$ = 1~M\euro = 100~MJPY.
The simulations reported in the text, if not specified otherwise, have used the Belle II Analysis Software Framework BASF2~\cite{Kuhr:2018lps}.

\begin{landscape}
\begin{table}[b]
	\centering
	\begin{tabular} {llll}
	\toprule
	Subdector   &	Function 		& upgrade idea 	        & time scale	 \\
	\hline
	PXD 		&	Vertex Detector & 2 layer installation 	&   short-term \\
                &	                & new DEPFET            &  medium-term   \\
    \hline
	SVD 		& 	Vertex Detector & thin, double-sided strips, w/ new frontend & medium-term \\
    \hline
    PXD+SVD    	&	Vertex Detector	& all-pixels: SOI sensors & medium-term  \\
		    	&					& all-pixels: DMAPS CMOS sensors & medium-term \\
	\hline
	CDC		&	Tracking		& upgrade front end electronics     & short/medium-term \\
			&	        		& replace inner part with silicon 	& medium/long term \\
		    &					& replace with TPC w/ MPGD readout  & long-term \\
	\hline
	TOP   	& PID, barrel		& Replace conventional MCP-PMTs          & short-term  \\
			&					& Replace not-life-extended ALD MCP-PMTs & medium-term \\
			&					& STOPGAP TOF and timing detector        & long-term \\
	\hline
	ARICH  	& PID, forward	&  replace HAPD with Silicon PhotoMultipliers & long-term \\
			&				&  replace HAPD with Large Area Picosecond Photodetectors & long-term\\
	\hline
	ECL 	&  $\gamma$, $e$ ID	& add pre-shower detector in front of ECL & long-term\\
			& 					& Replace ECL PiN diodes with APDs & long-term\\
			& 					& Replace CsI(Tl) with pure CsI crystals & long-term\\
	\hline
	KLM		& $K_{L}$, $\mu$ ID	& replace 13 barrel layers of legacy RPCs with scintillators & medium/long-term\\
			&					&  on-detector upgraded scintillator readout & medium/long-term\\
			&					& timing upgrade for K-long momentum measurement & medium/long-term\\
	\hline
	Trigger	& 					& firmware improvements & continuos\\
	\hline
	DAQ		&					& PCIe40 readout upgrade	& ongoing \\
			& 					& add 1300-1900 cores to HLT& short/medium-term\\

	\bottomrule
	\end{tabular}
		\caption{Known short and medium-term Belle II subdetector upgrade plans, starting from the radially innermost.  The current Belle II subdetectors are the Silicon Pixel Detector (PXD), Silicon Strip Detector (SVD), Central Drift Chamber (CDC),  Time of Propagation Counter (TOP), Aerogel Rich Counter (ARICH), EM Calorimeter (ECL), Barrel and Endcap K-Long Muon Systems (BKLM, EKLM), Trigger and Data aquistion (DAQ). DAQ includes the high level trigger (HLT). }
\label{tab:upgrades}
\end{table}
\end{landscape}

\printbibliography[heading=subbibliography]
\clearpage
\end{refsection}

\section{Physics and detector performance}
\begin{refsection}
\label{sec:Physics}
\editor{P.Urquijo}

\newcolumntype{R}[2]{%
    >{\adjustbox{angle=#1,lap=\width-(#2)}\bgroup}%
    l%
    <{\egroup}%
}
\newcommand*\rot{\multicolumn{1}{R{90}{1em}}}

\subsection{Introduction}

The Belle II upgrade intends to ensure that Belle II can consistently deliver high impact results in heavy flavour and dark sector physics. Many of the  measurements of key decay channels will continue to be precision limited by statistical uncertainties with data sets in excess of 10 ab$^{-1}$, including both CKM unitarity triangle precision inputs and properties of rare decays. A consideration for very large datasets not typically factored into sensitivity projections is that the luminosity dependent background conditions may be severe, in terms of detector operations and physics performance.

The experiment will ideally maintain or improve upon its particle detection and identification efficiencies, even with higher instantaneous luminosity environments. This would imply robust performance in challenging but key reconstruction processes, such as low momentum tracking, calorimetry for reconstruction of photons and $\pi^0$ mesons, and particle identification.

A summary of projected measurement sensitivities is listed in Table \ref{table:precision}. The Belle II 5 and 50 ab$^{-1}$ projections are primarily based on Ref.~\cite{Belle-II:2018jsg}, with some recent updates. They correspond roughly to the luminosity that can be collected before LS2 and by end of the current Belle II program. Physics motivations for these channels can be found in the cited document and also in the white paper on Belle II physics studies. These extrapolations typically assume integrated luminosity scaling without degradation in measurement efficiencies or purity, as well as improvements to systematic uncertainties that are data driven. Belle II projections to 250~ab$^{-1}$, the possible target luminosity of a future ultra-high luminosity upgrade,  take into account discussions in Ref.~\cite{Belle-II:2018jsg} on integrated luminosity scaling of statistical and systematic uncertainties.
Analogous projections for some of these physics channels can be found for LHCb in Ref.~\cite{LHCbCollaboration:2776420, LHCb:2018roe}, showing that the Belle II's broad physics program is both complementary and competitive with LHCb.

Continued improvements in precision are key to reveal new phenomena, however the current detector configuration is not expected to maintain its performance level when facing high beam background levels or high rates. This section will step through the major impacts of the detector upgrades, the key performance requirements for precision flavour physics, and finally selected physics channels that may demand detector upgrades for ultimate precision.

\begin{table*}[htbp]
   \centering
    \begin{tabularx}{0.9\textwidth}{l  X  X  X  X }
        \hline
       Observable                                   &  2022 Belle(II), BaBar & Belle-II 5~ab$^{-1}$ & Belle-II 50~ab$^{-1}$ & Belle-II 250~ab$^{-1}$ \\ \hline
        $\sin 2 \beta/\phi_1$                       & 0.03          & 0.012         & 0.005                 & 0.002             \\ 
        $\gamma/\phi_3$ (Belle+BelleII)             &11$^\circ$     &4.7$^\circ$    & 1.5$^\circ$           & 0.8$^\circ$       \\ 
        $\alpha/\phi_2$ (WA)                        &4$^\circ$      &2$^\circ$      & 0.6$^\circ$           & 0.3$^\circ$       \\ 
        $|V_{ub}|$ (Exclusive)                       &4.5\%            &2\%            &1\%                    & $<1\%$            \\ 
        \hline
        $S_{CP}(B \to \eta^\prime K_{\rm S}^0)$     &0.08           &0.03           &0.015                  &0.007              \\ 
        $A_{CP}(B \to \pi^0 K_{\rm S}^0)$           &0.15           &0.07           &0.025                   &0.018               \\ 
        $S_{CP}(B \to K^{*0}\gamma$)     &0.32           &0.11           &0.035                  &0.015              \\ 
        \hline
        $R(B\to K^*\ell^+\ell^-)^\dagger$           &0.26           &0.09           &0.03                   &0.01               \\ 
        $R(B \to D^* \tau \nu)$                     &0.018          &0.009          &0.0045                 &$<$0.003           \\ 
        $R(B \to D \tau \nu)$                       &0.034          &0.016          &0.008                  &$<$0.003           \\ 
        ${\cal B}(B \to \tau \nu)$                  &24\%           &9\%            &4\%                    &$2\%$              \\ 
        $B(B \to K^* \nu \bar \nu)$                 &$-$            &25\%           &9\%                    &$4\%$              \\ 
        \hline
        ${\cal B}(\tau \to \mu \gamma)$ UL            &$42\times10^{-9}$ &$22\times10^{-9}$   &$6.9\times10^{-9}$   &$3.1\times10^{-9}$      \\ 
        ${\cal B}(\tau \to \mu\mu\mu)$ UL           &$21\times10^{-9}$  &$3.6\times10^{-9}$   &$0.36\times10^{-9}$   &$0.073\times10^{-9}$   \\ 
\hline
    \end{tabularx}
    \caption{Projected precision (total uncertainties, or 90\% CL upper limits) of selected flavour physics measurements at Belle II.(The $\dagger$ symbol denotes the measurement in the momentum transfer squared bin $1<q^2<6{~\rm GeV}/c^2$.) }
    \label{table:precision} 
\end{table*}
\subsubsection{Discussion of detector upgrades}

Here we summarise aspects of the upgrade that directly impact physics performance. The detector upgrades and their quantitative performance impacts are discussed in detail in the later sections of this whitepaper. 

\begin{itemize}
    \item VXD systems: The proposed upgrades all reduce occupancy levels, with higher robustness against tracking efficiency and resolution losses from beam background. This implies improved tracking efficiencies with $p_{\rm T}<200$ MeV/$c$. 
    
    \item CDC: The proposed electronics upgrades improve the quality of tracking through cross-talk reduction, and faster more reliable triggering. This affects general tracking efficiencies, as well as $dE/dx$ measurements. 
    
    \item TOP: The TOP detector's sensitivity to single photons, i.e. the quantum efficiency, will degrade under irradiation without  sensor replacement and upgrade. This directly impacts overall efficacy of the TOP system, as well as time resolution, which is critical for particle ID PDFs.
    
    \item ECL: Three upgrade options include new pure CsI crystals with APDs, a pre-shower detector in front of the ECL, and an option where the existing CsI(Tl) are read-out with APDs. The performance of the ECL will degrade with higher background rates without future upgrades.  At nominal luminosity, the efficiency may decrease by around 50\% for $\pi^0$ reconstruction, while extra energy ($E_{\rm ECL}$) and pulse shape discrimination techniques will degrade in performance. 
    
    \item KLM: The RPCs will be replaced with new scintillator layers to handle high rates, and an overall upgrade to read-out will be considered with better timing resolution.  The inner RPC layers of the KLM may suffer hit efficiency losses of order 10-30\%. While this can have 2-5\% efficiency losses for muons at momenta below 1 GeV$/c$, it may lead to much larger losses in $K_{\rm L}^0$ detection, due to the much lower penetration depth of hadrons through the iron yoke. 
    
    \item Solid angle coverage (e.g. STOPGAP): The current particle identification systems still lack full coverage, such as regions between TOP bars, and the backward endcap. This may adversely affect analyses that require strong vetoes based on particle identification. STOPGAP-like upgrades could remedy this.
\end{itemize}

\subsection{Performance challenges}

The conditions at higher luminosity lead to physics performance degradations. Here we discuss the potential size of these effects. A summary of the physics performance concerns for each subdetector is given in Table~\ref{table:concerns} and described in detail below.
\begin{table}[htbp]
    \centering
    \begin{tabularx}{0.75\linewidth}{X ccccc}
        \hline
    Topic &  \rot{VXD}~ &  \rot{CDC}~ &   \rot{PID}~ & \rot{ECL}~ &  \rot{KLM}~   \\ \hline
    Low momentum track finding & $\checkmark$ & $\checkmark$   & & & \\
    Track $p$, $M$ resolution & & $\checkmark$   & & & \\
    IP/Vertex resolution &  $\checkmark$ & & & & \\
    Hadron ID & &  $\checkmark$   & $\checkmark$ & & \\
    $K_{\rm L}^0$ ID & & & & $\checkmark$ & $\checkmark$\\
    Lepton ID & & $\checkmark$   & & $\checkmark$ & $\checkmark$\\
    $\pi^0$, $\gamma$ & & & & $\checkmark$ & \\
    Trigger &  $\checkmark$ & $\checkmark$ & & & \\
     
\hline
    \end{tabularx}\caption{Key performance requirements vs subdetector upgrades.}\label{table:concerns} 
\end{table}

\begin{itemize}
    \item  Tracking at low momentum (i.e. 50-200 MeV/$c$ slow pions from $D^{*+}$ decays).
    \begin{itemize}
    \item  Upgrades expect to recover the approximately 15-25\% efficiency loss for tracks near 100 MeV/$c$ under nominal beam background levels.
    \item This affects a large range of precision semileptonic, and rare missing energy decays both directly and through impacts on $B$ full reconstruction algorithms. It is key for $V_{cb}$ studies at low hadronic recoil, vetoes in rare decay searches, and hadronic decays. Furthermore improvements to veto suppression power can make substantial statistical precision impacts.
     \end{itemize}

    \item Vertex and IP resolution. 
    \begin{itemize}
    \item Upgrades expect to see up to around 30-50\% impact parameter (IP) resolution improvements over the current system at nominal luminosity, owing to  better mitigation of beam background effects and higher detector resolution.
  \item These improvements reduce crossfeed background from tracks from different $B$ mesons, e.g. in $b\to s \ell\ell$ processes. It is also important for time dependent CP violation measurements with large data sets, where vertex resolution related systematic uncertainties can dominate, e.g. $B\to J/\psi K_S^0$.
    \end{itemize}

     \item  Calorimeter energy resolution and particle identification 
     \begin{itemize}
    \item The detector should be robust against background levels, with reliable rejection of beam background and hadronic clusters using timing and wave form based pulse shape discrimination.
     \item   The proposed ECL upgrades, such as a preshower detector, could recover much of the efficiency and resolution losses for photon reconstruction, particularly $\pi^0$.
     \item Low momentum lepton identification critically relies on calorimeter performance, and is used in the reconstruction of $b\to \tau + X$ channels. Under nominal conditions, cluster energy resolution could degrade by a factor of 2-3 with potential losses in track-cluster matching efficiency and lepton identification efficacy. 
    \item Any new calorimeter detector elements that improve solid angle coverage will improve vetoes in rare and dark sector searches.

    \end{itemize}

    \item Triggers. 
     \begin{itemize}
     \item The CDC and VXD upgrade programs both support track trigger development.
     \item The physics program can exploit high trigger efficiencies and the full Belle II data set, for both precision measurements and rare and forbidden decay searches in low multiplicity, dark sector, and $\tau$ physics.
    \end{itemize}

    
    \item $K/\pi$ separation. 
    \begin{itemize}
    \item It is crucial for a flavour experiment to have a robust and high performance particle ID system, particularly in  hadronic $B$ decays to isolate Cabibbo suppressed transitions. It affects most flavour studies, and  can be a limiting factor where kinematics information is not sufficiently constraining, for example in the separation of $b\to d\gamma$ and $b \to s \gamma$.
    \item A great challenge for the TOP and ARICH systems is robustness under beam background. In the TOP PMT upgrades will ensure QE losses are minimised, while faster more robust photon sensor technologies for the ARICH are being investigated to also maintain efficiencies and to improve timing resolution.
    \item Proposals to improve particle identification acceptance, e.g. STOPGAP, may provide step-change improvements where vetoes are needed. This will impact inclusive analyses such as $|V_{ub}|$ from inclusive semileptonic $B$ decays. 
    \end{itemize}

    \item $K_{\rm L}^0$ detection. 
    \begin{itemize}
    \item $K_{\rm L}^0$ mesons are identified via their interactions in the first few layers of the KLM, and through low energy hadronic showers in the ECL. Without detector upgrades, $K_{\rm L}^0$ detection efficiencies will adversely suffer, which is a significant loss to the physics program. $K_{\rm L}^0$ mesons are difficult to detect at LHCb.
    \item In addition to time dependent CP violation measurements of CP eigenstate modes, improved $K_{\rm L}^0$ detection is critical to step change improvements in background suppression in semileptonic, leptonic, electroweak penguin di-neutrino, and dark sector analyses.
    \end{itemize}

\end{itemize}

\subsection{Physics channels}

There are numerous flagship physics channels that would stand to benefit from the Belle II upgrade scenarios, summarised in Table \ref{table:benchmarks} and discussed in detail below.
 
 \begin{table}[htbp]
    \centering
    \begin{tabularx}{0.75\linewidth}{X ccccccc}
        \hline
    Topic  &  \rot{VXD}~ &  \rot{CDC (incl. Trigger)}~ &   \rot{PID}~ &\rot{PID $\Omega$} & \rot{ECL}~ &  \rot{KLM}~   \\ \hline
    ${\cal B}(B\to \tau \nu, B\to K^{(*)}\nu\bar\nu)$ & $\checkmark$& & & $\checkmark$& $\checkmark$ & $\checkmark$ \\
    ${\cal B}(B\to X_u \ell \nu)$ &$\checkmark$ & & $\checkmark$ & $\checkmark$& & $\checkmark$\\
    $R$, Polarisation$(B\to D^{(*)} \tau \nu)$ &$\checkmark$  & & & & $\checkmark$& \\
    FEI &$\checkmark$  & $\checkmark$& & $\checkmark$ & & \\    
    $S_{\rm CP},C_{\rm CP}(B\to \pi^0\pi^0,K_S^0\pi^0)$ &$\checkmark$  & $\checkmark$& & & $\checkmark$& \\
    $S_{\rm CP},C_{\rm CP}(B\to \rho \gamma)$ &  & $\checkmark$& $\checkmark$ & & $\checkmark$& \\
    $S_{\rm CP},C_{\rm CP}(B\to J/\psi K_{\rm S}^0, \eta^\prime K_{\rm S}^0)$ &$\checkmark$  & $\checkmark$& & & & \\
    Flavour tagger &$\checkmark$ & & $\checkmark$& &  & \\   
    $\tau$ LFV & & $\checkmark$& & & $\checkmark$ & \\   
    Dark sector searches & & $\checkmark$  & & & $\checkmark$ & $\checkmark$ \\

\hline
    \end{tabularx}\caption{Selected key physics channels and high-level analysis algorithms with the subdetector upgrades that would make substantial impacts to measurement reach. The symbol $\Omega$ refers to solid angle coverage of the particle identification systems.}\label{table:benchmarks} 
\end{table}

\subsubsection{Rare and missing energy decays}

Most analyses with missing energy in the final state utilise hadronic or semileptonic $B$ full reconstruction techniques. The performance of these methods is dependent on most key performance factors, most notably low momentum track finding for finding $B\to D^* +nh$, where $n\ge1$ and $h$ denotes hadron. MC simulations indicate efficiency losses of order 30-50\% when comparing early data and nominal luminosity scenarios. It is expected that the upgrade should mitigate these losses.

Rare and leptonic decay searches such as $B\to \tau \nu$ and $B\to K^{(*)}\nu\bar\nu$ rely on $b\to c\to s$ background suppression based on the presence of zero extra tracks in the event, and minimal excess energy in the calorimeter. These analyses often require the detection and veto of $K^0_{\rm L}$ mesons (in the ECL and KLM), as the majority the remaining background in analyses of this type contains undetected $K^0_{\rm L}$. Taking into account potential tracking efficiency loss, impacts of higher beam background in the calorimeter, and losses to KLM hit efficiencies, such analyses would have greatly reduced reach without detector upgrades. The effect could imply a further reduction of approximately 50\% in statistical power, leading to total losses of order 75\%.

\subsubsection{Semileptonic decays}

Semi-tauonic decay measurements, like rare missing energy decays rely on efficient and pure tag side $B$ full reconstruction. On the signal side they require  efficient detection of slow pions for signal efficiencies and low energy photons for signal efficiency, and to constrain feed-down e.g. $B\to D^{*,**}\tau \nu$ down to $B\to D\tau \nu$. The latter reduces correlations between measurements, and potentially large biases in the $B\to D\tau \nu$ channel. The detection of $b \to \tau \to \ell$ transitions requires good lepton identification below 700 MeV/$c$. Calorimeter energy resolution degradation adversely affects electron identification, particularly at lower momenta.  Measurements of $\tau$ polarisation with hadronic tau decays require good control of hadronic $B$ decay background. The ability to use $K_{\rm L}^0$ vetoes may be critical to achieve step change improvements in precision in these measurements. 

With a dataset in excess of 5 ab$^{-1}$ without loss of efficiency, Cabibbo suppressed $B\to \pi\tau\nu$ will also come within reach, probing $b u \tau$ couplings. However a high impact measurement will hinge on controlling statistical power loss, and purity under high beam background.

Inclusive $B \to X_u \ell \nu$ is still partly in tension with exclusive methods and can only be measured by Belle II. This analysis typically requires $nK^0 + nK^\pm =0$. Of the remaining $b\to c$ background, ~43\% have a $K_{\rm L}^0$ (a veto is not typically used due to detector inefficiencies and modelling uncertainties). It also requires effective $D^{*\pm}$ slow pion vetoes. Again, the ability to use $K_{\rm L}^0$ vetoes may introduce a step change in precision for this channel, specifically in controlling background modelling uncertainties. 

Without mitigating beam background effects on efficiencies and resolutions, many semileptonic channels may suffer from statistical power losses of order 50-75\%. In precision studies, control of systematics may be challenging if purity levels degrade, for example measurements of $|V_{ub}|$.

\subsubsection{CP Violation}

Ultimate measurements of time dependent CP violation  require a combination of data set size and high precision on the measurement of $\Delta z$, the distance between the tag and signal $B$ meson decay vertices. Beam background adversely affects flavour tagging efficiencies, hence statistical power, even with a full 2-layer PXD configuration. The effect was estimated to be as much as 10\% efficiency loss even at 2022 luminosity values, and could be much higher at nominal luminosity. Similarly tag-$Z$ resolution will degrade, but can be partly mitigated if cleaner tag side channels are used. This comes at the cost of statistical power. 

The UT angle $\phi_1$ will be systematics limited, demanding improved vertex resolution performance. New sources of CP violation in gluonic penguin decays on the other hand will typically require maximal statistical power, hence will be affected by flavour tagging efficiency losses.

CP violation measurements in Cabibbo suppressed decays, including $\phi_2$, through channels such as $B\to K_{\rm S}^0 \pi^0$ and $B\to \pi^0 \pi^0$ are affected by $\pi^0$ efficiency and resolution as well as flavour tagging efficiency losses. These channels are currently highly limited by statistical power and will be very demanding of the full Belle II statistical power. Losses of order 50-75\% or more due to beam background, in the absence of an upgrade, would be detrimental to reach.

\subsubsection{Radiative decays}

Radiative decays, particularly Cabibbo suppressed modes and inclusive modes require similar or better performance than currently achieved in particle ID and calorimetery at Belle II. A high luminosity golden mode measurement is that of time dependent CP violation in  radiative $b \to d \gamma$ transitions. Particle ID performance is key for suppression of $B \to K^* \gamma$ in the $B\to \rho\gamma$ channel. Similarly, good photon energy resolution is essential for background separation in $\Delta E$. 
Inclusive and exclusive radiative channels additionally require good suppression of neutral hadrons mimicking photons, typically based on pulse shape discrimination.

\subsubsection{Tau and dark sector}
The foremost consideration for high luminosity tau and dark sector physics is trigger efficiency. Both sectors aim to probe forbidden or ultra-rare transitions in low-multiplicity final states with as large a dataset as possible. A system that is not robust enough to efficiently trigger on dark sector and tau physics would lose valuable data for these channels. Many of these processes can only be accessed at Belle II, due to the presence of missing energy or neutrals.

\subsection{Summary}
Most of the Belle II physics program will continue to improve in precision with data set size. Without intervention, the statistical power of analyses may be reduced under nominal background levels. In golden mode measurements of the full Belle II dataset, such as
in missing energy or CP violation rare modes, the proposed upgrades can stand to improve statistical power as much as four fold. This is based on estimates of performance degradation with nominal beam background levels, and detector ageing effects. This is a critical consideration for the experiment as it aims to provide world-leading results results that are complementary to and competitive with LHCb.

\printbibliography[heading=subbibliography]
\clearpage
\end{refsection}

\section{Vertex Detector}
\begin{refsection}
\label{sec:VXD}

\subsection{Summary of VXD replacement options} 
\editor{F.Forti}

The Vertex Detector is particularly subject and sensitive to machine backgrounds. 
Extrapolations to full luminosity are affected by large uncertainties due to limitations 
of the models and by the as yet undefined design of the interaction region. 
Certain sources such as injection background and beam dust events are inherently 
difficult to model, leading to a general requirement of large safety factors against backgrounds, especially for the the VXD. In addition, in case of a major redesign of the interaction region, 
a completely new detector might be required, with the possibility of taking advantage of the more 
recent technology developments, leading to possible performance improvements, such as: 
\begin{itemize}
    \item Better impact parameter and vertexing resolution
    \item Improved tracking performance for low p$_T$ tracks
    \item Longer trigger latency
    \item L1 trigger capabilities
\end{itemize}
The new VXD will occupy the same volume of the current detector, roughly between \SI{14}{\milli\meter} and \SI{135}{mm} in radius. A good but not extreme spacial resolution below \SI{15}{\micro\meter} is required, along with a low material budget below about $0.2\% X_0$ for the inner layers and $0.7\% X_0$ for the outer layers.
To ensure good background robustness the reference radiation levels  used in the design of the systems are the following values for the innermost layer:
\begin{itemize}
    \item Hit rate capability: \SI{120}{\mega\hertz/\cm\squared}
    \item Total ionizing dose: \SI{10}{\mega rad/year}
    \item NIEL fluence: \SI{5 e 13}{n_{eq}/cm^2/year}
\end{itemize}

The DEPFET and Thin Strip proposals keep the current separation in pixels and strip detectors,
providing better performance. The SOI and CMOS MAPS proposals, on the other hand, plan to 
replace the entire VXD with a fully pixelated system.
\subsection{DEPFET}
\label{sec:DEPFET}
\editor{L.Andricek}
\subsubsection{Introduction}
The current PXD detector has been designed and constructed for operation of up to 10 years in Belle II at the luminosity of \SI{8 e 35}{cm^{-2}s^{-1}}
The sensor and the read-out scheme have been optimized in pixel size, thickness of the sensitive layer, overall material budget, integration time, and all other aspects to meet the challenging requirements at a collider operating at the intensity frontier.
As far as the sensor occupancy is concerned, the PXD has a safety factor of 3 for the well understood luminosity background contribution, and $>$10 for the less well manageable beam related background contributions. 
Since its installation and commissioning in Belle II, the PXD system is on operation and performs up to expectation. However, with progressing data taking and ramp-up of the luminosity, PXD experienced a few unexpected issues with the sensor and with one of the ASICs.
\begin{itemize}
\item Beam incidents lead not only to damaged collimators of the beam line but also to an increasing number of inefficient or dead gate lines (“rows”) on almost all PXD modules. The root cause was traced back to a high photocurrent in an on-chip voltage regulator of the Switcher chip due to the high instantaneous dose.

\item On the sensor side, the TID received in the bond oxide between the sensor part of the SOI wafer and the support wafer results in an electron accumulation layer between at the p-type guard rings on the back side of the sensor. The resulting high electric field leads to an avalanche current from the back side contact to the bulk node at the edge of the top wafer. This current does not compromise the sensor performance but requires a redesign of the power supply to provide enough current.

\item A more system related issue is the performance of the “gated mode” of the PXD. Although the DEPFET sensors were not designed for this operating mode, the PXD can nevertheless be operated to blind the sensors during the passage of the injected bunches. However, the achievable minimal length of the blind phase is about \SI{2}{\micro\second} due to ringing of the pedestal baseline upon restoring of the normal operation mode. To minimize the deadtime, it would be desirable to reduce the gate length to below \SI{1}{\micro\second}.
\end{itemize}

\subsubsection{Upgrade Proposal}
The aim of the proposed upgrade of the PXD is to provide a higher safety factor for the allowed occupancy and to remove the root cause of the previously described issues. To minimize risk and costs of the project, the proposal is to preserve the general layout of the PXD system. This includes as far as possible the module and ladder design and construction, mechanical support, cooling, services, backend electronics, DAQ, and slow control. The installation of the upgraded PXD will profit from the experience acquired with the current one. 

Parts of the system to be upgraded are the sensor, the DCD, DHP, and Switcher ASICs. The aim of the proposed upgrade of the PXD is to provide a higher safety factor for the allowed occupancy and to remove the root cause of the previously described issues.

\begin{itemize}
\item Due to the rolling shutter read-out mode of the sensor with the integration of two superKEKB revolutions in one frame (\SI{20}{\micro\second}), the sensor occupancy scales linearly with the read-out speed of the sensor. The current frame rate translates to 100~ns per read-out row. The time is limited by the settling of the signals needed to address and clear the row as well as the time required to digitize and process the read-out signal current. We propose to reduce the read-out time per row from \SI{100}{\nano\second}  to \SI{50}{\nano\second}. By keeping the optimized pixel size and number of the current PXD the same, the frame time and the background occupancy would be reduced by factor 2. For this, signal transmission on the pixel matrix as well as signal processing in the read-out ASICs needs to be improved.

\item The robustness against beam losses or other incidents will be greatly enhanced by adding protection circuits in the on-chip voltage regulator of the Switcher.

\item The TID effect in the bonding oxide leading to the unexpected avalanche current would be eliminated by removing the SiO$_2$ where the guard rings are at the backside of the sensor. A second approach is to provide a negative bias to the handle wafer in order to compensate the accumulated radiation induced oxide charge. 

\item The current PXD sensors are not specifically designed for gated mode operation. By optimizing the DEPFET geometry and implantation parameters towards shorter and narrower gate length, the bias voltage needed for gated mode operation will be reduced. Lower operating voltages for the clear node together with the already mentioned improvements of the signal transmission properties on the matrix including integrated termination resistors will improve the gated mode operation.
\end{itemize} 

The most time consuming R\&D is on the sensor side where both the DEPFET pixel cell as well as the on-module interconnect have to be improved. Simulations of the new DEPFET pixel layout with reduced gate length and width and optimized doping profiles of the drain implants show extremely promising results. The internal amplification is enhanced by  more than a factor of 2 to 1.3~nA/e-. At the same time, the required voltage for a complete clear was cut in half to about 10 V. After the experimental confirmation of the simulation results by evaluation of the ongoing test production, a dedicated test run with large pixel sensors and improved on-chip interconnect will be launched. The development of the new ASIC versions can be done in parallel.

\subsubsection{Cost Estimate and Person Power}

For a DEPFET based PXD upgrade, most of the existing support and mechanical structures can be re-used. The biggest contributions to the upgrade costs are new sensors and a new generation of read-out and control ASICs. Based on the experience of the current PXD and taking the yield and required contingency into account, 18 sensor wafers would be required of a PXD upgrade, yielding an overall cost of about 2 M\$ for the sensor part. Production of commercial CMOS ASICs for read-out and control, interconnect, and auxiliary electronics would add another 2~M\$. 

At the time of writing there are only two institutions within the Belle II collaboration interested in contributing to a DEPFET based PXD upgrade. While these two institutions can cover a large part of the sensor and ASIC development and production, both are open for new collaborators to join the effort. 

\subsection{Thin Strips}
\label{sec:TFP}
\editor{K.Nakamura}

The Double-sided Silicon Strip Detector (DSSD) is a prime candidate for a tracking device in the inner and middle detector volume since a single sensor can cover a large dimension of about 100$\times$100~mm$^{2}$ and the technology of the sensor production is already well established. In the existing Belle~II detector system, the DSSD technology is used in the SVD, which performs the vertex measurement and low-momentum tracking together with the PXD. Still, there are rooms for the improvements that are desirable for the future Belle~II operation. Also, it is expected that the CDC hit rate may exceed the limit to maintain the detector performance in the future beam background condition. The TFP-SVD (Thin and Fine-Pitch SVD) is a new detector concept featuring the DSSD, proposed to improve the SVD, as well as the CDC by replacing its inner part.

\subsubsection{Detector concept and improvements}

One improvement in the TFP-SVD is the reduction of the material budget. The small material budget improves the track reconstruction performance especially for the low momentum tracks. It also helps to improve the vertex measurement using $K_{\rm S}^{0}$ mesons decaying into two charged pions. The material budget of the current SVD is about 0.7\%$X_{0}$ per layer with the two largest contributors being the silicon sensor (about 0.3\%$X_{0}$) and flexible circuits (about 0.2\%$X_{0}$). The TFP-SVD reduces the sensor thickness to 140~\si{\micro\metre} to cut down the material budget by 0.19\%$X_{0}$. We also plan to replace the layers of copper with aluminium on the flexible circuits, which reduces the material budget by more than 0.1\%$X_{0}$. 

The small sensor thickness also reduces the voltage necessary for full depletion which will increase due to the radiation damage after the type inversion, because the full depletion voltage is proportional to square of the sensor thickness. Based on a relation between the full depletion voltage and the radiation damage~\cite{NIEL:R.Wunstorf:1992}, the necessary voltage after a 1-MeV equivalent neutron fluence of $2.5\times10^{13}$~neq/cm$^{2}$ is expected to be less than 40~V. Therefore, the acceptable neutron fluence of the detector system will be improved.

The hit rate limit of the TFP-SVD is expected to be similar to the current SVD limit, about 4~MHz/cm$^{2}$, assuming the similar detector geometry and current track reconstruction. The limits of both detectors will be improved by the further optimizations of the hit time filtering that rejects the background hits. For the improvement, the shorter signal duration of the shaping circuit possibly becomes important to avoid performance degradation due to the pile-up effect which may be sizable in a hit rate higher than 4~MHz/cm$^{2}$. The signal duration at a quarter of the signal height is expected to be about 55~ns in the TFP-SVD, while is more than 200~ns in the current SVD. The final hit rate tolerance with the optimization of the hit time filtering is to be evaluated with simulation studies, which are now under preparation.

The front-end ASIC of the TFP-SVD, so-called SNAP128, has 128 input channels, and generates the binary hit information sampled with a 127~MHz clock in each channel. The digitization in the SNAP128 itself offers a reduction of the amount of cables, which is necessary to fit them in the limited cable space. The 127~MHz binary hit information is stored in 2k-depth ring memory. When the SNAP128 receives the level-1 trigger, it outputs the data in the memory at the addresses corresponding to the trigger timing. The output data of the TFP-SVD hit information are recorded by the data acquisition system.

The TFP-SVD offers a new level-1 track trigger, which is generated by the TFP-SVD hit information. The SNAP128 also outputs another data for the trigger decision. The data are generated by applying the logical OR operation to the 127~MHz binary information of the 128 input channels. We name it as the VXD level-1 trigger. The new trigger finds the tracks with a production point resolution of about 1~cm, which is better than the existing CDC track trigger. The extraneous tracks due to the beam background originate close to the interaction point, resulting in frequent fake CDC triggers. They can be efficiently rejected by the new VXD level-1 trigger. The background rejection reduces the prescale factors of the track triggers, and thus improves the trigger efficiency especially for the low-multiplicity physics events such as the tau-pair production and some of dark sector searches.

Another attractive option is the replacement of the inner part of the CDC with the TFP-SVD. The CDC hit rate may exceed the limit to maintain the detector performance in the future beam background condition. The replacement with the silicon strip sensor can improve the tracking performance of the future Belle~II detector system. 

\begin{figure}[htbp]
  \centering
  \noindent
  \begin{minipage}{0.5\textwidth}
  \begin{annotationimage}{width=7cm}{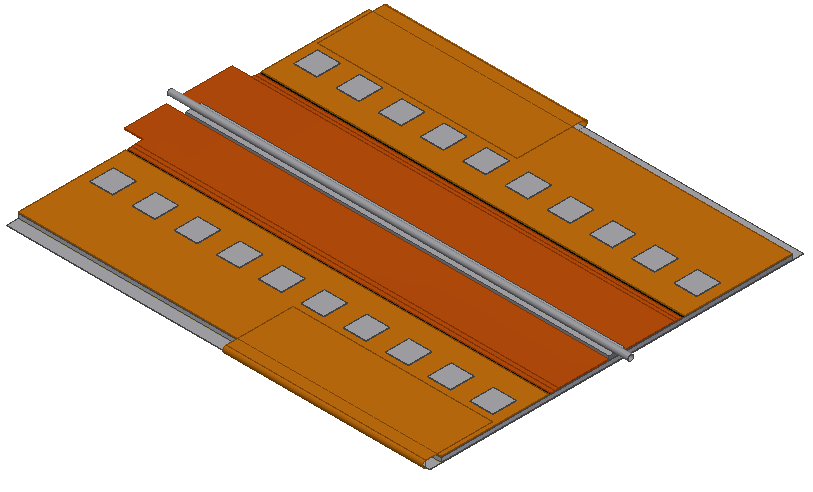}
  \imagelabelset{
  coarse grid color = red,
  fine grid color = gray,
  image label font = \sffamily\bfseries\small,
  image label distance = -3mm,
  image label back = none,
  image label text = black,
  coordinate label font = \sffamily\bfseries\scriptsize,
  coordinate label distance = 0mm,
  annotation font = \normalfont\small,
  arrow distance = 1.5mm,
  border thickness = 0.6pt,
  arrow thickness = 0.4pt,
  tip size = 1.2mm,
  outer dist = 0.5cm,
  }
  \draw[image label = {a) at north west}];
  \end{annotationimage}
  \end{minipage}
  \begin{minipage}{0.38\textwidth}
  \begin{annotationimage}{width=2.5cm}{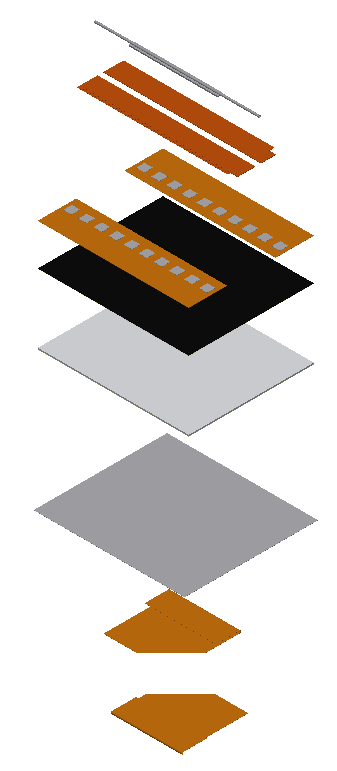}
  \imagelabelset{
  coarse grid color = red,
  fine grid color = gray,
  image label font = \sffamily\bfseries\small,
  image label distance = -2mm,
  image label back = none,
  image label text = black,
  coordinate label font = \sffamily\bfseries\scriptsize,
  coordinate label distance = 2mm,
  annotation font = \normalfont\small,
  arrow distance = 1.5mm,
  border thickness = 0.6pt,
  arrow thickness = 0.4pt,
  tip size = 1.2mm,
  outer dist = 0.5cm,
  }
  \draw[image label = {b) at north west}];
  \draw[annotation right = {Cooling pipe at 0.9}] to (0.55,0.9);
  \draw[annotation right = {Long FPC w/ Al layers at 0.8}] to (0.75,0.8);
  \draw[annotation right = {FPC and SNAP128 at 0.7}] to (0.85,0.7);
  \draw[annotation right = {Graphite sheet at 0.6}] to (0.75,0.6);
  \draw[annotation right = {Insulation sheet at 0.48}] to (0.70,0.48);
  \draw[annotation right = {DSSD sensor at 0.32}] to (0.85,0.32);
  \draw[annotation right = {Pitch-adapter at 0.18}] to (0.65,0.18);
  \draw[annotation right = {Pitch-adapter at 0.08}] to (0.65,0.08);
  \end{annotationimage}
  \end{minipage}
  \caption{a) Design of the TFP-SVD module. b) Exploded view of the module.}
  \label{fig:tfp_svd_module}
\end{figure}
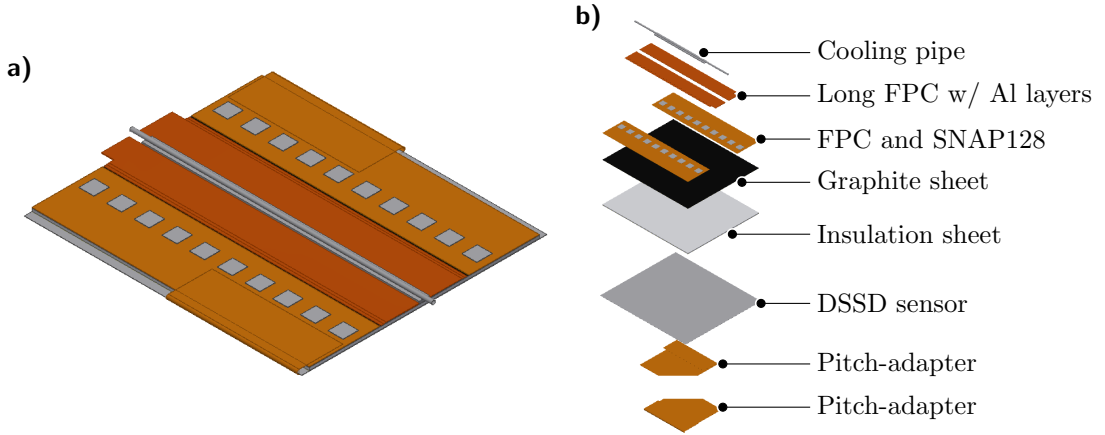

The module design and its exploded view are shown in Fig.~\ref{fig:tfp_svd_module}. Two to ten modules are aligned in a line to construct a ladder, and the ladders are cylindrically integrated to assemble the TFP-SVD. In each module, the strips of the DSSD sensor are connected to the SNAP128 chips via pitch adapters. The SNAP128 chips are mounted on the flexible printed circuits (FPC) which is also connected to another long FPC to communicate with the back-end electronics and receive electric power from the power supplies. The power and ground layers of the long FPC are made of aluminum to reduce the material budget. The cooling pipe is thermally connected to SNAP128 via a thermal-conductive pyrolytic graphite sheet, which has a thickness of 0.1~mm and a thermal conductivity of 700~W/mK, to absorb the heat dissipation from the chips. The DSSD sensor and SNAP128 are thermally isolated with an insulation sheet.

\subsubsection{Technology development}

The realization of the thin DSSD with a sensor size of about $100\times100$~mm$^{2}$ is one of the key challenges in the TFP-SVD project. One concern is the feasibility and yield of the sensor production. Another is the small signal charge due to the short path length of the particle in the sensor. 

In our project, the prototype DSSD sensors with 140~\si{\micro\metre} thickness were successfully produced by Micron Semiconductor Ltd (UK). The prototype sensor has a size of $52.6\times 59.0$~mm$^{2}$ and strip pitches of 50~\si{\micro\metre} and 75~\si{\micro\metre} in the P-side and N-side, respectively. The measured I-V and C-V curves are consistent with the expectation, and the full depletion voltage of the thin sensor is confirmed to be $14\pm 1$~V. We plan to increase the sensor size to about $100\times100$~mm$^{2}$ in the next prototype version.

The electrical noise in the readout signal must be enough small to keep a high signal-to-noise ratio and thus hit efficiency. The requirement for the electrical noise on the strip signal is less than 1000 ${\mathrm e}^{-}$. The SNAP128 is being developed to achieve the small strip noise, keeping the short signal duration and moderate power consumption. The 180~\si{\micro\meter} process technology is adopted. The SNAP128 chip has 128 input channels. The power consumption is about 350~mW per chip.

The analog circuit of each input channel consists of a charge sensitive amplifier with pole-zero cancellation and a CR-RC shaper. The shaper is optimized so that the output pulse has about 55~ns pulse width at quarter of the signal height. The shaper output is digitized to the binary hit information with a discriminator, in which the threshold is adjusted by an internal 8-bit DAC. The 127~MHz system clock samples the binary information to convert it to 127~Mbps hit data. Thus, the hit time resolution should be similar to the clock period 7.9~ns. The polarity of the discriminator is selectable to deal with both the positive and negative signals in the p-side and n-side strips.

The digital part of each channel has a 2k-depth memory, which holds the binary hit information up to about \SI{16}{\us} (= 1/127~MHz$\times$2048) until the chip receives the trigger signal. When the trigger signal arrives, the binary information in the corresponding time window is read out from the memory. The size of the time window is adjustable. Since the trigger timing has a jitter of $O(10~{\rm ns})$ in Belle~II, the time window needs to cover the timing fluctuation. The readout binary data from the 128 channels are serialized and sent as the SNAP128 output as a 127~Mbps output data.

In addition to the output that sends the 128-channels hit data per trigger reception, the SNAP128 outputs online hit information in the SNAP128 level, which is supposed to be used for the level-1 trigger generation. The 127~Mbps comparator output in each channel is converted to another 127~Mbps data by keeping only the rising-edge information. (For instance, if the comparator output is '00011110...', it is converted to '00010000...'.) The converted 127~Mbps data of all 128 channels are OR'ed to represent the chip-level hit information, and used to extract the track information in the VXD volume for the level-1 trigger decision.

\subsubsection{Human resources and necessary development}

Currently, IISER Mohali, PU Chandigarh, Kavli IPMU, KEK, TIFR, and University of Tokyo are the primary contributors to the TFP-SVD project. Still, there are several development to be done. For example, the detailed designing of the detector, the development of the back-end electronics, the development of the trigger algorithm and its firmware implementation, and the data taking software development. 

\subsubsection{Cost estimation}

A rough estimation of the necessary cost for the TFP-SVD production is listed in Table~\ref{tab:tfp_svd_cost}. The numbers are based on the experience of the Belle~II SVD development and production.

\begin{center}
\begin{table}[h]
\caption{Order of magnitude cost estimation for the TFP-SVD production}
\centering
\begin{tabular}{ | l | l | l | c | } \hline
{\bf Item} & {\bf Unit price (k\$)} & {\bf Estimated number} & {\bf Cost (k\$)} \\ \hline
 DSSD sensors    & 5 /sensor  & 350 sensors & 1,750 \\ 
 ASIC production & 200 /batch & 2 batches   & 400 \\
 Detector parts  & 1 / sensor & 350 sensors & 350 \\
 Detector support &           &             & 100 \\
 Power supply    & 500 /unit  & 1 unit      & 500 \\
 Readout electronics & 10 /board & 60 boards & 600 \\
 Cables          &            &             & 200 \\
 Servers         &            &             & 100 \\ \hline 
 {\bf Total } & & & {\bf 4,000} \\ \hline
\end{tabular}
\end{table}
\label{tab:tfp_svd_cost}
\end{center}

\subsection{CMOS MAPS}
\label{sec:MAPS}
\editor{C.Marinas, J.Baudot}

The VTX proposal consists in exploiting demonstrated performance of current CMOS monolithic active pixel sensor (MAPS) to replace the VXD system with 5 detection layers uniformly equipped with the same fast, light and granular sensor. The VTX solution reduces by at least two orders of magnitude the current PXD integration time and replaces the double-sided strip sensors of the SVD by real 2D-pixel detectors, with the following expected benefits:
\begin{itemize}
\item Ability to cope with larger background levels than the current VXD, thanks to a low detector occupancy alleviating any data transmission bottleneck and
allowing a resilient track finding efficiency with respect to hit-rate.
\item Improved momentum resolution and impact parameter resolution at low transverse momentum, performance being robust against scaled backgrounds. In general, the tracking chain is simplified with all layers involved in the track finding algorithm, allowing also for natural tracking alternatives (inside-out tracking).
\item Simplified and lighter services due to smaller cross sections of the data cables and less complex cooling system. The detector integration is expected to be simple allowing an installation over one long summer shutdown. Fewer material in front of the endcaps gives additional room for extra shielding in the interaction region and bellows to protect the outer detectors. Current machine-detector interface boundaries can be kept or slightly changed if required by modifications of the final focusing system. 
\item Direct connection to the standard Belle II data acquisition path due to the low transmission bandwidth and input source to the high level and level 1 triggers. Completely standalone 5-layer tracker available for pattern recognition.
\item Operation are simplified since no special mode of operation (e.g. Gated Mode) nor data reduction (e.g. ROI selection) are required. Due to the usage of the same sensor all over, the control and power system becomes unique.
\end{itemize}
Though all layers use the same sensor, their designs differ due to their active length varying from 12 to 70 cm to cover the required acceptance. While the two innermost layers follow an all-silicon ladder concept (iVTX), the outermost (oVTX) layers require carbon fiber support structures and flex print cables.
The VTX targets an overall material budget in the range 1.5-2.0\%~$X_0$ and will be the first light and compact MAPS-based vertex-tracker operating at an high-rate $e^+e^-$ collider.  

\begin{figure}[htb]
\begin{center}
\includegraphics*[width=12cm]{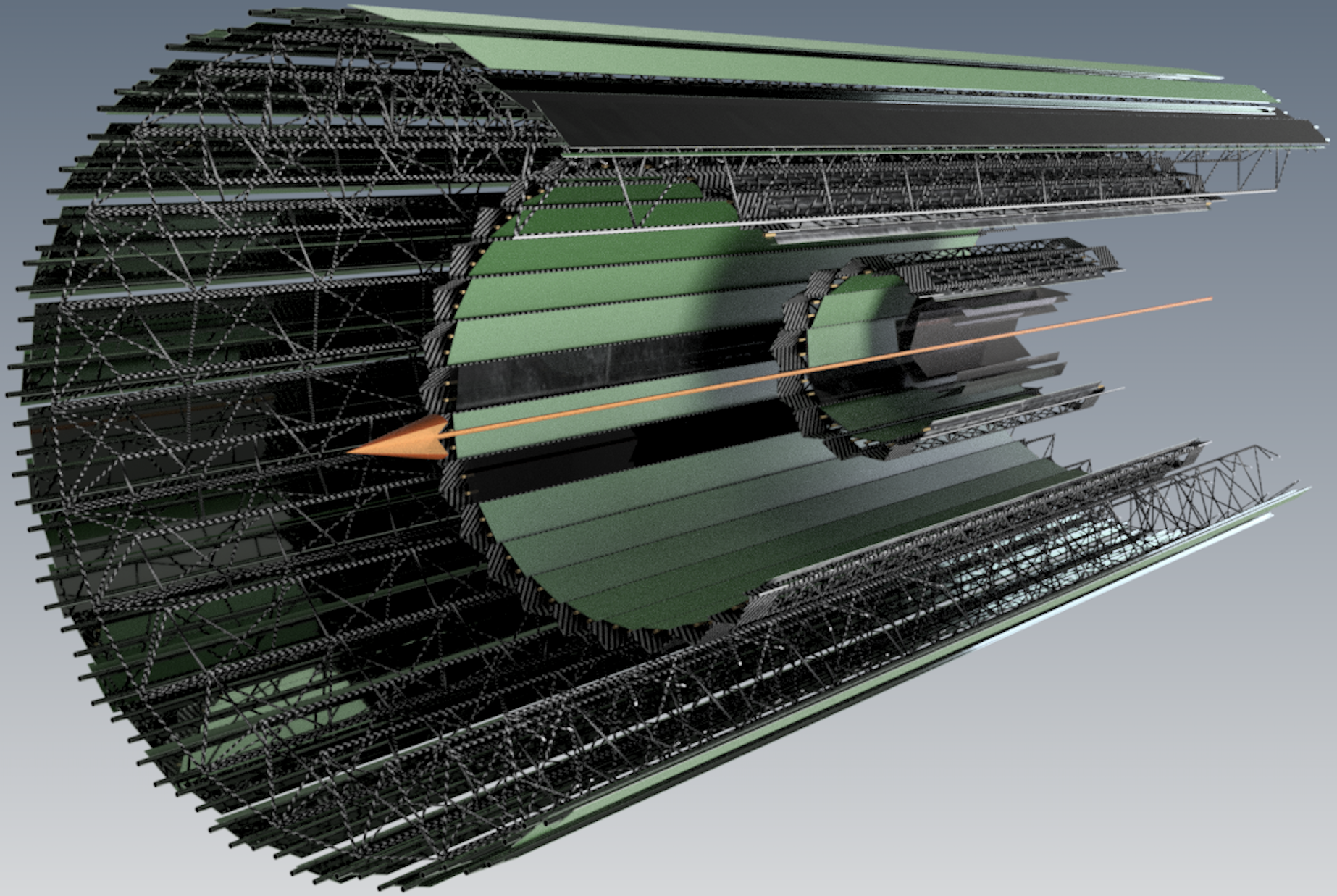}
\caption{Overall VTX layout.}
\end{center}
\label{figure:VTX}
\end{figure}

\subsubsection{Technological solutions}

The CMOS sensor for the VTX project relies on the CIS-180 nm process by Tower Semiconductor 
(https://towersemi.com/technology/cmos\_image\_sensor/). Intense R\&D has been carried out with this process since about ten years in the high-energy physics community and lead to major successes \cite{aglieri_rinella_alpide_2017,bespin_dmaps_2020,deveaux_observations_2019}. Especially the depleted MAPS TJ-MONOPIX-2 \cite{moustakas_design_2021} offer essential features already matching the requirements for Belle II VXD sensors.
\begin{itemize}
\item The sensitive volume is made of an epitaxial layer with thickness between 18 and 40~$\mu$m and a
high resistivity (in excess of 1 k$\Omega$.cm) for a sizeable signal generation and an additional shallow
layer (so-called modified process) to ensure full depletion, which is critical for radiation
tolerance.
\item Four wells and six metal layers are available, allowing for a design of complex circuitry within the
pixel and over the matrix, thus facilitating the implementation of advanced read-out
architectures needed to cope with the high data rate expected.
\item The maximal dimensions of the process reticle implemented on 8” wafers reaches $25\times 31$~cm$^2$ matching well the expected size for the VTX sensor of about $20\times 30$~cm$^2$.
\item The thinning of MAPS is actually one of their key benefits. The possibility to thin sensors down to about
60 or 50~$\mu$m has been demonstrated in various projects and for vast quantities of chips over the past 10
years.
\end{itemize}
The sensor dedicated to the Belle II VTX is called OBELIX and its main design features are summarised in table \ref{table:OBELIX}. \\

\begin{center}
\begin{table}[htb]
\caption{Design features of the OBELIX sensor.}
\centering
\begin{tabular}{ | l | r | } 
\hline
Pixel pitch & 30 to 40 $\mu$m \\
\hline
Matrix size & 512 rows $\times$ 928 to 752 columns \\
\hline
Time stamping & 25 to 100 ns precision over 7 bits \\
\hline
Signal Time over threshold & 7 bits \\
\hline
Output bandwidth & 320 to 640 Mbps \\ 
\hline
Power dissipation & 100 to 200 mW/cm$^2$ \\
\hline
Radiation tolerance & 100 MRad and $10^{14}$ n$_{\mathrm{eq}}$/cm$^2$\\
\hline
\end{tabular}
\label{table:OBELIX}
\end{table}
\end{center}

The two innermost layers, namely iVTX, targets an individual material budget of about $0.1$\%~$X_0$. Such light layer is possible due to the moderate overall surface of these layers, below 400~cm$^2$, the low sensor power dissipation and the few connections needed for the sensor operations. First such conditions makes air cooling a viable solution. Then in addition, the all-silicon ladder is essentially made of four contiguous OBELIX sensors diced out of production wafers thinned to 50~$\mu$m except in some areas still featuring few hundreds of $\mu$m thickness to insure mechanical stability. A post-processing step etches additional metal strips to interconnect sensors along the ladder and provides a unique connector at the ladder end.\\
   
\begin{figure}[htb]
\begin{center}
\includegraphics*[width=12cm]{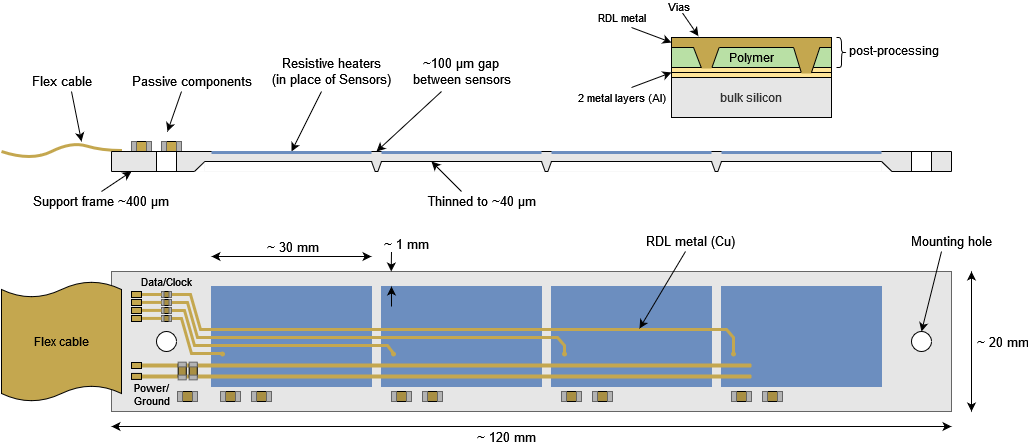}
\caption{Schematic view of the iVTX ladder design.}
\end{center}
\label{figure:iVTX}
\end{figure}

A more traditional approach drives the oVTX ladder concept for outer layers, strongly inspired by the design successfully developed for the ALICE ITS2 \cite{fantoni_upgrade_2020}. Each ladder is made of a light carbon fiber support structure, a cold plate including tubes for water leakless circulation, two rows of sensors glued on the cold plate and finally two flex print cables connecting each half the sensors to a single end-ladder connector. Depending on their radius, the material budget of individual oVTX ladders ranges from 0.3 to 0.5\%~$X_0$. \\

\begin{figure}[htb]
\begin{center}
\includegraphics*[width=12cm]{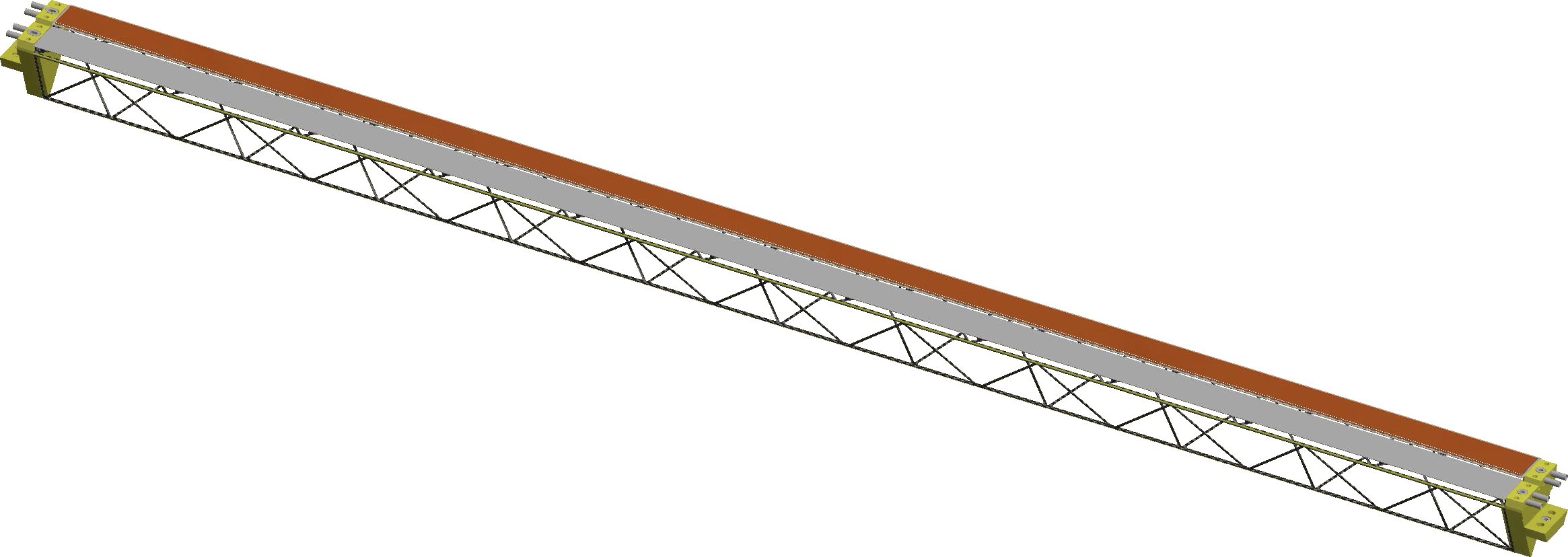}
\caption{Schematic view of the oVTX ladder design.}
\end{center}
\label{figure:oVTX}
\end{figure}

The proposed VTX detector is expected to be operated at room temperature. The low sensor occupancy leads to moderate data rates to be extracted. These aspects result in a dramatic reduction of services with respect to the current VXD. Consequently, similar or simpler overall mechanical support, cabling and acquisition system are required and are not further described here. In particular, the standard PCIe40 \cite{daq:pcie40} acquisition boards used in Belle~II match well the data throughput requirement.

\subsubsection{Design work and schedule}

Assuming a potential installation for the new VTX subsystem in 2026, the coarse project schedule is to develop all technical solutions till early 2023, then produce all parts and assemble them to have a complete detector in early 2026. Additional 10 to 12 months will be needed to ship all parts to KEK and perform the required re-assembly and tests before inserting the complete instrument into the Belle II system. Though the deadline for the development phase might appear ambitious, it has to be noted that numerous solutions are already pre-existing as described below.\\

The OBELIX sensor is based on the pixel matrix of the TJ-MONOPIX-2 chip, which is itself an optimised successor of an earlier version TJ-MONOPIX-1 \cite{bespin_dmaps_2020}. 
 The design of the first CMOS MAPS dedicated to Belle~II, OBELIX-1, has started in 2021 and targets a fabrication mid-2022, allowing for the full characterisation by 2023. The main design work to complete OBELIX-1 consists in extending the TJ-MONOPIX-2 pixel matrix, adding power regulators for easier power distribution system and of course develop an end-of-column digital adapted to the Belle~II triggering scheme. In addition, simulation and test studies on the front-end used in TJ-MONOPIX-2 shall indicate how much power can be spared if the integration time is enlarged. Currently, the front-end allows hit-tagging at 25~ns, while physics simulations demonstrated that 100~ns ({\it reference to section where simulation performance described}) is sufficient with respect to the expected hit rate. Operating the front-end at a lower current and/or lower frequency will lower the power dissipation toward 100~mW/cm$^2$ compared to the expected current baseline of 200~mW/cm$^2$.\\
 
 The iVTX full-silicon concept is currently being assessed with industrial partners, first using dummy silicon wafers and in the near future with real sensors. Key parts of the oVTX ladder, the longest carbon fiber support structure and flex cable have been prototyped and fabricated. Their tests shall allow a validation of the full ladder design. For both concepts initial proof of concept results are expected in 2022.
 
\subsubsection{The VTX collaboration}

The VTX proposal encompasses the main critical hardware aspects of the development required for a vertex detector: sensors, detection layers, acquisition system. The work is currently conducted by a still open collaboration of seventeen groups from five countries: Austria (HEPHY-Vienna), France (CPPM-Marseille, IJCLab-Orsay, IPHC-Strasbourg), Germany (UNiversity of Bonn, University of Dortmund, University of Goettingen, KIT-Karlsruhe), Italy (University of Bergamo, INFN-Pavia, INFN \& University of Pisa), Spain (IFAE-Barcelona, IMB-CNM-CSIC-Barcelona, IFCA-CSIC-UC-Santander, IMSE-CNM-CSIC-Seville, IFIC-CSIC-UV-Valencia, ITAINNOVA-Zaragoza).

\subsubsection{Cost estimation}

Based on the experience from other projects exploiting the CMOS-MAPS technology, especially the recent ALICE ITS2 \cite{fantoni_upgrade_2020}, an initial rough cost for the VTX was established. This estimation assume a VTX with five detection layers, corresponding to 16 iVTX ladders, 64 oVTX ladders and about 2200 CMOS sensors. Table \ref{tab:VTXcost} breaks down the various component costs (without the beam pipe) including the development and production phases.

\begin{center}
\begin{table}[htb]
\caption{Initial cost estimation in k\$ for the VTX project}
\centering
\begin{tabular}{ | l | r | r | r | } \hline
{\bf Component} & {\bf Development } & {\bf Production} & {\bf Total (k\$)} \\ 
\hline
 CMOS sensors & 600 & 900 & 1,500 \\ 
 Ladders & 200 & 600 & 800 \\
 Assembly & 100 & 500 & 600 \\
 Mechanics & 50 & 150 & 200 \\
 DAQ \& Services & 300 & 1,000 & 1,300 \\
\hline
 {\bf Total } & 1,250 & 3,150 & {\bf 4,400} \\ 
\hline
\end{tabular}
\label{tab:VTXcost}
\end{table}
\end{center}

\subsection{SOI}
\label{sec:SOI}
\editor{A.Ishikawa}
To cope with the higher background at upgraded SuperKEKB with a luminosity of $4 \times 10^{36}$cm$^{-2}$s$^{-1}$, DuTiP~(Dual Timer Pixel) sensor concept was invented~\cite{Ishikawa:2020gpo} based on PIXOR striplet detector~\cite{Ono:2013rka} to replace whole VXD system.
This DuTiP concept with a Silicon on Insulator~(SOI) technology can be also used for VXD upgrade around 2026.

\subsubsection{Concept}
Under the higher background environment, a fast vertex detector is required to reduce the occupancy. 
A global shutter readout mode based on level~1 trigger should be adopted to realize faster readout, lower occupancy, smaller data size and smaller data transfer rate. 
Hits should be stored somewhere in the detector during the trigger latency of 4.5~$\mu$s at Belle~II, and another hit in the same pixel during the trigger latency is desired to be kept to avoid information loss for inner layers at high background environment. 
To satisfy these requirements, DuTiP concept was invented. 

\begin{figure}[htbp]
\begin{center}
\includegraphics[width=1.0\textwidth]{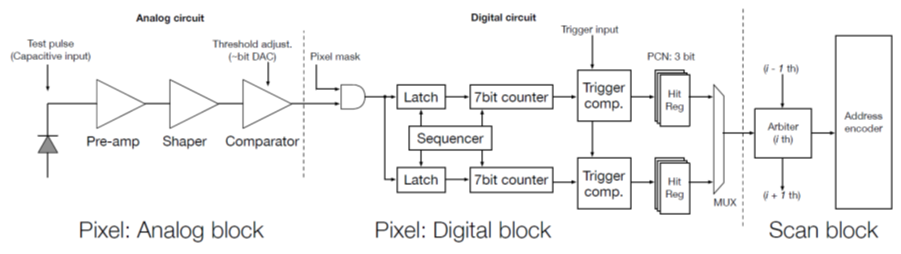}
\end{center}
\caption{Analog, Digital and Scan blocks for DuTiP detector. The analog and digital circuits are on a pixel while the Scan block is on a periphery.}
\label{fig:DuTiP}
\end{figure}
Figure~\ref{fig:DuTiP} shows the block diagram for the DuTiP. 
The analog part consists of a pre-amplifier, shaper, and comparator which are usual configuration for the binary detector. 
When the binary hit signal is sent to digital part, one of the timers starts counting down. 
The starting time is set as trigger latency plus one clock. 
If the trigger signal is received when the time is 1~(2/0), the signal is readout as current~(next/previous) timing~(PCN timings). 
If trigger signal is not received at the PCN timings in the pixel, the timer is reset. 
To take into account the second hit in a pixel during trigger latency, a sequencer and two timers are equipped. 

The concept does not specify the technology however the complicated digital circuit should be fabricated on each pixel, thus an SOI technology is chosen as a baseline for the development of DuTiP detector. 

\subsubsection{SOI Technology}
The SOI technology is suitable for pixel vertex detector for elementary particle physics due to monolithic, thin, low power, and low parasitic capacitance.
We choose the Lapis semiconductor 0.2~$\mu$m FD-SOI CMOS technology for the development since many pixel detectors are already developed, for example, SOFIST for the ILC~\cite{Ono:2019yrh}, and pixel detectors for CEPC~\cite{Wu:2019nqx} and CLIC~\cite{Bugiel:2018ckn, Bugiel:2021nag}. 



The active merge technique can reduce the circuit area, for example, the area of a standard D-type flip-flop becomes 16.5\%~\cite{Arai}.
The SOI is tolerant against neutrons and single event upset. 
The most important issue is TID. Charged particles create holes in the SiO$_2$ insulator layer which causes the backgate effect. 
This had been solved by double SOI structure up to 1~MGy~\cite{Hara:2019dfo}.

\subsubsection{Sensor Design}
We choose pixel size of $45~\mu$m and a sensor layer thickness of $50~\mu$m,
which gives about $11~\mu$m of intrinsic resolution in $z$ direction averaging over incident polar angle.
For analog circuits, ALPIDE analog circuit with some modifications for optimizing to the SOI is adopted~\cite{Suljic:2016bmm}. To achieve faster shaping time, the supply current is increased to 200~nA which gives about 500~ns pulse width for the output signal. The equivalent noise charge~(ENC) is expected to be less than 100 electrons. 
The clock in the DuTiP digital circuit is determined by SuperKEKB clock 509~MHz divided by $32$, 15.9~MHz~(62.9~ns). To hold hit signals during the trigger latency of 4.5~$\mu$s, two 7~bit counters are equipped which allow at most 8~$\mu$s trigger latency with 15.9~MHz clock.

The size of the SOI mask for the Lapis Semiconductor is $ 24.6 \si{\mm} \times 30.8 \si{\mm}$ which determines the size of one chip. Table~\ref{tab:chip} summarizes the size of small and large chips. 
The periphery circuits are collected to one edge of the sensor and the width of 2 mm is expected. 
By using stitching technique, multiple chips can be combined into a single one. 
The stitching technique had been established with a SOPHIAS X-ray chip with the boundary width of 10~$\mu$m. 

\begin{table}[htb]
\centering
\caption{The size of Small~(S) and Large~(L) DuTiP chips.}
\label{tab:chip}
\begin{tabular}{c|c|c|c|c|c|c}
\hline
sensor & layer & pitch    & row $\times$ column  & array $r$-$\phi$ $\times$ $z$ & array area & chip $r$-$\phi$ $\times$ $z$\\
type     &  & [$\mu$m] &   [pixels]   & [mm$^2$]                      & [cm$^2$] & [mm$^2$] \\
\hline
 S & 1--3 & 45       &  320    $\times$     640            & 14.4 $\times$ 28.8 & 4.15 & 17.2 $\times$ 29.6 \\
 L & 4--7 & 45       &  480    $\times$     640            & 21.6 $\times$ 28.8 & 6.22 & 24.4 $\times$ 29.6 \\

\hline
\end{tabular}
\end{table}

DuTiP readout circuit is designed so that the physics hit signal is efficiently acquired even at the high background rate of $> 113$~MHz/cm$^2$ and high trigger rate of 100~kHz. 
We already designed 1.8V LVDS and PLL circuits with a speed of 300MHz~(600Mbps) which will be fabricated in the periphery of DuTiP and can transfer binary information from chip to outside.

Issuing input signals to the track trigger system is desired for the outer layers of VXD system.
Since the DuTiP is a binary detector, digital OR can be easily taken with the circuit on the pixel array. 
For an L type chip, the circuit can be fabricated. 



\subsubsection{Mechanical Structure}
The geometry of the DuTiP pixel detector is designed to cover the VXD acceptance with 7-layer ($14{\rm mm} < r < 135{\rm mm}$ and $17 \si{\degree} < \theta <  150 \si{\degree}$ ).  
The layers 1-3~(4-7) will be covered with the S~(L) type sensors~(Tab.~\ref{tab:chip}). 
With the stitching technique, longer chips in the $z$ direction can be produced to minimize the dead region between chips in a ladder. 

The detailed structure of the ladders has not been determined yet.
However, we plan to use the peripheral part for supporting with CFRP material.
To operate the DuTiP chips, the polyimide flexible cable is necessary.  
At present the copper is the standard conductor material, however, development of the aluminum flexible circuit is in progress. 
If the aluminum flex technology is established, the material budget by the conductor will be reduced by 30--50~\%.

The power consumption of the DuTiP is estimated to be 0.3~W for an S-type chip located at the innermost layer (in total 7.2~W for layer~1). The cooling by airflow at room temperature is possible for inner layers. Outer layers can be cooled by a combination of air and water flows.
Inside the VXD region, a high-density flex cable will be used, just outside of the VXD region, the flex cables will be connected to the normal copper cables and transferred to the DOCK region, 2-3 meters away from the DuTiP sensor, and then sent to the backend electronics outside the Belle II volume.

\subsubsection{Occupancy, Data Transfer Rate and Data Size}
Since the environment for innermost layers is severe, here we consider layer 1.
We calculated pixel occupancy with the trigger latencies of 8.0$\mu$s.
For offline analysis, hit occupancy in a frame, PCN timing~(189ns), is important for tracking and vertexing. 
The both occupancies are enough small, ${\mathcal{O}}$(10$^{-4}$) or less, thus stable tracking and vertexing are expected. 
Note that if we do not adopt two timers in a pixel for layer 1, the signal loss probability with the trigger latency of 4.5$\mu$s~(8.0$\mu$s) is about 0.2\%~(0.4\%) which is not negligibly small. If the background rate is higher or trigger latency is longer, the signal loss probability becomes higher, thus dual timer is essential for the upgrade.
Since the DuTiP is binary detector and relatively small amount of digital data is sent with LVDS lines based on level~1 trigger, the data acquisition is not so difficult and a region of interest selection used for current DEPFET to reduce data at HLT, which might cause information loss especially for hits from low momentum tracks, is not needed.

%

The data rate per chip and data size in a smy for layer 1 with an assumption of trigger rate of 30kHz without high level trigger~(HLT) filtering are calculated as 11.3~Mbps and 340~TB, respectively.
Thanks to the global shutter mode with fast shaping time, the data size is much less than 1~PB. With the HLT filtering, the data size should be reduced by a factor of 3. 


\subsubsection{Prototype Chips and Schedule}
\begin{figure}[htbp]
\begin{center}
\includegraphics[width=0.35\textwidth]{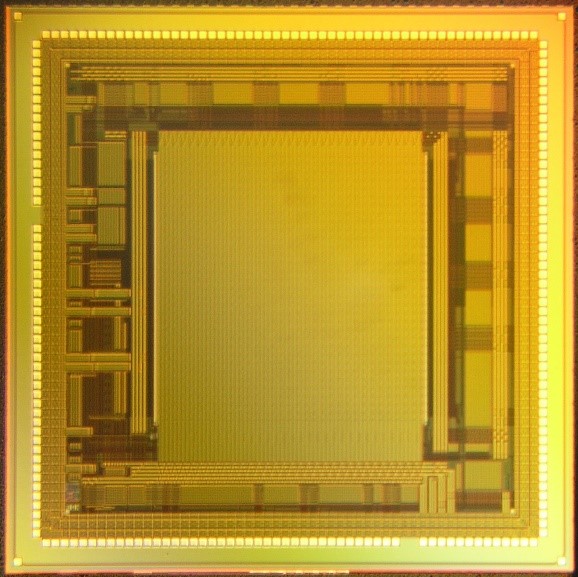}
\includegraphics[width=0.44\textwidth]{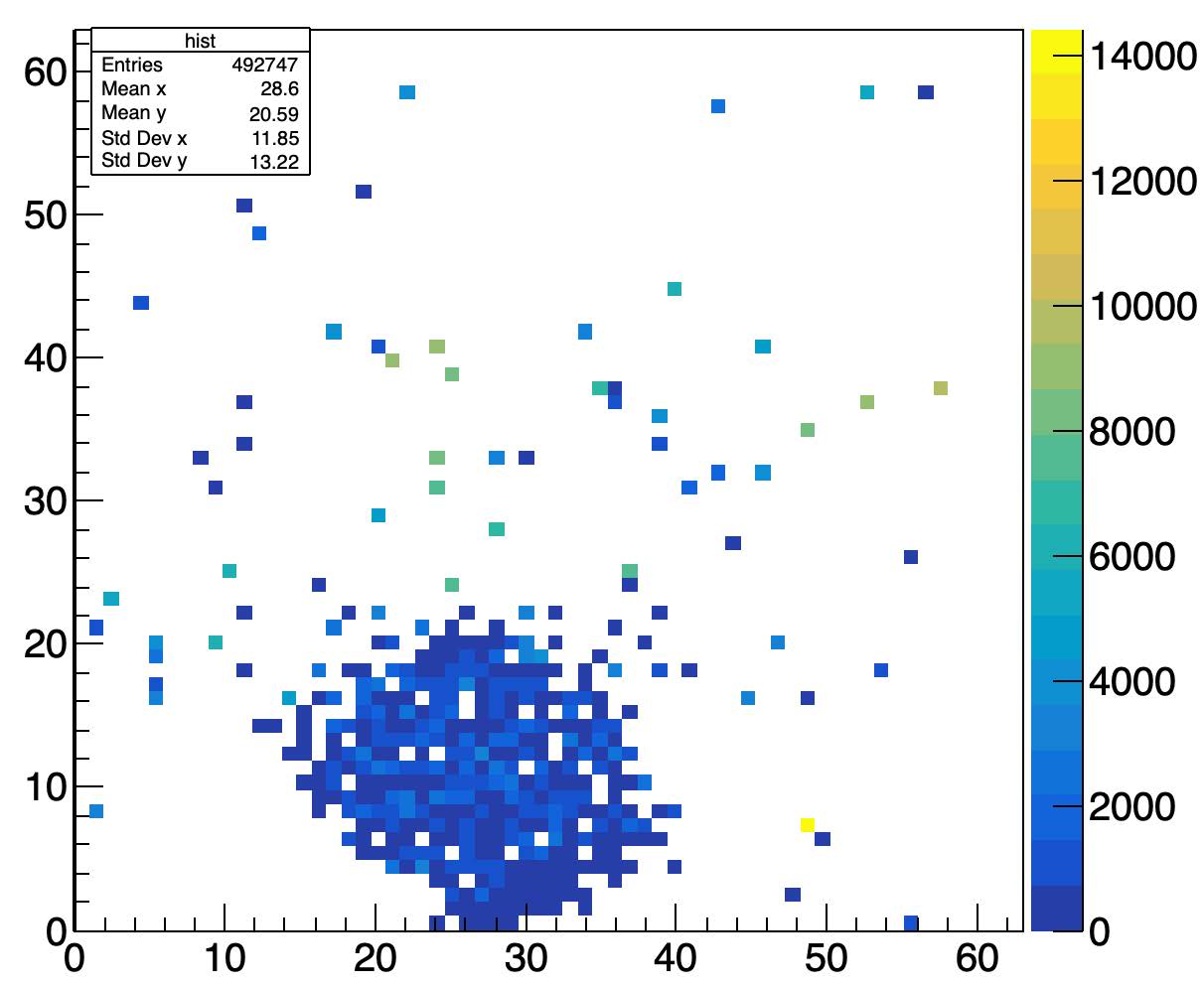}
\end{center}
\caption{(Left)~Picture of the DuTiP1~(6$\times$6mm${}^2$). (Right)~DuTiP1 sensed the red laser signal.}
\label{fig:DuTiP1}
\end{figure}

The first prototype DuTiP1 had been delivered in June 2021. 
This prototype was developed for demonstration of DuTiP concept thus all functionalities in pixel were fabricated while scan block and fast readout system were not implemented. 
The chip size is 6$\times$6mm${}^2$ with a pixel array of 64$\times$64~(Fig.~\ref{fig:DuTiP1} left).
Characterization of DuTiP1 chip is on-going. 
Digital circuit perfectly worked as expected using test pulse with 25~MHz clock.
The chip could sense the electron from ${}^{90}$Sr source and photons from a red laser~(Fig.~\ref{fig:DuTiP1} right).

\begin{figure}[htbp]
\begin{center}
\includegraphics[width= 0.75 \textwidth ]{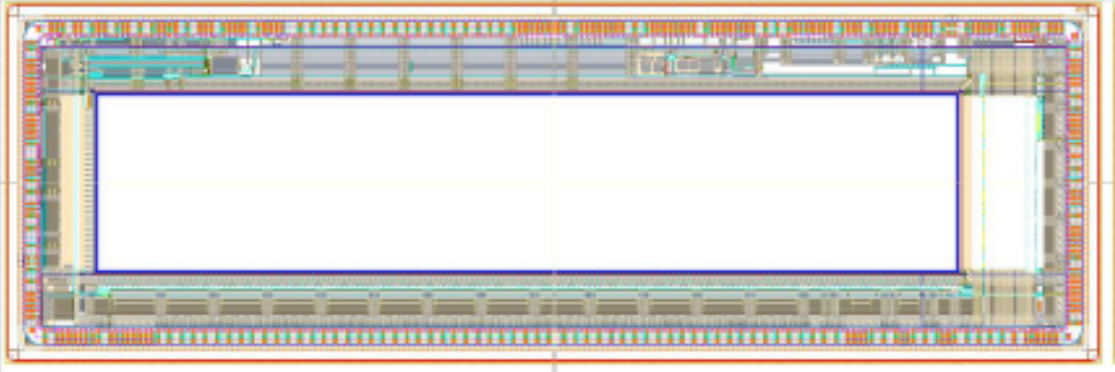}
\end{center}
\caption{The layout of DuTiP2~(18$\times$6mm${}^2$).}
\label{fig:DuTiP2}
\end{figure}

The second prototype DuTiP2 had been submitted in December 2021 and will be delivered in March 2022.
This prototype has all functionalities and is full size in column direction~(Fig~\ref{fig:DuTiP2}), chip size of 18$\times$6mm${}^2$ with a pixel array of 64$\times$320, thus we can perform the full characterization needed for operation (extending the chip size to row direction is trivial). 
If the characterization of DuTiP2 has been finished successfully in 2023, the final chip is designed in 2023 and the mass production of the final chip will be stated in 2024.

\subsubsection{Cost}
Table~\ref{tab:cost} shows the very rough estimation of the cost for DuTiP construction. 

\begin{table}[htb]
    \centering
\caption{Cost estimation.}\vspace{2mm}
    \label{tab:cost}
    \begin{tabular}{lr}
    \hline
    item &  cost [k\$]\\
    \hline
    sensor &  1860\\ 
    ladder &  1600\\ 
    ladder assembly and mechanics& 650\\
    backend electronis and DAQ & 1720\\ 
    \hline
    total & 5830\\
    \hline
    \end{tabular}
\end{table}

\subsubsection{Potentially interested community}
Table~\ref{tab:community} shows the possible contributions to DuTiP development. 
The DuTiP team is the world wide community from Asia~(China, Japan and Korea), America~(USA) and Europe~(France and Poland).
The DuTiP can be used for layer 7 and 8 of ILD vertex detector with minor modifications, which should provide bunch ID timing information. Thus contributions from ILC community might be expected.

\begin{table}[htb]
    \centering
\caption{Possible contributions to DuTiP development.}\vspace{2mm}
    \label{tab:community}
    \begin{tabular}{lc}
    \hline
    study item &  affiliations  \\
    \hline
    sensor development & IHEP, IPHC, AGH, KEK, PAN, TMCIT, Tohoku \\
    sensor characterization & Hawaii, IHEP, KEK, KNU, TMCIT, Tohoku, Tokyo, Tsukuba \\
    mechanics and assembly & KEK, KNU, Tokyo  \\
    backend and DAQ & Hawaii, IHEP \\
    trigger capability & KEK, Tohoku\\
    performance study &  KEK, Tohoku\\
    \hline
    \end{tabular}
\end{table}

\printbibliography[heading=subbibliography]
\clearpage
\end{refsection}


\newpage

\section{CDC}
\begin{refsection}
\label{sec:CDC}
\editor{N.Taniguchi}
\subsection{Upgrade of Readout electronics}
\def\II{\rm I\hspace{-.1em}I}

\subsubsection{introduction}
The front-end electronics which readout 14336 signals from CDC is located inside detector.

\begin{figure}[h]
\centering
\includegraphics[height=6.0cm]{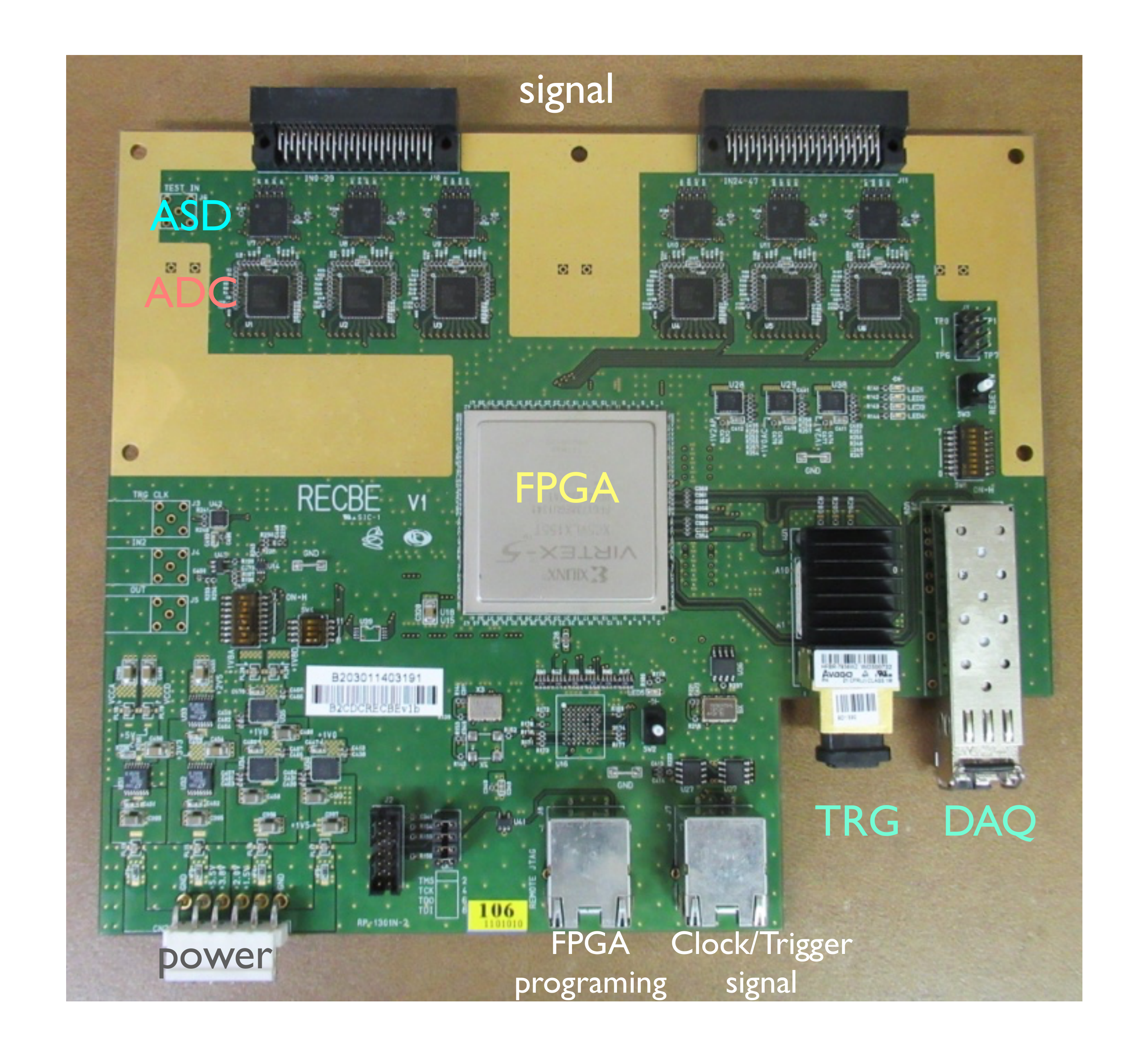}
\caption{The present CDC readout electronics board.} 
\label{fig:cdc_ele_00}
\end{figure}

Radiation damage in the optical transceiver on the front-end electronics may require replacement 
before the run end of Belle-$\II$ experiment.  We plan to replace not only optical transceiver but also entire readout board with new ASIC chip of Amp-Shaper-Discriminator.
A new ASIC chip have developed including new function of double threshold in discriminator to suppress cross talk which particularly affects online track trigger performance.  
Basically, applying cut using charge($dE/dx$) is effective in online analysis since pulse height of cross talk hit is lower than that of signal, while only way to suppress cross talk hit in offline system is to apply higher threshold in electronics at the present readout system, however, timing resolution get worse due to time walk.
Cross talk hit inside of ASD ASIC which is induced by charged particle increases fake rate of track trigger as background increases.  
Not only trigger efficiency but also tracking performance will be improved by reduction of cross talk at higher luminosity which we aim to.

\subsubsection{new ASD-FADC ASIC}
In the new ASIC chip, double threshold is applied to signal to determine wire hit.
It is expected to be useful to suppress cross talk hit by higher threshold level, 
while timing is determined by lower threshold level so that time walk is reduced.
Several tests have proved that new ASIC chip is better than the present one in terms of cross talk.
In addition, it is expected that information of charge($dE/dx$) can be used for track trigger firmware logic in the new readout electronics.
Data handling of FADC has $\sim$\SI{300}{\ns} delay from timing information in the present electronics.
The delay is not acceptable for the level-1 trigger decision in the present TRG system.
For new readout, FADC is designed with combined into ASD-ASIC chip by an expert of the electronics system group in KEK.
The delay of FADC is expected to be decreased to be acceptable level and lower power consumption is also expected by combining. By selecting wire hit which is used for track trigger using charge information, efficiency of track trigger is expected to be improved.

\subsubsection{Radiation hardness}
Radiation tolerance should be studied via gamma-ray and neutron irradiation.
As for optical transceiver, we will perform gamma irradiation test for several candidates of consumer product.
New readout electronics is also FPGA(field-programmable gate array)-based.  Radiation hardness against especially neutron is important for operation at higher luminosity in terms of single event upset(SEU).
In the JFY 2021, a prototype of readout board has been produced. We will study performance with test beam and perform radiation test 
via gamma-ray and neutron irradiation.

\subsubsection{Potentially interested community in Belle II}
The new ASIC of ASD and layout of a prototype of readout board is also don by experts in KEK Esys group.
National Taiwan University(NTU) had contributed on quality checks of all ASD ASIC chips for the present readout electronics and all readout boards. Mass production was done in Taiwan.  
It is anticipated that KEK and NTU will contribute on upgrade.

\subsubsection{Order of magnitude cost estimate}
An estimation of the main cost is shown in Table.~\ref{cdc_tb1}. We include $10\%$ spare. 
Cost for mass production of electronics is based on a cost of FPGA.

\begin{table}[h]
\begin{tabular}{|c|c|c|}
\hline
item & number & cost \\
\hline\hline
mass production of ASIC & $\sim$ 2,000 chip  & 15 M JPY\\
\hline
mass production of FE board & $\sim$330 FEs (including spare) & 100 M JPY\\
\hline
\end{tabular}
\caption{Estimated cost for mass production of ASD ASIC and readout electronics board.}\label{cdc_tb1}
\end{table}

\printbibliography[heading=subbibliography]
\clearpage
\end{refsection}

\section{TOP}
\begin{refsection}
\label{sec:TOP}
\editor{E.Torassa}
\subsection{Introduction}

With three years of running experience at Belle II, the world-record luminosity of SuperKEKB has been accompanied by higher than anticipated backgrounds.  For the TOP subsystem this means it is looking increasingly likely that even the life-extended Atomic Layer Deposition Micro-Channel Plate Photomultiplier Tubes (ALD MCP-PMTs) may not survive until the target integrated luminosity.  R\&D has started on a possible silicon photomultiplier (SiPM) replacement option. This option looks promising, although the required cooling will reduce the space available for the readout electronics and limit its acceptable heat load. Anyway thermal gradients and potential thermal and mechanical stresses need to be carefully evaluated. Moreover, the projected high rates of Single Event Upsets (SEUs) is problematic, as some number of them are not recoverable and will lead to increasing data taking loss.  Perhaps of greater concern is the known/measured lack of radiation tolerance of the fiber optic transceiver modules. The readout electronics upgrade has been considered as well.

\begin{figure}[ht]
\begin{center}
\includegraphics[scale=0.5]{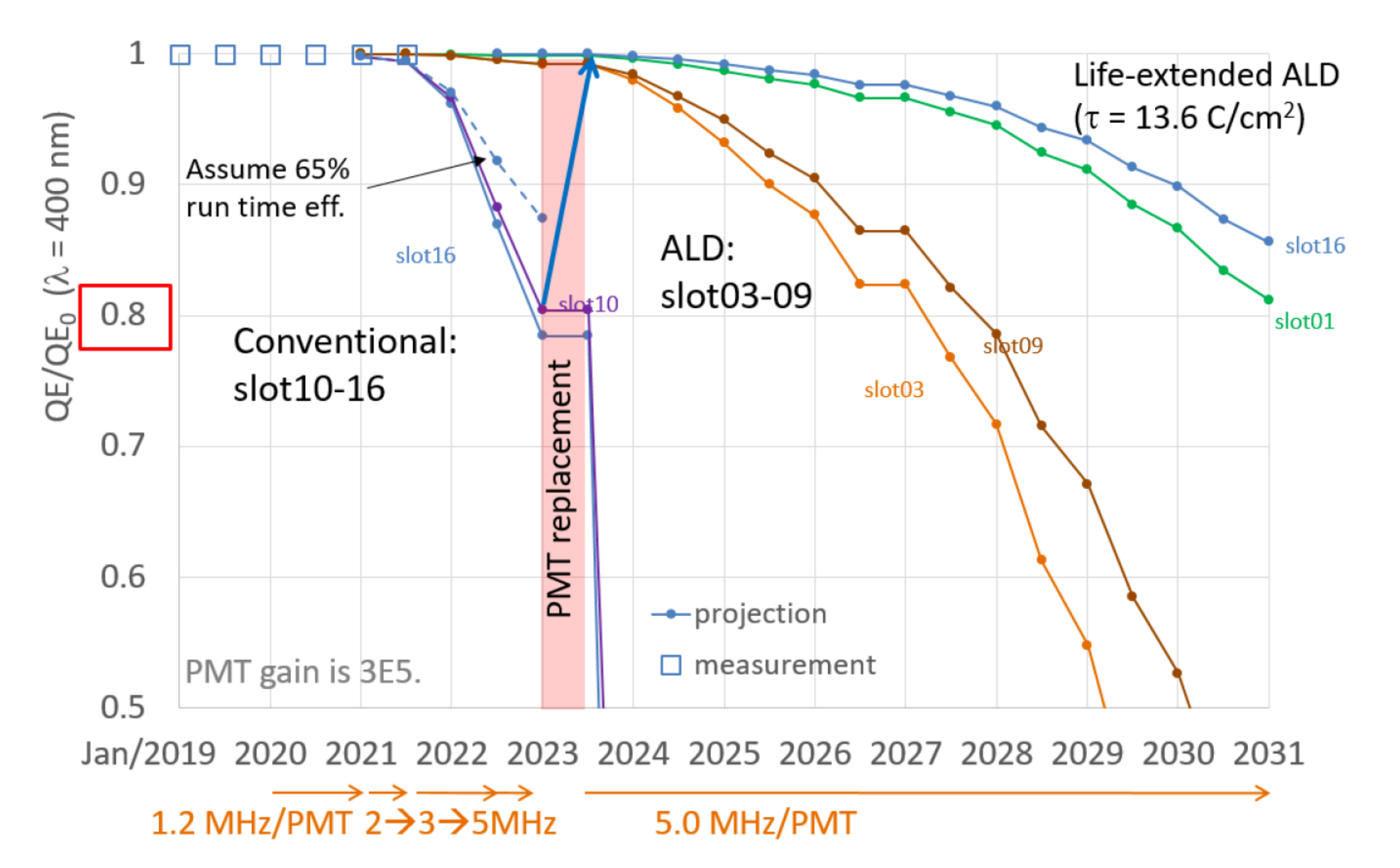}
\caption{\it Expected projection of QE degradation for the three different generations of MCP-PMTs installed in the TOP detector: conventional, ALD and life-extended ALD.  Horizontal axis shows the year and vertical axis is the relative QE.}
\label{fig:QEprojection}
\end{center}
\end{figure}

The timescale for replacement of the TOP photosensors is related to the level of background during the next years. Fig.~\ref{fig:QEprojection} shows the expected projection of PMT quantum-efficiency (QE) degradation. The Monte Carlo expected background was considered for the components related to the luminosity (Bhabha scattering, two-photon processes) and a maximum limit of 5 MHz/PMT for the other components (beam-gas, Touschek). In this case, the QE of normal ALD type PMTs will become less than 80~\% starting from 2027 and all PMTs will reach this level of QE degradation at around the end of the current luminosity target.

Planning has started for these upgrades happening during a possible long shutdown to rebuild the interaction region on the timescale of roughly 5 years from now. 

\subsection{Executive Summary}

The rudimentary schedule and cost estimates for the electronic readout are based upon production and delivery of the 8k channels of TOP readout and 20k channels of KLM scintillator readout, as well as the KLM data concentrators.  This endeavor leverages heavily two sizeable past and new investments: re-use of existing infrastructure and firmware, wherever possible; and DOE supported "Small Business Innovation Research" (SBIR) funding for development of ASICs appropriate to these upgrades.  

In both cases the pre-production plus production timescale is approximately 2 years, with 6-9 months for installation and commissioning.  Based upon the preliminary prototyping being performed and costed separately from this, the total project cost is estimated to be 1.1-1.3 M$\$$ and 1.4-1.8 M$\$$ for upgrades to the TOP and KLM readout systems, respectively, depending upon amplifier choices.

For the photosensors the primary option for the upgrade in 2026 is the replacement of normal ALD PMTs with life-extended ALD type PMTs.
With this option, we need ~220 new MCP-PMTs to replace normal
ALD type PMTs, which costs is estimated to be 1.5 M$\$$ and requires to start the mass-production from JFY2021. The production of life-extended ALD MCP-PMT already restarted, 30 MCP-PMT have been ordered and will be delivered within JFY2021. The backup option is the replacement of MCP-PMTs with cold SiPMs. For further luminosity upgrade, SiPMs operated at low temperature is the primary option. 


\subsection{Upgrade Proposal}

The goals of baseline upgrade to the readout electronics are twofold: reduce significantly their size, as well as the power required.  Both goals facilitate the implementation of the requisite cooling needed to operate the SiPMs in single-photon detection (though with significantly increased dark-count) mode.

In order to reduce cost and development time, much of the existing software, firmware and hardware infrastructure can be retained.  The boundary conditions considered for this costing exercise are:

\begin{itemize}
    \item Retain the Standard Control Read-Out Data (SCROD) board, the existing cables and low-voltage power system
    \item Remove the HV divider chain/delivery board and associated cables
    \item Employ 32-channel TOP System on Chip readout ASICs, each with multiple Gbps connections to the SCROD FPGA
    \item 2 different amplifier options for the Si-PMs are considered and costed
\end{itemize}


 
Based upon this proposed modification, the upgraded readout system still consists of 64 "board stacks", though there is now a single "Carrier" card containing four 32-channel readout AISCs replacing the current four "Carrier" cards containing four 8-channel readout AISCs.  In the first amplifier choice, the SiPM overvoltage trim and amplification are done inside the TOPSoC ASIC, greatly reducing component count and board space requirements.  Should this prove to be technically unfeasible, a second amplifier method requiring an additional circuit board, requisite components, and engineering effort is costed.
For the photosensors, the Hamamatsu production line of life-extended ALD MCP-PMT is still operating with an expected output of 10 MCP-PMTs/month. The possible further increase of lifetime can be investigated for example by an optimization of the atomic layer thickness.
The option of MCP-PMT replacecment requires several studies and possibly further SiPM developments.
It's important to find best balance among different SiPM's characteristics:
single device size, single cell dimension, operating temperature.
We need to be able to distinguish single photons also after an irradiation of \SI{10 e 11}{n/cm^2} 
but keeping a good photon efficiency, a good time resolution with an acceptable thermal and mechanical stress for the detector.



\subsection{Design Development and Human Resources}

The design of the new electronics is still needed.
The design of the new photosensors for the primary option is available, it is the same of the current PMT modules. 
It is anticipated that for the photosensor upgrade, Nagoya U., INFN/U. Padova and INFN/U. Torino;  while for the electronics upgrade, Indiana U, U. Hawaii, U. Pittsburgh and Nalu Scientific personnel will be the primary contributors.

\subsection{Cost and Schedule}

%
With either amplifier option it is anticipated that one engineer will be involved in the design of the ASIC Carrier board, at 100\% FTE for year 1 and dropping to 50\% FTE for year 2.  Full-time funding for 1 postdoc and 1 graduate student to assist in assembly, test and documentation is requested, under the oversight of faculty members Nishimura and Varner.  Board fabrication and assembly costs are based upon those realized during earlier production. An increase of 20\% is added to account for inflation. As noted earlier, it is assumed that the development costs for the TOPSoC ASIC will have been provided by SBIR funding of Nalu Scientific activities.  Costs are based upon engineering runs in 130~nm CMOS technology.  A yearly firmware development engineering support is costed during the 2 years of production, and at a reduced level during commissioning.
%
The estimated costs of the electronics upgrade are described in Table~\ref{iTOP:table1} for two options.
In the first option it is assumed that 80 new Carrier boards will be prepared.  The electro-mechanics and cooling of the photosensors are to be provided off-project (INFN + Nagoya U.).  In the second option  an additional "front" board has been considered. The estimated costs for the replacement of the 220 ALD MCP-PMTs are described in Table~\ref{iTOP:table2} for life-extended ALD MCP-PMT option and Table~\ref{iTOP:table3} for the SiPM option.
\begin{table}[hp]
\caption{TOP Readout upgrade Amplifier cost estimate.}
\centering
\begin{tabular}{ | l | l | l | c | } \hline
{\bf Phase} & {\bf Resource} & {\bf Basis of est.} & {\bf Cost (k\$)} \\ \hline
\multicolumn{4}{|c|}{\bf option 1} \\ \hline
 Pre-production & Carrier Engineering \& test & 3.0 FTE  & 255 \\ 
 & Parts procur. \& assembly & cost summary & 113  \\ 
  \hline
 Production & Carrier Engineering \& test & 2.5 FTE  & 185 \\
 & Parts procur. \& assembly & cost summary & 376  \\ 
 \hline 
 Commissioning & Engineering \& installation & 1.5 FTE & 105 \\  
 & TOPSoC ASIC support & Nalu Scientific & 80  \\  \hline 
 {\bf Total option 1} & & & {\bf 1,114 k\$} \\ \hline 
\multicolumn{4}{|c|}{\bf option 2 (additional costs)} \\ \hline
 Pre-production & Front Engineering & 0.2 FTE engr. & 40 \\ 
 & Parts procur. \& assembly & Engr. est. & 5  \\ \hline
 Production & Front Engineering & 0.15 FTE engr. & 30 \\  
 & Parts procur. \& assembly & Engr. est. & 67 \\ \hline 
 Commissioning & Integration & 0.05 FTE engr. & 10 \\  \hline 
 {\bf Total option 2} & & & {\bf 152 k\$} \\ \hline 
{\bf Total opt. 1+2} & & & {\bf 1,266 k\$} \\ \hline \hline
\end{tabular}
\label{iTOP:table1}
\end{table}
\begin{table}[ht]
\caption{Life-extended ALD MCP-PMT upgrade cost est.}
\centering
\begin{tabular}{ | l | l | l | c | } \hline
{\bf Phase} & {\bf Resource} & {\bf Basis of est.} & {\bf Cost (k\$)} \\ \hline
 Production & MCP-PMT Production & Engr. est. & 1500 \\  
           & MCP-PMT Test & Engr. est. & 15 \\
           & Assembly parts & Engr. est. & 50  \\  \hline 
 Commissioning & Integration & 0.5 FTE & 50 \\  \hline 
{\bf Total} & & & {\bf 1615 k\$} \\ \hline \hline
\end{tabular}
\label{iTOP:table2}
\end{table}
\begin{table}[ht]
\caption{SiPM upgrade cost est.}
\centering
\begin{tabular}{ | l | l | l | c | } \hline
{\bf Phase} & {\bf Resource} & {\bf Basis of est.} & {\bf Cost (k\$)} \\ \hline
 Pre-production & Engineering & 0.5 FTE & 50\\ 
 & SiPM R\&D & Engr. est. &  10 \\ 
 & Cooling R\&D & Engr. est. & 30 \\ \hline
 Production & Engineering & 0.5 FTE & 50 \\  
 & SiPM + assembly parts & Engr. est. & 700 \\
 & Cooling system & Engr. est. & 50  \\  \hline 
 Commissioning & Integration & 0.5 FTE & 50 \\  \hline 
 {\bf Total} & & & {\bf 930 k\$} \\ \hline \hline
\end{tabular}
\label{iTOP:table3}
\end{table}

\printbibliography[heading=subbibliography]
\clearpage
\end{refsection}

\section{ARICH}
\begin{refsection}
\label{sec:ARI}
\editor{R.Pestotnik}
\subsection{Introduction}
The Aerogel RICH particle identification subsystem (ARICH) located in a perpendicular 1.5~T magnetic field in front of electromagnetic calorimeter consists of several key elements: aerogel radiator plane, photon detector plane with single photon sensors and front-end readout electronics. Cherenkov photons emitted  in the 4~cm aerogel radiator  are propagated through  16~cm  expansion gap and detected on the photon detector consisting of Hybrid Avalanche Photo detectors (HAPD) produced by Hamamatsu \cite{ari:hapd}. After the sensor production, the producer disassembled its production line. 
The HAPD is a hybrid vacuum photon detector consisting of an entrance window with a bi-alkali photo-cathode with  28\% peak quantum efficiency and 4 segmented avalanche photo diodes in proximity focusing configuration. 420 HAPD sensors with 144 channels are mounted in seven concentric rings surrounding the beam-pipe in the forward end-cap region of the spectrometer.
The ARICH was designed to operate up to the
nominal design luminosity of 8$\times10^{35}$~cm$^{-2}$ s$^{-1}$. Due to high background environments its operation poses new challenges when operation is extended beyond the design luminosity.
   At high luminosity the neutron background radiation and gamma radiation levels increase. They can affect the functionality of all detector components. 
 
The silica aerogel  is produced in a super critical drying process and is not affected by the irradiation \cite{ari:radaerogel}. Its performance was measured  during the irradiation with a dose of 98~kGy and no significant degradation was found. 

HAPDs however, are sensitive to a gamma and a neutron irradiation. Already during the development of the sensor different modifications in the APD structure have been implemented. Different designs have been tested including the film on active area and different thickness of p+ layer. The design with an intermediate electrode, a thin p+  layer and a film on the active area was finally selected for ARICH. The HAPD were demonstrated to work up to the design luminosity, but not after its 5 $\times$  longer-term upgrade. Also, due to its high Boron content introduced during the chip production, a Xilinx Spartan-6 FPGA, part of the front-end electronics board, is prone to single event upsets. By using a custom FPGA memory scrubbing \cite{ari:Giordano} and triple redundancy logic, we will be able to operate the electronics in the short and in the mid-term period. 

We have already started to investigate different possible upgrade scenarios for the long-term upgrade with the focus on the possible photon detector candidates.

\subsection{Photon detector Candidates}

The choice of  photon detectors, capable of operating in the magnetic field of 1.5~T, is limited to several different technologies, which we already considered during the R\&D of the current photo detector \cite{ari:sipmrich,ari:mcppmtrich}. 

\subsubsection{Silicon Photo-Multipliers}
The first technology are silicon photo multipliers, that have several advantages over the hybrid avalanche photo detectors. They have a very high photon detector efficiency reaching 60\% in the peak. They are very easy to operate as they only require reverse biasing in the order of 30 to 70 V, much less than standard or hybrid vacuum detectors. In addition they have a very good timing resolution which opens new possibilities for particle identification in ARICH. As a semiconductor device, they are inherently insensitive to magnetic fields.

They have however also several drawbacks. First is the large dark count rate, even more problematic one is their sensitivity to radiation, which was a limiting factor a decade ago, when we were selecting the photon detector candidates for the current ARICH photon detector.

The radiation damage increases the sensor currents, affects its breakdown voltage and increases the dark count rate. The dark counts affect the measurement in two ways. First, as they produce equal signals as single photons, they cannot be distinguished from them, therefore they produce a high background on the detectors plane. The resulting effect on the separation capabilities for an increased level of background can be seen on Fig.~\ref{fig:lappd}. Second, the increase of the dark count rate, proportional to the size of the device, is also responsible for the sensor baseline loss, which happens above $10^{11}$~n$_{eq}/$cm$^2$ or less for small 1~mm$^2$ devices. The limit depends on many factors related to the SiPM design and the operation conditions. E.g., the dark count rate increases exponentially with increasing the operating temperature. However, the silicon photo-multipliers and their operating range should be tested for each specific application. 
The loss of baseline affects single photon counting much more than the use of silicon photo-multipliers in the multi-photon regime, e.g., in a calorimetry. 
 
We are following the technology and evaluating currently available samples. We are also working together with the producers to reduce the neutron sensitivity and to find the operation parameters where the damage might be under control. We will consider different engineering changes in the layout, substrate, and the sensor SPAD size \cite{ari:aidainnova}.

There are several measures that can contribute to operation of Silicon photo-multipliers: the reduction of their operating temperature, the use of timing information for a background rejection, the reduction of sensor sensitive surface by using a light collection system, and the high temperature annealing. We are studying the response of the silicon photomultipliers at low temperatures and study the multi-channel timing electronics to read out the sensors. 
The key point in our studies of using the silicon photo-multipliers is to demonstrate their capability to measure single photons and to reduce the dark count rate to an acceptable level.

In the case of the RICH with the aerogel radiator where neutron fluences of up to \SI{5 e 12}{n/cm^2} are expected, the operation at lower temperatures requires the use of an additional cooling system, which introduces more material in front of the electromagnetic calorimeter.

\subsubsection{Large Area Picosecond Photodetector}
The second possible photon detector technology we are considering for the upgrade of the photon sensors are micro-channel plate photo-multipliers (MCP-PMT). 
Compared to silicon photo-multipliers, the sensors have several drawbacks. As they are based on bi-alkali photo-cathodes and suffer from additional collection inefficiency, their photon detection efficiency is similar to the current HAPDs, much lower than the one of SiPMs. As a vacuum detector with an internal amplification structure, they also suffer from the gain drop in the magnetic field and lifetime issues due to charge collection on the photocathode. 

\begin{figure}[h]
\centering
\includegraphics[height=4.0cm]{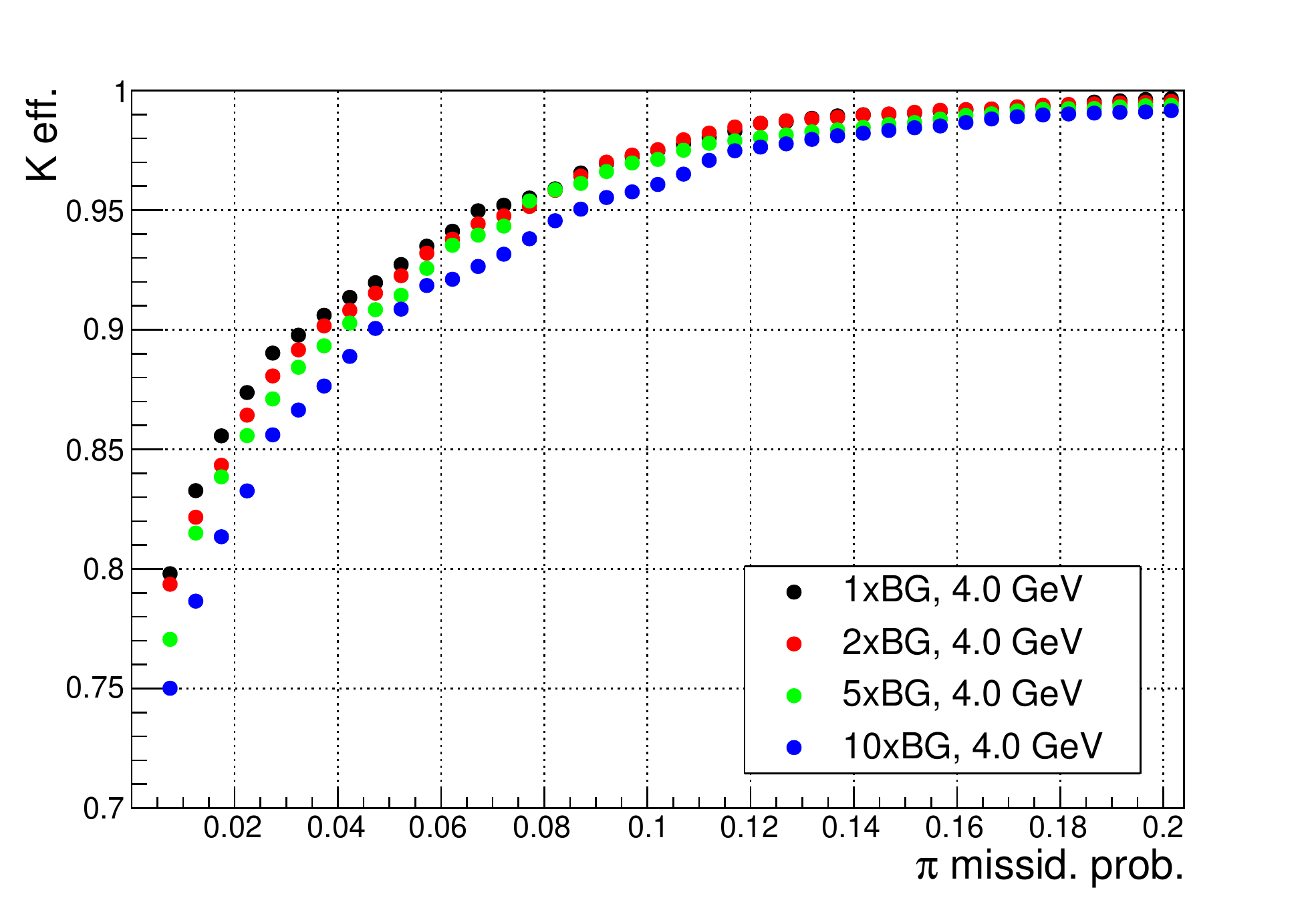}
\includegraphics[height=4.0cm]{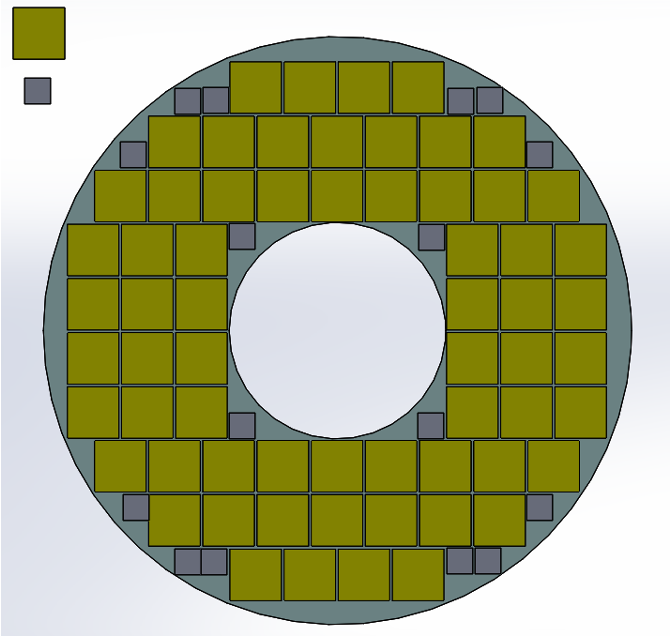}
\caption{\sl Left: Simulated Kaon identification efficiency versus pion misidentification probability at a momentum of 4~GeV/c for various levels of nominal background. Bi-alkali photocathode was assumed during the simulation.  Right: Conceptual layout of ARICH photon detector plane populated with 20$\times$20~cm$^2$ and 10$\times$10~cm$^2$ LAPPDs.}
\label{fig:lappd}
\end{figure}

Due to its high production price, we are considering the Large Area Picosecond Photodetector (LAPPD) technology, where only inexpensive glass is used in the production process, leading to a significantly lower volume pricing. The LAPPDs, produced by Incom USA, come in two sizes  20$\times$20~cm$^2$ and 10$\times$10~cm$^2$ . Large size reduces the complexity of the photon detector, however it introduces additional inactive areas on the photon detector. A conceptual design layout of the ARICH photon detector plane equipped with LAPPDs is shown on Fig.~\ref{fig:lappd}. We are studying possible charge sharing of the signal spread over many pixels in  segmented devices, which may result in a limited range of operation, when signals from photons start to overlap.   

\subsection{Sensor read-out electronics}
Both possible sensors require to read out the signals in a narrow time window of several ns with respect to the trigger signal, requiring optimized internal design and high integration with the read-out electronics at the back side of the sensor. We are studying two potential front-end chip options. The first is based on the upgrade of the current ASIC used for Hybrid avalanche photo-diodes \cite{ari:asic}. The chip has been designed; first samples are currently under functional tests. This option allows to use proven technology, however it is not yet demonstrated, how it will handle high dark count rates. The second option is the use of novel FastIC ASIC, featuring low noise fast timing capabilities, developed in collaboration of CERN and University of Barcelona \cite{ari:fastic}. As the new ASIC will be used the coming upgrades of the LHC experiments, the synergies in common development (e.g., with LHCb RICH) are expected to reduce the development costs.

\subsection{Radiator}
A study has already been launched to estimate the possible gains with using  more transparent aerogel tiles and possible higher refractive index. Currently the aerogel production technology has been transferred from Chiba University to a local company. We are evaluating the quality of aerogel after the technology transfer by measuring the characteristics of new samples.

\subsection{Potentially interested community in Belle II}
The following collaborators expressed their interest in the hardware upgrade activities: KEK, Tokyo Metropolitan University, and Niigata University, all from Japan,  University of Naples, Italy, and  Jožef Stefan Institute, Slovenia.

\subsection{Cost and schedule}
The current Aerogel RICH will be able to operate efficiently until the design luminosity is reached. Therefore all the sub-detector changes considered above are projected for the long-term upgrade. Preliminary cost estimates are given for two photon detector options: silicon photomultipliers with 250.000 1~mm$^2$ channels and 64 20$\times$20~cm$^2$ and 16  10$\times$10~cm$^2$ LAPPDs.   The estimates are based on the current understanding of the technologies under development. 
\begin{table}[t]
\caption{Preliminary cost estimates in kUSD for photon detector upgrade based on SiPM  and LAPPD. These include the costs for the associate R\&D.}
\vspace{-0.6cm}
\begin{center}
\begin{tabular}{|l| c| c|}
    \hline
    {\bf Option} & {\bf SiPM  } & {\bf LAPPD}  \\
    \hline
    {\bf Description} & {\bf k\$ } & {\bf k\$}  \\
    \hline
    \hline
     \multicolumn{3}{|l|}{\it Photo-sensor} \\
    \hline
    Research and development  & 500 & 500 \\
    \hline
    Production &  1.500 & 2.300 \\
     \hline
    \multicolumn{3}{|l|}{\it Electronics} \\
    \hline
    ASIC including TDC & 700 & 700  \\
    \hline
    Front-end board & 200 & 200  \\
    \hline
    Merger boards & 200 & 200 \\
    \hline
    \multicolumn{3}{|l|}{\it Services and detector infrastructure} \\
    \hline
    Photon detector mechanical frame & 200 & 200\\
    \hline
    Cooling system & 300 & re-use current system\\
    \hline
    Power supplies & 50 & re-use current system\\
    \hline
    \multicolumn{3}{|l|}{\it Radiator} \\
    \hline
    
    Aerogel upgrade & 700 & 700\\
    
    \hline \hline
    Total & 4.350 & 4.800\\ 
    \hline
    Contingency & \multicolumn{2}{|l|}{\it +15\% }  \\
    \hline
    
    {\bf Grand total} & {\bf 5000}  & {\bf 5520}  \\
    \hline
\end{tabular}
\end{center}
\label {tab:ARICost}
\end{table}
\printbibliography[heading=subbibliography]
\clearpage
\end{refsection}

\section{ECL}
\begin{refsection}
\label{sec:ECL}
\editor{C.Cecchi}

This section contains a summary of the upgrade of the Belle II Electromagnetic Calorimeter (ECL). The current detector as some limitation due to being insensitive to the incident photon direction and pileup effect by beam backgrounds will result in deterioration of the energy resolution. Some upgrade options of the Electromagnetic Calorimeter have been discussed in order to cope with the high occupancy and background rate to maintain good performances in fundamental object reconstruction.

\subsection{Anticipated performance limitations}
\begin{figure}[htb]
    \centering
    \includegraphics[width=0.47\textwidth]{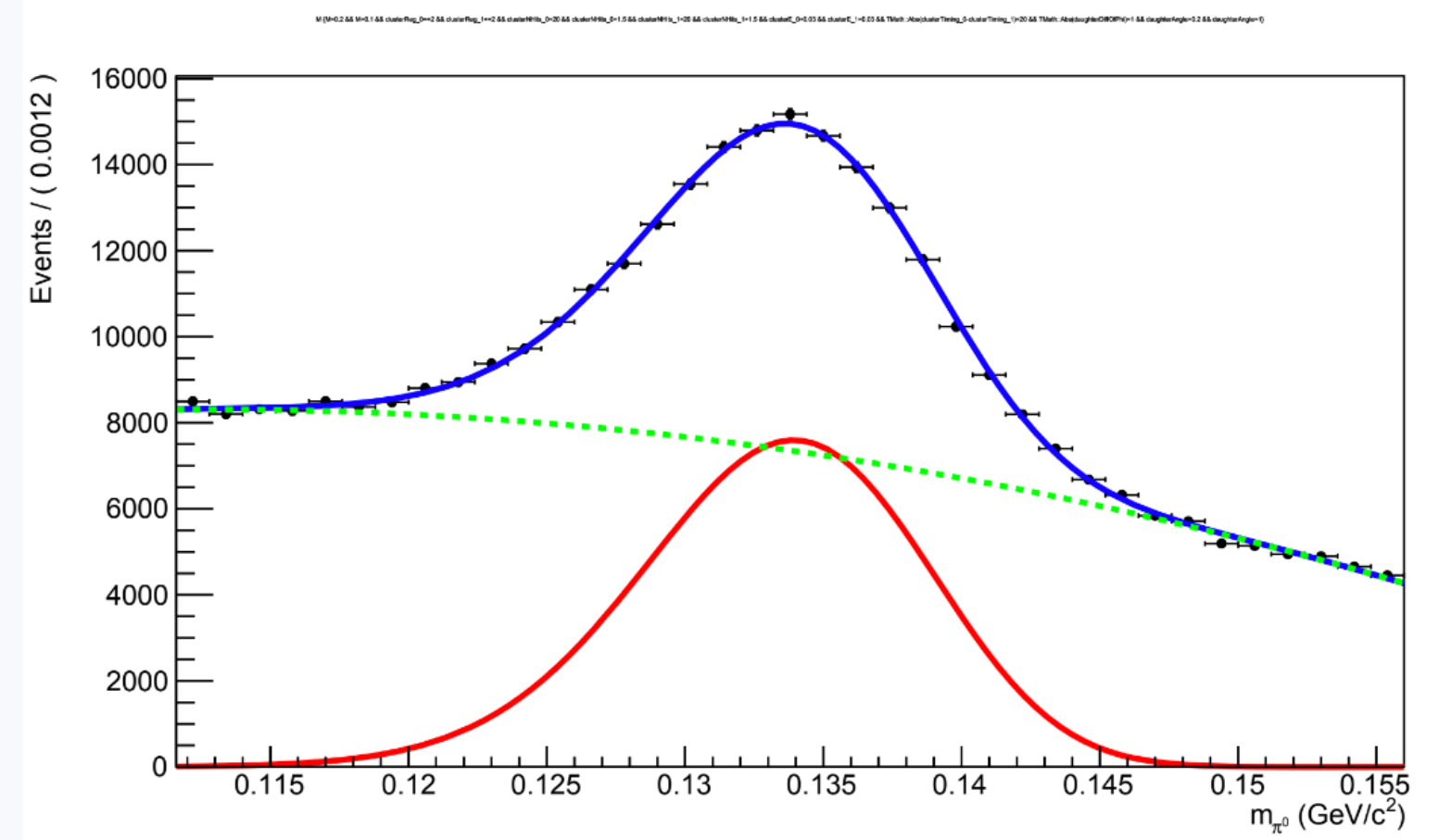}
    \includegraphics[width=0.45\textwidth]{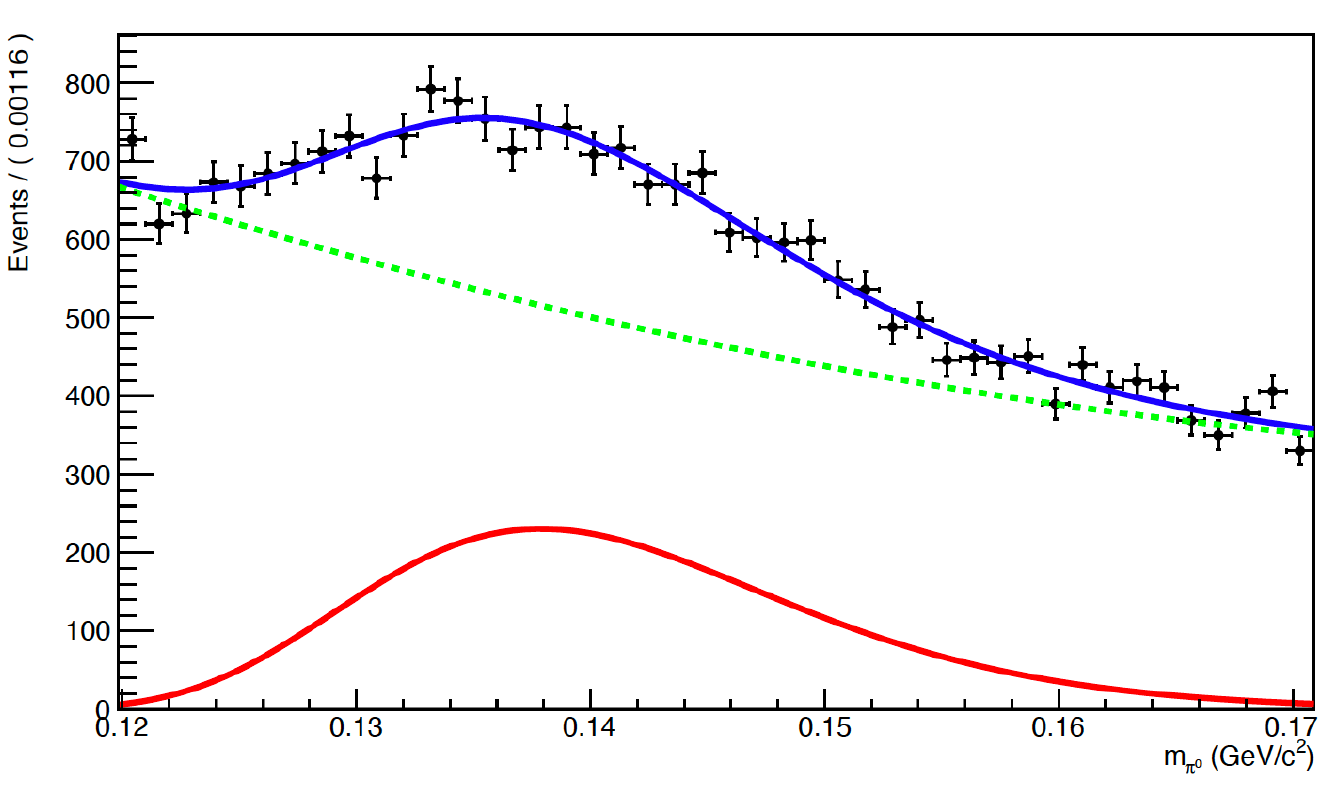}
    \caption{Reconstructed $\pi^0$ mass for the Early Phase3 BGx1 background simulation (left) and for the Nominal Phase3 BGx1 background simulation (right). The signal peak has been fitted with a Novosibirsk function while the background is described by a Chebichev distribution.}
    \label{fig:pizero-mass-fit}
\end{figure}

Since CsI(Tl) crystals scintillation decay time is relatively long, $\sim 1~\mathrm{\mu}s$, increase of beam background results in the larger pileup noise that deteriorate energy resolution for neutral particles. Simulated beam background has been superimposed on generic Monte Carlo (MC) $B\bar B$ events,  and studies on the $\pi^0$ mass resolution have been done for the Early Phase3 ($L = 1 \times 10^{34}\mbox{cm}^{-2}\mbox{s}^{-1}$) with nominal, ×2 and ×5 beam background as well as the ultimate Phase3 ($L = 8 \times 10^{34}\mbox{cm}^{-2}\mbox{s}^{-1}$) with nominal beam background cases. The $\gamma \gamma$ invariant mass spectrum is fitted with a Novosibirsk and a Chebichev functions for the $\pi^0$ signal and background, respectively. Resultant distributions are shown in Figure  \ref{fig:pizero-mass-fit}, Early Phase3 nominal background (left) and Phase3 (right). In Early Phase3 case, the $\pi^0$ mass resolution and efficiency are found to be 6 MeV and 30$\%$, while those become 9 MeV and 15$\%$ in Phase3 case. The Phase3 environment with higher background deteriorates the matching of the ECL deposit with the true photon because of the higher combinatorial in the number of ECL clusters. This could represent a strong limiting factor for physics analysis with rare B decays where high efficiency is required.

\begin{wrapfigure}{R}{0.5\textwidth}
\centering
\vspace{-15pt}
\includegraphics[width=0.4\textwidth]{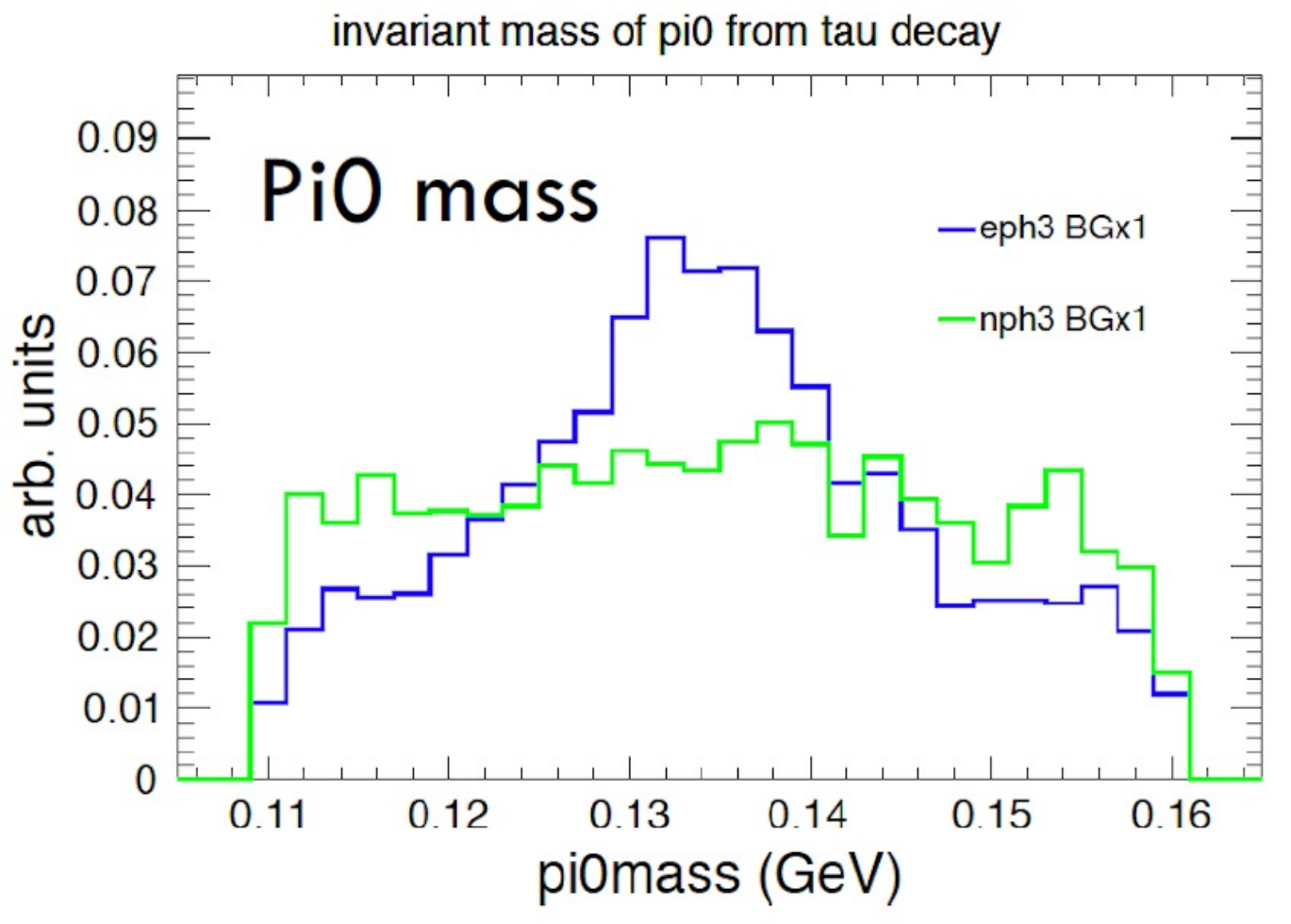}
\caption{$\pi^0$ mass distribution in $B\rightarrow \tau \nu$ events for Phase3 background (green) compared to the Early Phase3 background distribution (blue).}
\label{fig:pizero-mass-nominal}
 \vspace{-15pt}
 \vspace{1pt}
\end{wrapfigure}

In the $B\rightarrow \tau \nu$ analysis, one of the $B$ mesons in the event is fully reconstructed and the rest of the event is required to have $\tau$ daughters. Since substantial portion of $\tau$ decays contain $\pi^0$, a good reconstruction of its invariant mass is crucial. The $\gamma \gamma$ invariant mass distribution is shown in Figure \ref{fig:pizero-mass-nominal}. In Phase 3, the $\pi^0$ signal peak is unseen, it indicates a serious limit in reconstructing fundamental physics objects.

\begin{figure}[htb]
\centering
\includegraphics[width=0.45\textwidth]{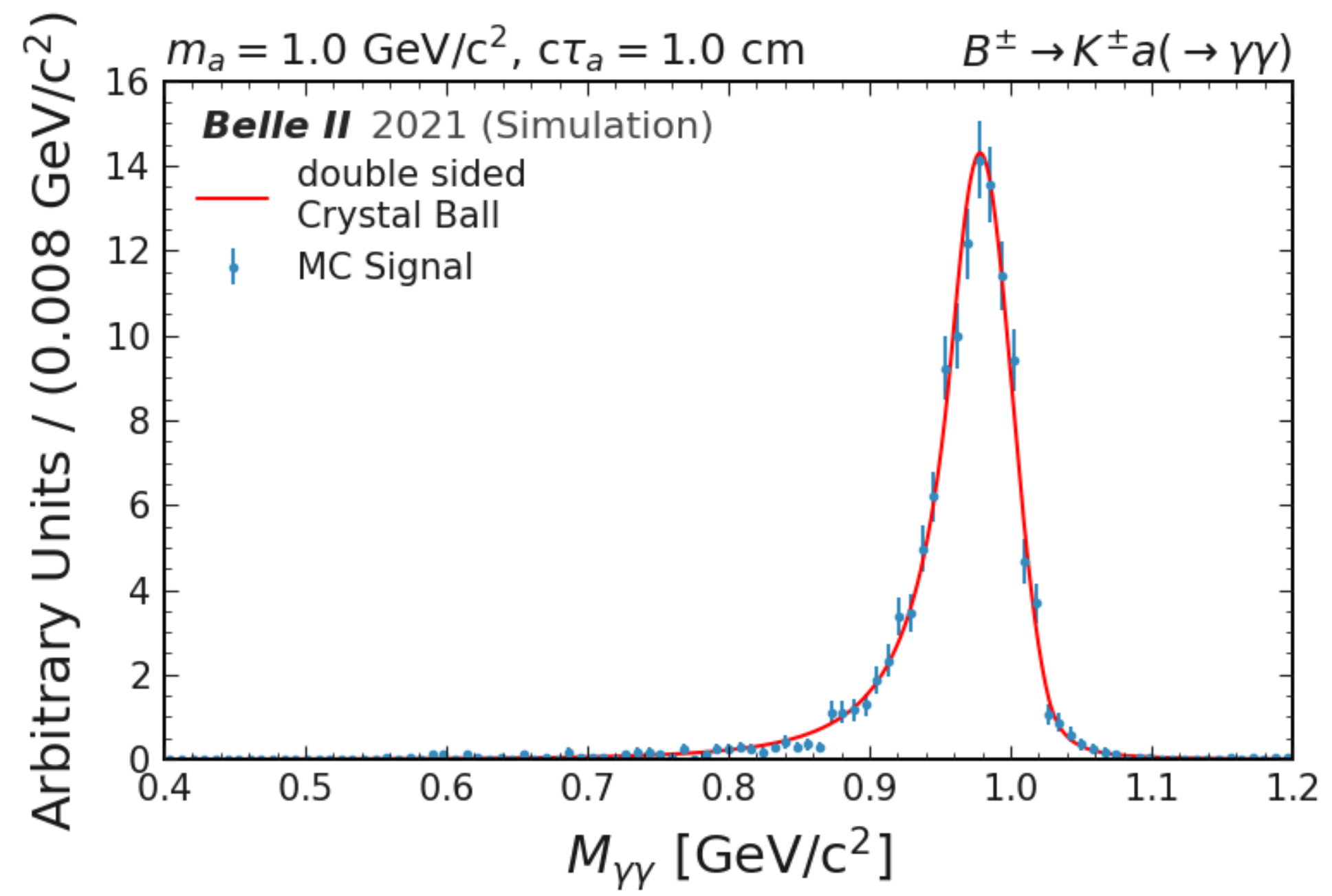}
\includegraphics[width=0.45\textwidth]{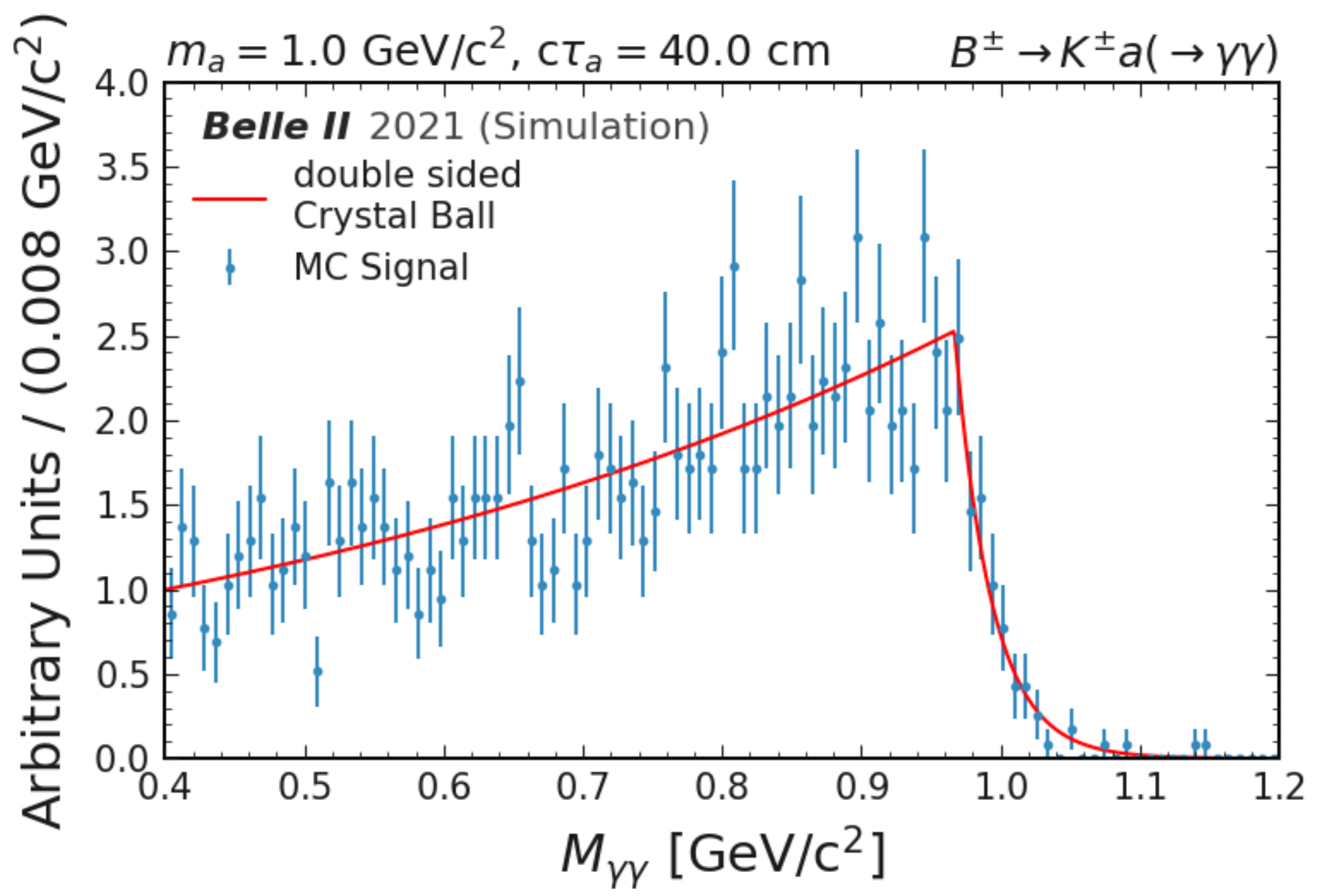}
\caption{The reconstructed photon pair's invariant mass distribution in $B \to Ka$, $a \to \gamma \gamma$ MC sample for the axion-like-particle (ALP), $a$ mass $m_a=1$ GeV/$c^2$ and its life $c\tau=1$ cm (left) and 40 cm (right) cases.}
\label{fig:LonglivedALP}
\end{figure}

To explore dark-sector particles in Belle II data, importance to search for
a long-lived particle is increasing. One of very important process is production of an axion-like-particle (ALP), ${a}$, in the B decay of $B \rightarrow Ka$. Such ALP is likely to be a long-lived particle decaying into a photon pair. Since current ECL configuration has very little sensitivity for incident photon direction relying only by subtle shower shape difference, its mass determination for the long-lived ALP can only be done at the edge of the reconstructed invariant mass distribution when its lifetime ($\tau$) is several tens of cm in $c\tau$ , as shown in Figure \ref{fig:LonglivedALP}. Adding an extra device to become sensitive to incident photon direction is expected to significantly increase the mass determination power for such physics
case.

\subsection{Upgrade hypothesis}
\subsubsection{Replacement of CsI(Tl) crystals with Pure CsI scintillator counters}
In order to eliminate pile up noise, faster scintillation crystal gives a fundamental solution. Pure CsI is suitable for this purpose because its fast light component decay time is about 30 ns, a factor of 40 faster than CsI(Tl). Pure CsI light output is about 20 times smaller than CsI(Tl) and its peak emission of 310 nm is in the near-UV region where photosensors are less sensitive. That is why the use of CsI crystals for the Belle II calorimetry is very challenging and requires the identification of performing photodetectors in the near-UV region and a fast electronics for the signal readout. To match this condition, avalanche photodiode (APD) readout based on the Hamamatsu APD S8664-55 \cite{ecl:hama2} and S8664-1010 \cite{ecl:hama1} has been studied. The S8664 series APD is usually operated at the gain $\times 50$, using it at higher gain of $\times 100$ is found to be still stably operational. Attaching a wavelength shifter (WLS) to emit about 500 nm wavelength light is also effective to get larger signal pulse by a factor 2$\sim$3. Combining these methods, the best one has been identified as 4 Hamamatsu APD S8664-55 placed on the edge of the crystal with the photodetector coupled to a WLS to satisfy the requirement on the equivalent noise energy at sub MeV level. 

\subsubsection{A preshower detector in front of ECL}

\begin{wrapfigure}{R}{0.5\textwidth}
\centering
\vspace{-18pt}
\includegraphics[width=0.4\textwidth]{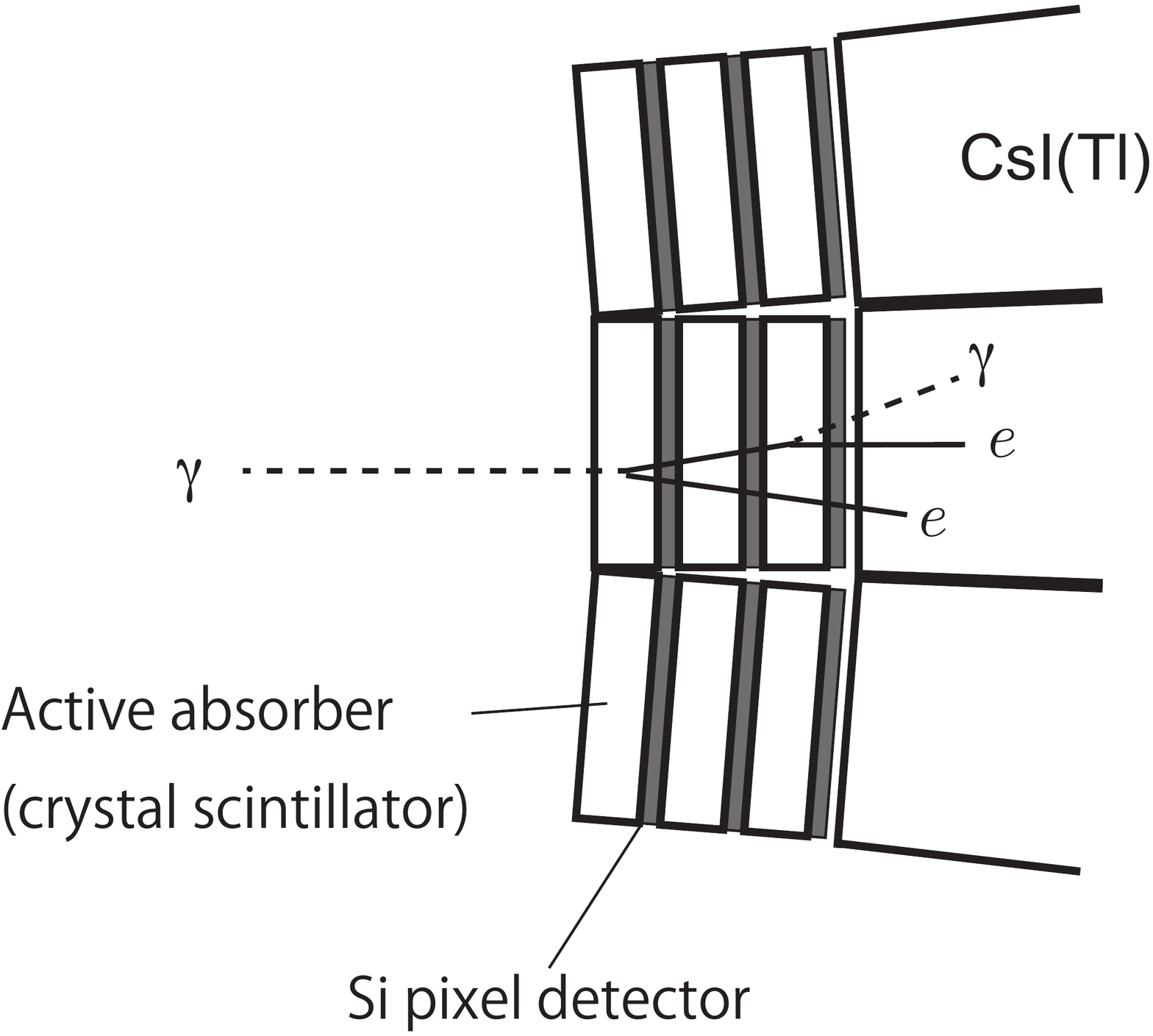}
\caption{Concept of the preshower detector. It is placed just in front of ECL and having capability to detect energy deposition, shower position and thus becomes sensitive to the photon incident direction.}
\label{fig:concept}
 \vspace{-18pt}
 \vspace{1pt}
\end{wrapfigure}

A second option considered for the ECL upgrade is a small detector system just in front of the ECL acting as a preshower detector as shown in Fig.\ref{fig:concept}. 
That has a sensitivity for photon incident direction, better incident position resolution. Superior timing resolution can be featured. One layer consists of the active absorber with 1 radiation length thick crystal scintillator and the Si pixel detector to detect electron and positron passage for the created electromagnetic shower. 
A GEANT4 based simulation with BGO/LYSO as the active absorber material and 1 mm$^2$ pixel shows that the incident direction can be reconstructed for 80\% of the 1 GeV photons because they have two or more layers Si pixel detector hits, position and angular resolutions are found to be 1.9 mm and 0.08 rad., respectively, for normal incident. For the long-lived ALP decay to a photon pair with $m_a=2$ GeV/$c^2$ and 40 cm flight, the photons from IP can be separated about 2 $\sigma$. We also see that significant portion of events' energy deposition is a few hundred MeV, thus the absorber part must be active with a reasonably good energy resolution. As for introduction of such a preshower detector, (1) it can act as a shield to stop significant fraction of beam background soft photons to prevent their incident into the main ECL, thus mitigate pile up noise effect, (2) longitudinal sampling may help to distinguish photons and neutral hadrons by their different shower development characteristics, such possibility is worth to pursuit in addition to the already implemented pulse shape discrimination (PSD)~\cite{LONGO2020164562} technique to improve particle identification capability inside ECL, (3) it is interesting to identify physics cases due to better photon incident position resolution, timing resolution and capability to reconstruct incident photon direction, and (4) geometrical activity match between the preshower and the main ECL also may help to reduce backgrounds.

\subsubsection{Replacement of the PiN diode photosensors with Avalanche Photodiodes APD}

\begin{figure}[ht]
    \centering
    \includegraphics[scale=0.4]{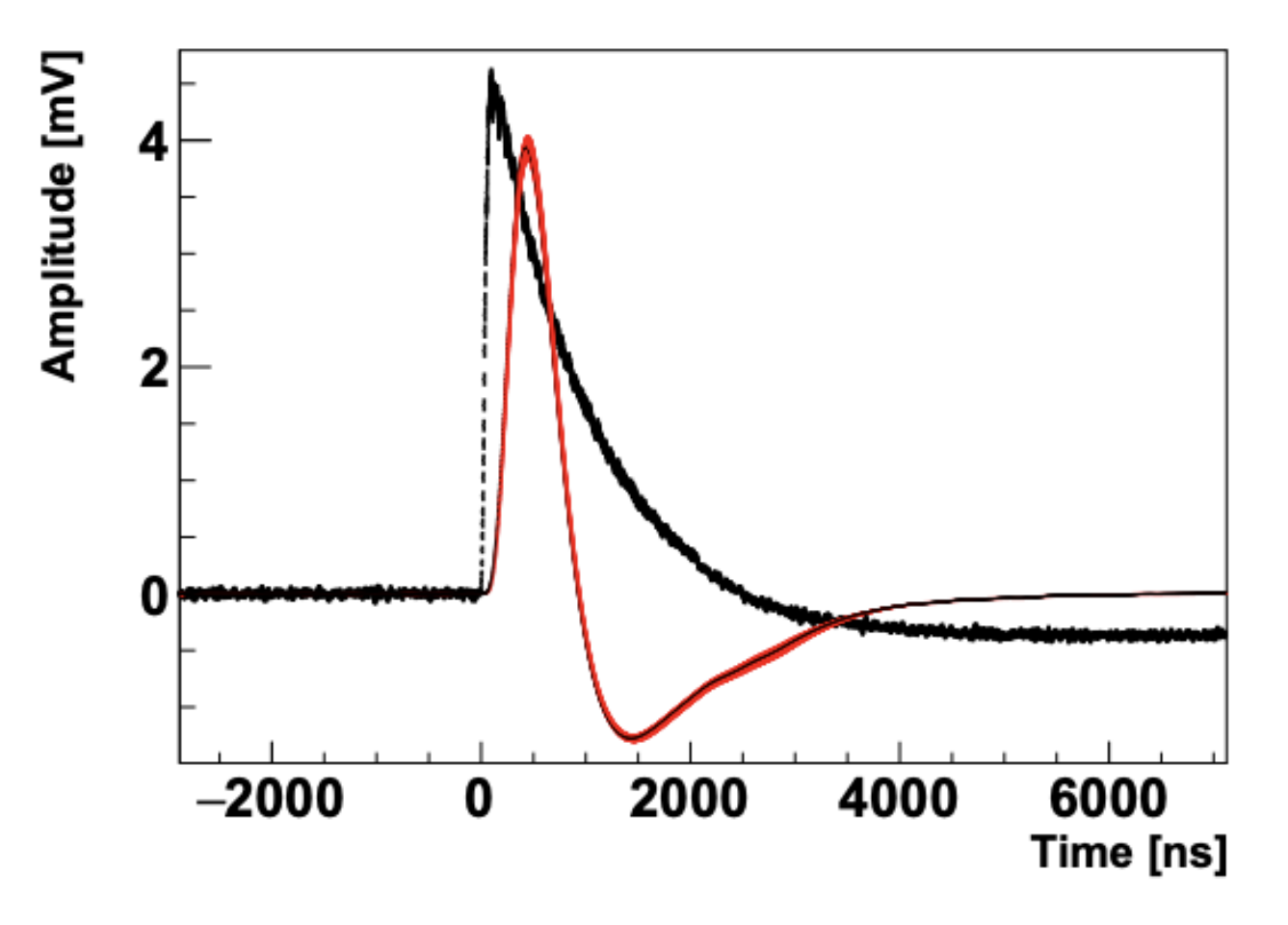}\includegraphics[scale=0.5]{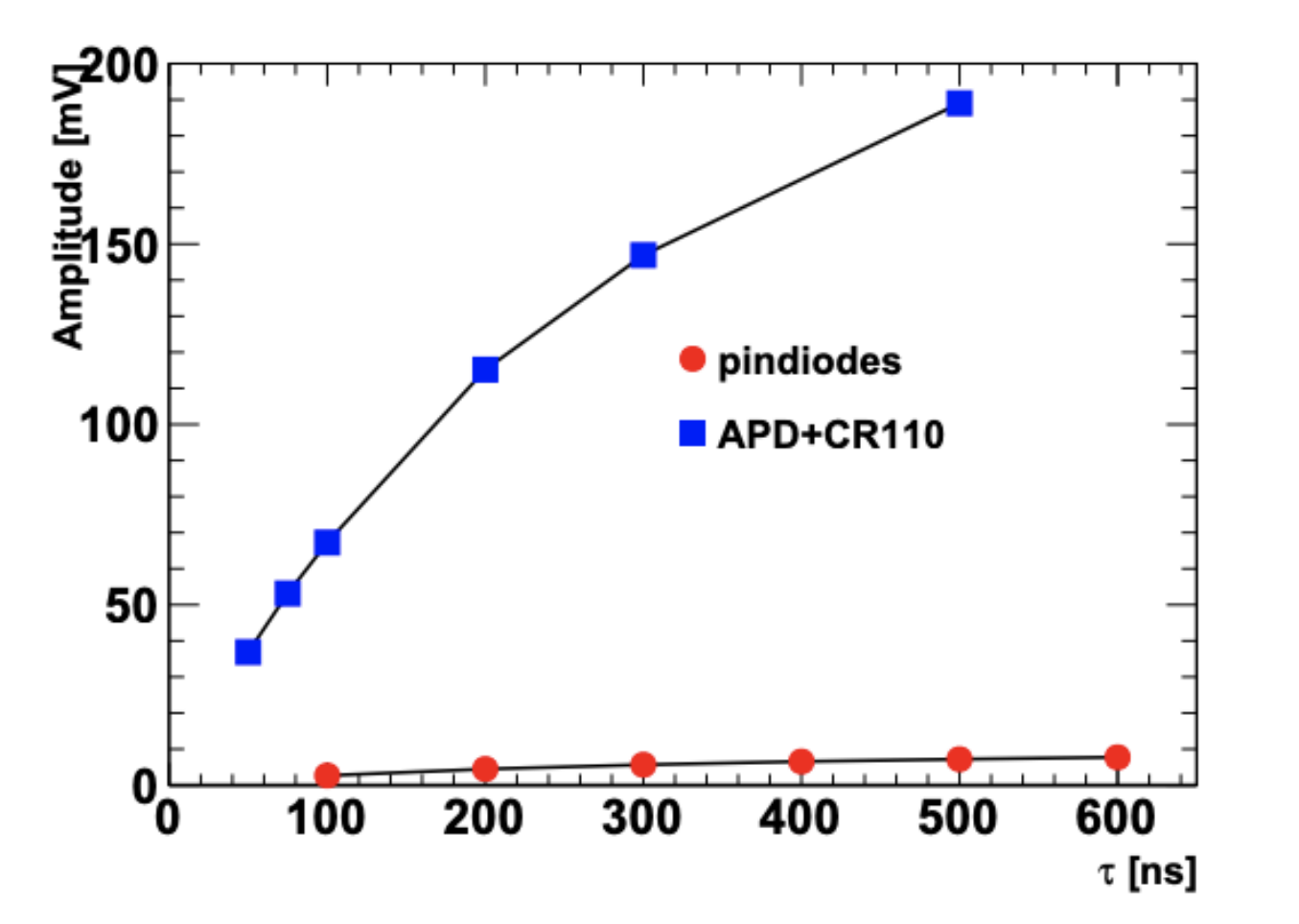}
    \caption{Signal amplitude for the APD plus transimpedence preamplifier configuration (left). Signal amplitude after shaping in the case of APD plus CR110 preamplifier (right).}
    \label{fig:signal-CsI(Tl)-APD-TZA-CR110}
\end{figure}

In this section the possibility of complementing the Belle photodiodes with
two high-gain APDs for the to readout of the CsI(Tl) crystals will be described. The APDs were readout with either transimpedence amplifiers (TZA) [13] or with the charge-integrating CR-110 amplifiers.
The APD+TZA signal is shown in Figure \ref{fig:signal-CsI(Tl)-APD-TZA-CR110} (left) while the signal amplitude, after the $CR-(RC)^4$ filter, from APDs and the charge-integrating CR-110 is shown on the right side of the same Figure.

\begin{figure}[ht]
    \centering
    \includegraphics[scale=0.5]{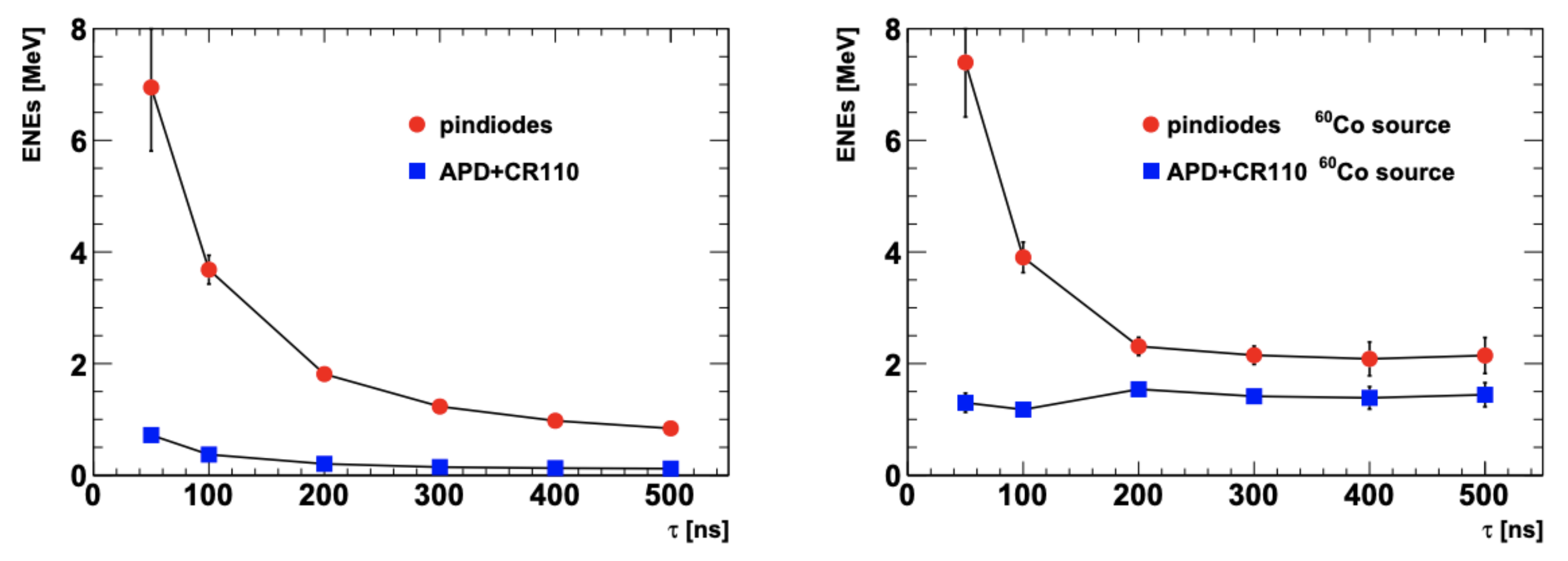}
    \caption{Equivalent Noise Energy as a function of the shaping time comparing pin diodes with APD with CR110 amplifier without (left) and with (right) $^{60}$Co source.}
    \label{fig:ENE-CsI(Tl)-APD-CR110}
\end{figure}

The Equivalent Noise Energy (the input-referred noise calculated from the signal to noise ratio at the shaper output, indicated as ENE) of the sum of two APDs with CR-110 readout, compared to the pin diodes on the same data run, is shown in Figure \ref{fig:ENE-CsI(Tl)-APD-CR110} with and without background overlay. 

 The large signal grants the APD's a very small electronic noise and a reduced pile-up noise with respect to the pin diodes.
 
\subsection{Cost and schedule}

An estimation of the main cost for the Pure CsI crystals with APD readout + WLS option is reported in Table \ref{tab:CsIPure-cost}.

\begin{table}[htb]
    \centering
    \begin{tabular}{|l|c|c|} \hline \hline 
                           & Cost/unit (k\euro  ) & Total cost (M\euro)\\ \hline
         Crystals (8736)   & 4.9                 & 42.5  \\
         APD's (4/crystal) & 0.1                 & 3.5 \\
         WLS   (1/crystal) & 0.05                & 0.4        \\
         FE    (4/crystal) & 0.3                 & 2.6  \\ \hline \hline 
         TOTAL             &                     & 49 \\ \hline
         
    \end{tabular}
    \caption{Estimated cost for the Pure CsI crystal calorimeter with APD + WLS readout.}
    \label{tab:CsIPure-cost}
\end{table}

A rough estimation of the main cost for the pre-shower otpion has been evaluated (see Table \ref{tab:pre-shower-cost}) considering the following dimensions:
each endcap has 4.4 m$^2$ (Outer radius 1.25 m, Inner radius 0.42 m),
the Barrel Inner cylinder surface is 11.7 m$^2$ wich results in 20 m$^2$ in total.
\begin{table}[htb]
    \centering
    \begin{tabular}{|l|c|c|} \hline \hline 
                    & Cost/unit (k\euro) & Total cost(M\euro)\\ \hline
Si sensors          & 100/m$^2$                & 2  \\
LYSO                & 600/m$^2$                  & 12 \\
Photosensor (3 layers, 2pc, 8736 crystal) & 100 & 5.2       \\
Readout                  &                 & 2  \\ 
Mechanics                  &                 & 2  \\ \hline \hline 
TOTAL                &                     & 23.2 \\ \hline
         
    \end{tabular}
    \caption{Estimated cost for the pre-shower calorimeter.}
    \label{tab:pre-shower-cost}
\end{table}

Finally, an estimation of the main cost for the CsI(Tl) crystals with LAAPD readout option is reported in Table \ref{tab:CsI(Tl)-cost}.

\begin{table}[htb]
    \centering
    \begin{tabular}{|l|c|c|} \hline \hline 
                           & Cost/unit (k\euro  ) & Total cost (M\euro)\\ \hline
         APD's (2/crystal) & 0.5                 & 4.4 \\
         FE    (2/crystal) & 0.2                 & 1.7  \\ 
         ECL extraction    &                     & 2.0 \\ \hline \hline 
         TOTAL             &                     & 8.1 \\ \hline
         
    \end{tabular}
    \caption{Estimated cost for the CsI(Tl) crystal calorimeter with APD readout.}
    \label{tab:CsI(Tl)-cost}
\end{table}

\subsection{Potentially interested community in Belle II}
The following collaborators expressed their interest in the hardware upgrade activities: Budker Institute of Nuclear Physics (BINP), Russia, KEK, Japan, Nara Women's University, Japan, University of Naples and INFN-NA, Italy, University of Perugia and INFN-PG, Italy.

\printbibliography[heading=subbibliography]
\clearpage
\end{refsection}

\section{KLM}
\begin{refsection}
\label{sec:KLM}
\editor{L.Piilonen, X.Wang, G.Varner}

A performance upgrade has been proposed for the Belle II KLM subdetector, which identifies $K_L$ mesons and muons with momenta up to about 4.5 GeV/$c$, to add the capability of energy measurement for the $K_L$ mesons via a time-of-flight measurement. This is described in Sec.~\ref{timing-upgrade}.

An independent upgrade has been proposed to refactor the existing scintillator readout by moving the analog and 
analog-to-digital elements of the readout into the detector panels, simplifying immensely the external connections. This 
upgrade would rebuild each detector panel and replace all existing resistive plate counters (RPCs) with scintillators. 
This is described in Sec.~\ref{readout-upgrade}.

\subsection{KLM geometry and instrumentation}

The KLM subdetector occupies the volume outside the superconducting solenoid, spanning radii of 200--$340~\si{\centi\metre}$ in the 
octagonal barrel and 130--$340~\si{\centi\metre}$ in the forward and backward endcaps. Large-surface-area detector panels of about 
3.1-$\si{\centi\metre}$ thickness are sandwiched in the gaps between the steel plates of the solenoid's magnetic-field flux return. 
Each panel contains either RPCs (barrel layers 3--15) or scintillator strips (barrel layers 1--2, forward-endcap layers 
1--14 and backward-endcap layers 1--12). The RPC cathode readout strips are about $4~\si{\centi\metre}$ wide and each scintillator 
strip is $4~\si{\centi\metre}$ wide, measured orthogonal to a line from the SuperKEKB $e^+ e^-$ collider's interaction point (IP); 
this matches the Moli\`ere radius of muons originating at the interaction point and traversing the electromagnetic 
calorimeter located just inside the solenoid. Each detector panel provides two orthogonal measurements of a 
through-going particle's position via hits on these strips.  There are about 38,000 readout channels in the KLM, of 
which half are for the scintillators.

The cross section of each barrel scintillator strip is $4~\si{\centi\metre} \times 1~\si{\centi\metre}$, so each panel in layers 1--2 contains 54 
$z$-readout strips and between 38 and 69 $\phi$-readout strips, for a total of 93 to 124 strips. The cross section of 
each endcap scintillator strip is $4~\si{\centi\metre} \times 0.75~\si{\centi\metre}$; each panel contains 75 $x$-readout strips and 75 $y$-readout 
strips for a total of 150 strips. Photons from each scintillator strip are collected by an embedded wavelength-shifting 
optical fiber (Kuraray Y11(200)MSJ) and detected by a Silicon Photomultiplier (SiPM), also known as a Multi-Pixel Photon 
Counter (MPPC: Hamamatsu S10362-13-050C)---an avalanche photodiode operated in Geiger mode. This solid-state device is 
affixed directly to the scintillator strip at one end of the WLS fiber.  The fiber's other end is mirrored. The signals 
(preamplified internally by a factor of $\sim$10 for the scintillators) are routed via ribbon cables to external readout 
electronics mounted on the magnet yoke.

\subsection{Timing upgrade for $K_L$ momentum measurement}
\label{timing-upgrade}

There is no marked difference in the current particle-identification performance measures between RPCs and 
scintillators. However, the scintillator readout can be upgraded to provide a precise time-of-flight measurement for 
the hit clusters that arise from hadronic interactions of a $K_L$ meson in either the yoke steel or in the ECL (with 
shower leakage into the KLM). This measurement can be converted into the determination of the $K_L$ meson's energy, 
which is not measurable in the present KLM. Also, the precise hit-time measurement can be applied to a neutron's primary 
hit in a single one-dimensional scintillator strip to better reject out-of-time neutron backgrounds that would 
otherwise contaminate the prompt $K_L$-candidate population.

As in a typical time-of-flight (TOF) counter based on scintillators, it is possible to achieve a precise time 
measurement of a hadronic cluster created by a neutral hadron, such as $K_L$ or a neutron. Converting this to an energy 
measurement and then combining it with the present direction measurement (assuming that the hadronic cluster originated 
from the IP), the four-momentum of a neutral hadron can be determined.
%
A time resolution of 
$\delta t = 100~\si{\pico\second}$ translates into a momentum resolution of $\delta p = 0.19~\hbox{GeV}/c$ or a fractional momentum 
resolution of $\delta p/p \approx 13\%$. 

To measure the $t$ of a neutral hadron precisely, both the start and stop times have to be measured.
The start-time resolution is less than $32~\si{\pico\second}$, dominated by the $e^+ e^-$ collision-time determination.
The stop time will be measured to a comparable precision from the collection of hits in the hadronic cluster 
in the new KLM.

\begin{figure}[htb]
\begin{center}
\includegraphics[width=0.35\textwidth]{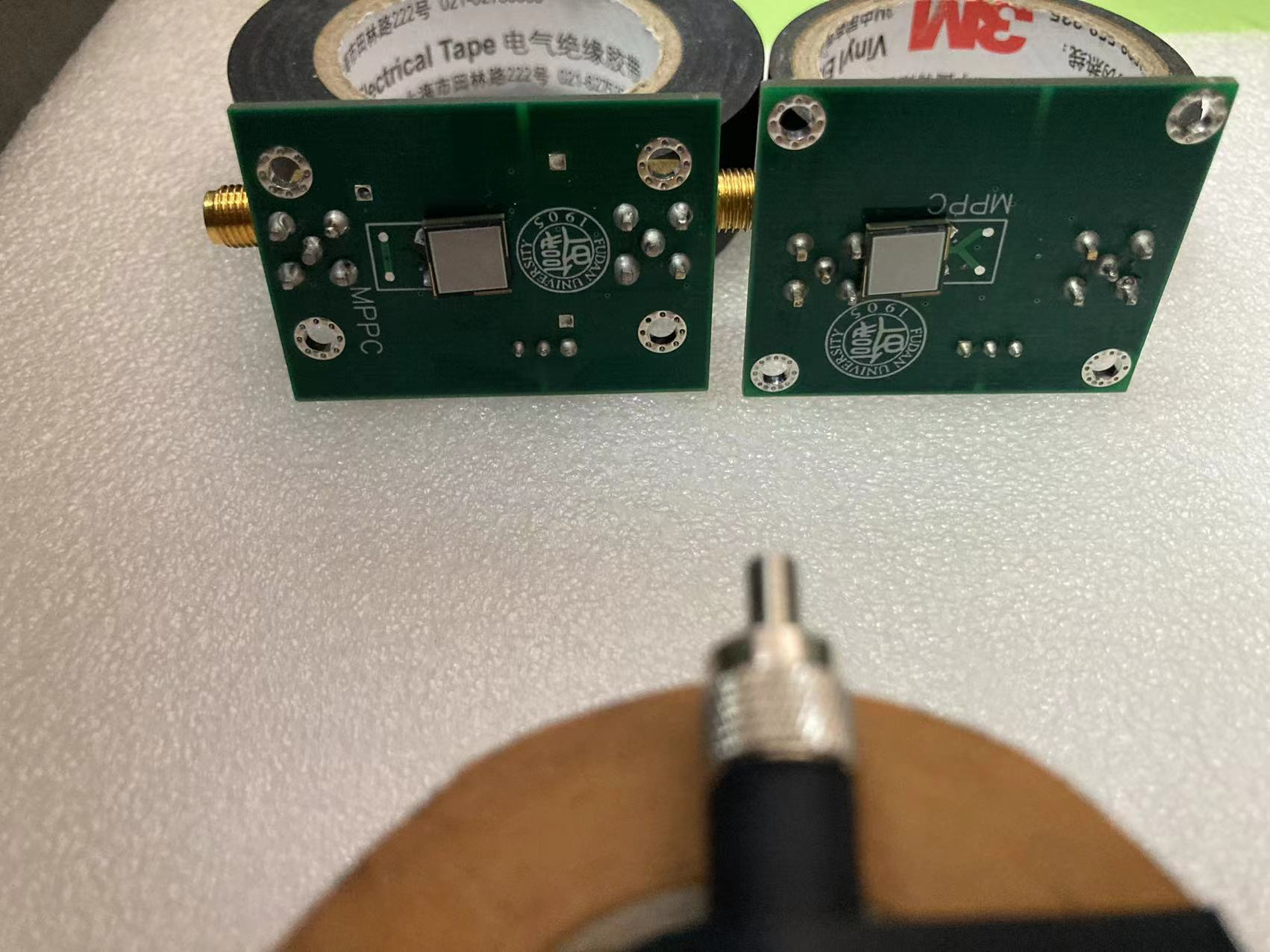}
\includegraphics[width=0.55\textwidth]{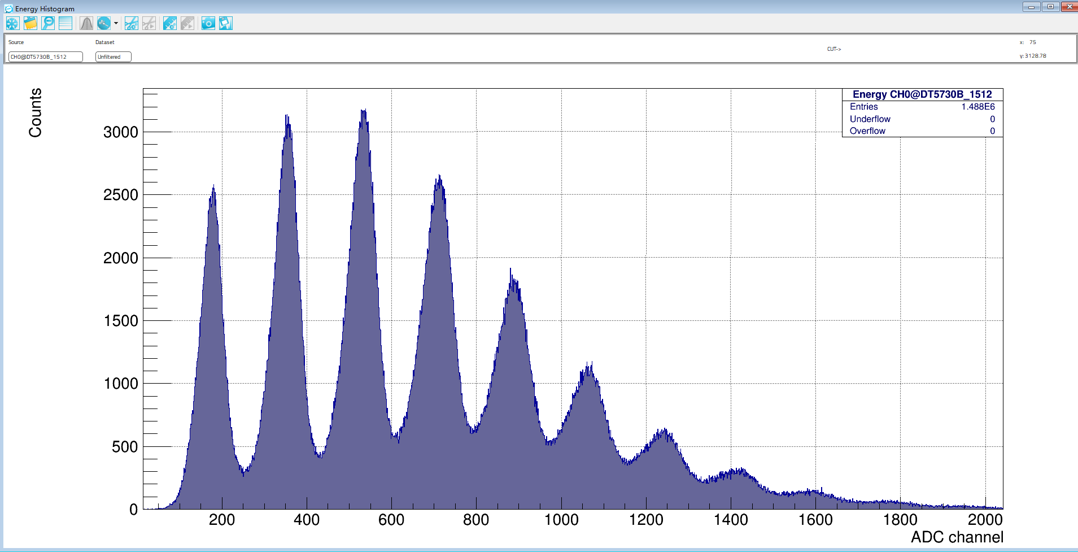}
\caption{\sl Left: two sets of new photosensors with newly designed preamplifier and a laser source. Right: the ADC 
distribution shows clear peaks of p.e.}
\label{fig:new_SiPM_readout}
\end{center}
\end{figure}

%
With several years of R\&D in hand, it is found that the combination of scintillator with long attenuation length
\textit{and without an embedded wavelength-shifting fiber}, 
large-area photosensor and a new preamplifier with improved timing is able to achieve the desired time resolution. 
Photosensors with $6\times 6~\si{\milli\metre}^2$ area from Hamamatsu and NDL (Beijing) are tested. Figure~\ref{fig:new_SiPM_readout} 
shows testing with the new Hamamatsu MPPC (S14610-6050HS, powered at $V_{\rm BR} \approx 38~\si{\volt}$) and a newly designed 
preamplifier. With a laser whose time resolution is $100~\si{\pico\second}$, clearly distinguished peaks of $n$ photoelectrons ($n = 
1,\ 2,$ ...) are seen in Fig.~\ref{fig:new_SiPM_readout} (right). This setup shows a time resolution of $45\pm 1~\si{\pico\second}$. 
Another test with the NDL photosensor (EQR15 11-6060D-S, powered at $V_{\rm BR} = 28.0\pm 0.2~\si{\volt}$) shows a time 
resolution of $32.9\pm 0.4~\si{\pico\second}$. 

With the new photosensor and preamplifier, a pair of 
$4~\si{\centi\metre} \times 1~\si{\centi\metre} \times 10~\si{\centi\metre}$
scintillator strips with new scintillation material having long attenuation 
length is tested.
Four $6\times 6~\si{\milli\metre}^2$ SiPMs are combined and mounted to one end of each strip. Preliminary testing with cosmic rays 
illuminating both strips shows a time resolution of $127.8\pm 3.0~\si{\pico\second}$ per strip. There is additional capacity to 
improve this overall time resolution by using a better ground in the readout electronics, increasing the visible-light 
yield from the scintillator and/or the collection of light by the photosensor(s) at the end of the strip.
Having multiple photosensors at the end of the scintillator strip is feasible, considering their low price.
The cross-sectional shape of the new scintillator can be re-optimized to improve the light yield per strip while not 
sacrificing the positional resolution for IP-originating muons. The coupling between the end of the scintillator strip 
and the photosensor active surface can be improved as well.
%

A typical hadronic cluster will contain one to $\approx 20$ 2D hits in the KLM, depending on the parent hadron's energy.
Combining the hit times of the multiple 
2D hits in a more energetic hadron-induced cluster can improve the time resolution further; such an improvement is 
needed at the higher $K_L$ or neutron momenta. A toyMC study shows that ten 2D hits (i.e., twenty strips) with a time 
resolution of $200~\si{\pico\second}$ for an individual strip yields an average time resolution of less than $50~\si{\pico\second}$.

To maintain the required timing precision,
a robust time-calibration protocol with 
frequent updates several times per day will be necessary for the TOF-like KLM.

\subsection{Scintillator Readout Upgrade}
\label{readout-upgrade}

Leveraging experience from the deployment and operation of the endcap and inner barrel KLM scintillators, there is a 
compelling reason to refactor the readout in a way that eliminates the many
kilometers of twisted-pair ribbon 
cables. 
%
%
As upgrade readout baseline, a more compact 
readout will be placed inside each detector panel. 
This internal readout interfaces with the external DAQ system via 
a few optical fibers.

Independently of this proposed readout-electronics upgrade for the newly installed scintillator panels in the barrel 
KLM, progress is underway to address some of the deficiencies of the present readout design for the existing 
scintillator panels in the endcaps and inner two barrel layers. New firmware for the ASICs has been written and tested 
to perform waveform sampling and provide a higher-resolution ($< 1~\si{\nano\second}$) time measurement than is possible with the 
existing latch (binary) readout. This may help determine the energy of slow $K_L$ mesons via time-of-flight 
but will only be of sufficient precision for momenta below about $0.7\,{\rm GeV}/c$. (This 
waveform sampling can also measure the pulse area, but this measurement is and will remain compromised by the saturation 
of the $\times$10 preamplifiers that reside within the scintillator panels.) At the same time, this waveform sampling 
algorithm will be able to sample the SiPM output just before and after the main pulse and thereby correct for the 
localized shift in the waveform baseline that is expected in the coming years from the accumulated direct dose of 
neutrons in the SiPMs and the resultant increase in the spontaneous single-pixel discharge noise from each SiPM.

Again, independently of this proposal's readout-electronics upgrade, work is underway to improve the algorithm in the 
KLM trigger firmware to more reliably detect and report activity in the KLM arising from track segments (cosmic rays or 
IP-originating), clusters, leakage from the ECL (particularly in the uninstrumented cracks between the ECL barrel and 
endcaps), and the forward polar-angular region outside the CDC coverage. KLM-trigger tracks can then be aligned with 
CDC-trigger tracks in the GRL/GDL. The goal is to be able to trigger on such KLM activity with sufficient purity that 
other conditions such as a back-to-back requirement on the KLM trigger output will not be needed, so that the KLM 
trigger can serve as a reliable veto for dark-matter studies.


Motivated by the considerations described previously, we propose a modified readout design that incorporates the 
following features:
\begin{itemize}
\item Embed the readout ASICs on circuit boards inside new detector panels (including those in the two innermost 
layers); 
\item Develop a compact SCROD (Control and Readout) that resides inside the detector panel;
\item 64-channel HDSoC readout ASICs are employed, each with high-speed serial connection to the new SCROD FPGA.
\end{itemize}

The main readout card inside the detector panel can accommodate a higher channel density ASIC. The baseline ASIC being considered for upgrade of the TARGET ASIC is the HDSoC ASIC, which has 4 times higher channel density, as well as streaming readout.


Employing the HDSoC ASIC, a readout system that integrates into the new barrel scintillator panels 
will consist of 480 ASICs on dedicated cards, and 240 SCRODs to service them.


\subsection{Cost estimation}

\begin{itemize}
\item {\bf Materials Cost Estimate}

Table~\ref{table:materialcost} provides the material-cost estimate for manufacturing new scintillator-strip panels for 
the outer thirteen layers of the barrel KLM as part of the scintillator-readout upgrade described in 
Sec.~\ref{readout-upgrade}. The material costs are based on the actual costs for the same items used to make
the scintillator-strip panels for the inner two barrel layers, scaled up by the relative sizes and numbers of the panels
and inflated at a rate of 3\% per year.
The panel-construction labor cost depends on where the panels are fabricated, since some of the personnel might already 
be on the institution's payroll. The low-end estimate is \$1.3M.

%
\item {\bf Readout Cost Estimate}

The total cost for 3 options is listed in Table~\ref{KLM:table4}. Based upon the preliminary prototyping being performed 
and costed separately from this, the total project cost is estimated to be \$1.4-1.8M for upgrades to the KLM readout 
system, depending upon amplifier choices.

\item {\bf Cost Estimate for TOF-like KLM}

To achieve a good timing resolution,
the innermost  
layers
should be replaced by new scintillators with long attenuation length and large-area 
photosensors.
The cost of scintillators would be about \$1.8M for 
three innermost barrel layers and about \$0.25M for five innermost endcap layers.
With four SiPMs per strip,
the cost for SiPMs will be \$0.3M for the barrel and \$0.17M for the endcaps. Therefore, the additional material cost
is about \$2.5M.

\end{itemize}

In summary, the rough estimation on the cost for KLM upgrade is about \$7.8M.

\begin{table}[htb]
\begin{center}
\begin{tabular}{lrcrcr}
\hline
                & {\bf L0--1} &                        & {\bf L2--14} & {\bf Inflation} & {\bf Cost} \\
                & {\bf Cost} & {\bf Geometry} & {\bf Cost} & {\bf 13 years} & {\bf Estimate} \\
{\bf Item} & {\bf (2012)} & {\bf scale factor} & {\bf (2012)} & {\bf (1.03/year)} & {\bf (2025)} \\
\hline
Scintillator strips & 50,650 & 7.60 & {\it 384,940} & 1.47 & 565,862 \\
WLS optical fiber & 36,564 & 7.60 & {\it 277,886} & 1.47 & 408,493 \\
Photosensors & 85,658 & 7.60 & {\it 651,001} & 1.47 & 956,971 \\
Aluminum, HDPE, etc  & 49,507 & 7.60 & {\it 376,253} & 1.47 & 553,092 \\
Shipping to KEK & 16,800 & 7.60 & {\it 127,680} & 1.47 & 187,690 \\
HV modules & 20,672 & 6.67 & {\it 137,816} & 1.47 & 202,590 \\
HV mainframes & 37,801 & 0.33 & {\it 12,600} & 1.47 & 18,522 \\
\hline
Subtotal & & & {\it 1,968,177} & & 2,893,220 \\
Contingency (20\%) & & & {\it 393,635} & & 578,644 \\
\hline\hline
\multicolumn{3}{l}{\bf TOTAL MATERIAL COST} & {\it 2,361,812} & & {\bf 3,471,764} \\
\hline
\end{tabular}
\caption{\sl Materials-procurement cost estimate (US dollars) for the outer thirteen layers of the barrel KLM.}
\label{table:materialcost}
\end{center}
\end{table}

\begin{table}[hbt]
\caption{\sl KLM Upgrade readout Amplifier Option total cost estimates (common + option additional cost).}
\centering
\begin{tabular}{ | l | c | c | c | } \hline
{\bf Amplifier Option} & {\bf Total cost [k\$]} & {\bf Rough Cost [M\$]}  \\ \hline \hline
 1) HDSoC ASIC internal amplifiers & 1,441 & {\bf 1.4}  \\
 2) discrete amplifiers & 1,762 & {\bf 1.8}  \\
 3) dedicated amplifier ASIC & 1,660 & {\bf 1.7}  \\  \hline \hline
\end{tabular}
\label{KLM:table4}
\end{table}

\printbibliography[heading=subbibliography]
\clearpage
\end{refsection}

\section{Trigger}
\begin{refsection}
\label{sec:TRG}
\editor{T.Koga}
\subsection{Introduction}

The hardware trigger system (TRG) was designed to satisfy the following requirements at the nominal design luminosity of $8\times10^{35}cm^{-2}s^{-1}$:
\begin{itemize}
    \item high efficiency for hadronic events from $ Y(S) \rightarrow BB $  and from continuum
    \item a maximum average trigger rate of 30 kHz
    \item a fixed latency of 4.5 µs
    \item a timing precision of less than 10 ns
    \item a trigger configuration that is flexible and robust
\end{itemize}

\begin{figure}[ht]
\begin{center}
\includegraphics[scale=0.5]{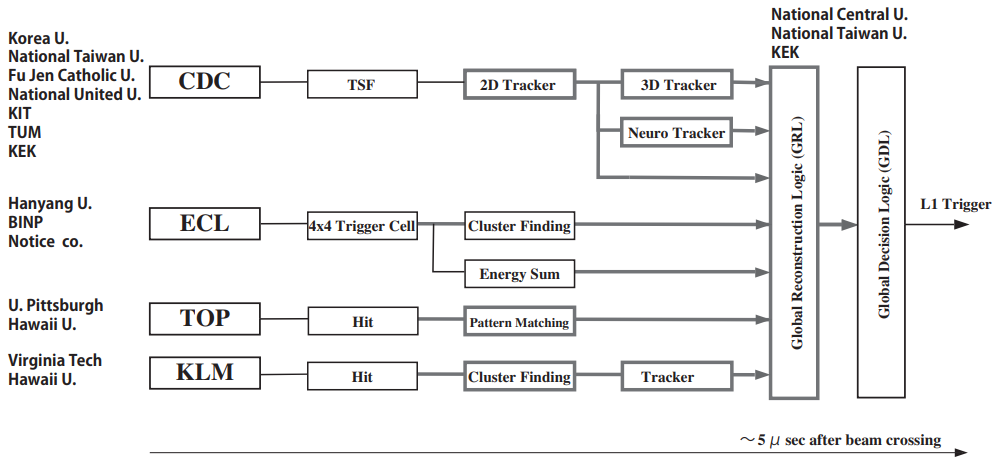}
\caption{\it Schematic view of the hardware trigger system}
\label{fig:trgview}
\end{center}
\end{figure}

Figure~\ref{fig:trgview} shows an overview of the system. The CDC and ECL take a major role to trigger charged particles and photon. The KLM triggers muon and the TOP measures event timing precisely. The information from sub detector triggers are send to the Global Reconstruction Logic (GRL) for the matching and final trigger decision is made at the Global Decision Logic (GDL). Although TRG is working well at present as originally designed, further upgrade is needed to satisfy the requirements with the nominal design luminosity or the higher luminosity, due to the higher beam originated background rate than the initial expectation. In addition, it is important to keep high efficiency of the non-B physics with low multiplicity of the final state particles, such as physics with tau, dilepton, and dark/new particles searches. In order to achieve that, several upgrades are planned for both hardware and firmware with various time schedules.

\subsection{Hardware upgrade}

TRG consists of several kinds of FPGA based electronic circuit board. Each board is connected with optical communications, as well as the detector front end boards. The universal trigger board (UT) has been developed as a main board, as shown in Table~\ref{tab:trg_ut}. Upgrade of the UT is on-going to improve the resource size of the FPGA and the optical transmission rate, in order to implement more complicated core trigger logic with increased information from the detector. At present, the 3rd (UT3) and 4th (UT4) generation of the boards are used for the TRG operation. The number of UT4 is planned to be increased in the short and medium term. Development and production of the next generation of the UT (UT5) are scheduled in the medium and long term. In addition, configuration of the system need to be modified when detector output from the front end board is upgraded. 

\begin{table}[hbt]
    \centering
    \begin{tabular}{c|ccc}
        \hline 
        \hline 
        UT generation & UT3 & UT4 & UT5 \\
        \hline 
        Main FPGA (Xilinx) & Virtex6         & Virtex Ultrascale &  Virtex Ultrascale$+$ \\
                           & XC6VHX380-565   & XCVU080-190       &                       \\
        Sub FPGA (Xilinx)  & $--$              & Artex7            &  Artex7, Zynq         \\
        \# Logic gate             &  500k & 2000k         &  8000k            \\
        Optical transmission rate &  8~Gbps  & 25~Gbps           &  32~Gbps              \\
        RAM                    & $--$ & DDR4 & DDR4, UltraRAM  \\
        \# UT boards             &  30 & 30  & 10 \\
        Cost per a board (k\$) & 15     & 30        & 50 \\
        Time schedule          & 2014-  & 2019-2026 & 2024-2032 \\
        \hline 
        \hline 
    \end{tabular}
    \caption{Specification of the universal trigger boards.}
    \label{tab:trg_ut}
\end{table}

\subsection{Firmware upgrade}

In order to improve signal efficiency and background rejection, several firmware upgrades are planned with the new UT, as shown in Table~\ref{tab:trg_firm}. Because the firmware modification can be done anytime with FPGA, the plan will flexibly change, depending on machine condition, physics discovery, performance of new logic, new idea and human resources etc. For each upgrade component, one or two full time workers are needed as the human resource.

%
%
\begin{table}[b]
	\centering
	\begin{tabular} {lp{0.3\linewidth}p{0.25\linewidth}lll}
	\toprule
        Component          & Feature & Improvement & Time & \#UT  \\
	\hline
        CDC cluster finder          & transmit TDC and ADC from all wires with the new CDC front end    & beamBG rejection & 2026 & 10 \\
        CDC 2Dtrack finder          & use full wire hit patterns inside clustered hit                   & increase occupancy limit & 2022 & 4 \\
        CDC 3Dtrack finder          & add stereo wires to track finding                                 & enlarge $\theta$ angle acceptance & 2022 & 4 \\
        CDC 3Dtrack fitter (1)      & increase the number of wires for neural net training              & beamBG rejection         & 2025 & 4 \\
        CDC 3Dtrack fitter (2)      & improve fitting algorithm with quantum annealing method            & beamBG rejection         & 2025 &  4 \\        
        Displaced vertex finder     & find track outside IP originated from long loved particle         & LLP search            & 2025 & 1 \\
        ECL waveform fitter         & improve crystal waveform fitter to get energy and timing          & resolution & 2026 & -- \\
        ECL cluster finder          & improve clustering algorithm with higher BG condition             & beamBG rejection & 2026 & 1 \\
        KLM track finder            & improve track finder with 2D information of hitting layers        & beamBG rejection & 2024 & -- \\
        VXD trigger                 & add VXD to TRG system with new detector and front end    & BG rejection & 2032 &  -- \\
        GRL event identification    & implement neural net based event identification algorithm         & signal efficiency  & 2025 & 1 \\
        GDL injection veto          & improve algorithm to veto beam injection BG                       & DAQ efficiency & 2024 & -- \\

	\bottomrule
	\end{tabular}
		\caption{TRG firmware upgrade plan.}
    \label{tab:trg_firm}
\end{table}

\printbibliography[heading=subbibliography]
\clearpage
\end{refsection}

\section{DAQ}
\begin{refsection}
\label{sec:DAQ}
\editor{S.Yamada}

We have near-term or already ongoing upgrade plans for the Belle II DAQ system~\cite{daq:daq1}~\cite{daq:daq2} to replace old hardware with new technology and keep up with increasing dataflow as the luminosity of SuperKEKB accelerator increases. In this section, we describe two components of the Belle II DAQ system, which are to be upgraded in near future. One is a readout system, and the other is a high-level trigger (HLT) system.

\subsection{Readout system}
The functionalities of the readout system~\cite{daq:readout} are receiving data from front-end electronics boards and performing partial event-building and a basic sanity check of data. The Belle II DAQ system has a common readout interface with front-end electronics (FEE) of all sub-detectors except for PXD, which needs special data-treatment for online data-reduction. In the current system, we are using COPPER readout boards~\cite{daq:copper} to receive data from FEE. The COPPER board was developed as a versatile board, on which one can attach daughter boards for various purposes. For the Belle II experiment, we installed three different types of daughter boards: a receiver board, HSLB (High Speed Link Board), an interface board with the trigger-timing distribution (TTD) system, TTRX and a CPU board to process data received by a receiver card. A home brew protocol, Belle2link~\cite{daq:belle2link} is used for bidirectional communication between HSLB and FEE boards, such as data transfer from a sub-detector, slow control and monitoring.  

We started to use the COPPER board in the previous Belle experiment. Recently, its maintenance became difficult as the number of discontinued parts for repairment was increasing. That is one concern for using the current readout system during the entire lifetime of the Belle II experiment. In addition to that, the throughput of the readout system is limited by various components, which are PCI bus used on a COPPER board for data-transfer, Gigabit Ethernet PHY chips used on a CPU board and a COPPER board for sending data to servers downstream and the performance of a CPU card in processing data. After considering those concerns, we decided to replace the current readout system with state-of-the-art technology.

To minimize the effect on the other part of the DAQ system in the upgrade, we will keep the same Input/Output(I/O) interfaces as in the current system. Those interfaces are serial data-transfer with an FPGA on a FEE board and a TTD module, and TCP/IP communication with the PC farm of the HLT system. To fulfill the boundary conditions, a new readout hardware would be equipped with a high-performance FPGA with transceivers for serial data-transfer, which can be used with the Belle2link protocol and an interface to Ethernet network such as PCIExpress interface to a server or an FPGA with TCP/IP protocol stack. Since electronics boards with such functionalities were already under development in other physics experiments for the basically same purpose, we decided to use a PCIe40 board~\cite{daq:pcie40}, which was originally developed for the LHCb and ALICE experiments. We then concentrated on the developments of firmware and software to make the hardware fit in with the Belle II DAQ system. A PCIe40 board is equipped with Intel Arria10 FPGA with 48 bi-directional serial links and 16 lanes of PCIExpress 3.0 for I/O interfaces. With the much larger number of links per board compared with a COPPER board, which has 4, we can make a more compact readout system with PCIe40 boards~\cite{daq:daq_upgrade}.

The firmware of the new PCIe40-based system takes care of formatting and data checks which has been done by the software on CPU board in the COPPER system. Those formatted data are read by software running on the host server of a PCIe40 board and are distributed to different HLT units for online data reduction. After implementing those functionalities in the new hardware, we performed a high-trigger-rate test with dummy data produced by KLM FEE boards and achieved the throughput of 630 MB/s per one PCIe40 board and readout PC. The throughput is currently limited by CPU usage of a readout PC when software is double-checking a checksum value of data, which is already checked by firmware of a PCIe40 FPGA. The check by software can be stopped after the operation of the new system becomes stable. To cover all sub-detectors except for PXD, we are going to use 21 PCIe40 boards, which we think should be sufficient to handle data even when the level-1 trigger rate reaches 30 kHz, which is the expected maximum trigger-rate at the designed luminosity of SuperKEKB. Since the bottlenecks of the new readout system is currently CPU and network performances of a host server for a PCIe40 board, we can improve its performance by updating server or network equipment, in case that event size or trigger rate becomes larger than expected which depends on the status of beam background.

For the replacement of the readout system, it can only be done in an accelerator shut-down period, because the Belle II experiment is a currently running experiment. There are two annual shutdown periods of SuperKEKB accelerator in summer and winter, whose length are usually two or three months. Since it is difficult to replace all sub-detectors’ readout system in one shutdown period, we are going to replace the system partially with the unit of sub-detectors. After the development and commissioning of firmware and software for the new system from 2019, we first replaced the TOP and KLM readout system in the summer shutdown of 2021. In the following physics run, we still used the COPPER-based system for other sub-detectors. Since the stability and performance of the new readout system was basically fine for TOP and KLM, we will continue the replacement of other sub-detectors’ readout system in coming shutdown periods in 2022.

\subsection{High level trigger system}
The HLT system~\cite{daq:hlt} of the Belle II DAQ system performs an online event reconstruction and issues a software trigger to reduce the number of events to be recorded on disks, which is necessary to save offline computing resources with the minimum effect on signal efficiency. The HLT system also sends region-of-interest (ROI) information to the PXD readout system, which is calculated by extrapolating a track obtained from other sub-detectors’ data. PXD data of the outside of the ROI can be discarded, as they are unlikely to contain hits of signal tracks. For these purposes, the HLT system collects event-fragments from all-subdetectors except for PXD and performs online event-reconstruction after event-building~\cite{daq:eb} is done, which requires powerful computational power and high network throughput.

Like the readout system, the throughput requirement for the HLT system also gradually increases with the luminosity of the SuperKEKB accelerator. In addition to the luminosity, the beam background, which is difficult to predict precisely at the beginning of the experiment, can also affect trigger rate and event size. Considering that, we adopted a staging plan of the reinforcement of the HLT system to keep up with increasing trigger rate and event size. With this staging plan, we can also benefit from the evolution of information technology to prepare computational resources with higher performance at a later stage.

To make the staging plan possible, the HLT system was designed to be a scalable system, consisting of multiple units, which can work somewhat in parallel, and adding other units will increase the total throughput of the system linearly. An HLT unit consisting of servers and network switches is stored in one rack and performs event building and online data reduction independently of other units. In the HLT unit, there are an input server, worker nodes and an output server. The input server receives event fragments from the read-out system, builds events, and sends them to worker nodes, which perform online event reconstruction. Since the online reconstruction of an event can be done independently of other events, events which are built in the input server are distributed to different worker servers to perform online reconstruction in parallel. On each worker server, there are multiple reconstruction processes running for parallel data-processing to make use of multiple cores in the server. Event selection is performed here with reconstructed information and events which do not meet high-level trigger conditions are discarded. The output server then sends raw data with track and cluster information to a storage system, which is dedicated for recording data only from a specific HLT unit.

From the start of the Belle II experiment, we gradually increased the number of HLT units and number of CPU cores in each HLT unit. In 2019, 10 HLT units were in operation with 2800 CPU cores in total and in the winter shutdown in 2020-2021, the total number of CPU cores was increased to 4800. With this computational power, reconstruction software can currently handle data at the Level-1 trigger rate of around 15 kHz. In the summer shutdown of 2022, we plan to install two or three more HLT units. After this upgrade, the total number of CPU cores in the HLT system reaches 6400, which was the designed value at the beginning of the experiment, and we estimate that the HLT system can process data up to the trigger rate of 20kHz. We will closely watch the increase of trigger rate and event size as the luminosity changes and add more HLT units if necessary.

\subsubsection{Cost and schedule}
The forseeen improvements on the DAQ and HLT systems should be completed by 2023. Table~\ref{tab:daq_costs} shows the estimated costs for the DAQ upgrades. 

\begin{table}[hbt]
    \centering
\begin{tabular}{|c|c|c|}
\hline
item & number & cost \\
\hline\hline
Readout upgrade : PCIe40 hardware & 31 boards  & $\sim$ 26.1 M JPY\\
\hline
Readout upgrade : readout server & 21 & $\sim$ 9.5 M JPY\\
\hline
Readout upgrade : network switch & 2  & $\sim$ 3 M JPY\\
\hline
HLT reinforcement : HLT unit & 3  & $\sim$ 60 M JPY\\
( $\sim$ 500 CPU cores per unit ) &  &  \\
\hline
\end{tabular}
    \caption{Estimated costs for the DAQ upgrade and HLT reinforcement }
    \label{tab:daq_costs}
\end{table}

\printbibliography[heading=subbibliography]
\clearpage
\end{refsection}

\section{Ideas for longer term upgrades}
\subsection{STOPGAP}
\begin{refsection}
\label{sec:STOPGAP}
\editor{O.Hartbrich, U.Tamponi}
The STOPGAP \cite{stopgap} upgrade proposal aims to increase the geometrical acceptance of the currently installed TOP system by instrumenting the gaps between the sixteen TOP quartz bars of about \SI{2}{\cm} width in $\phi$, accounting for about \SI{6}{\percent} of missing geometric coverage in the nominal TOP acceptance. Tracks passing through the outermost edges of a TOP bar also have reduced particle identification performance, effectively widening this gap to around \SI{16}{\percent} of tracks with no or degraded TOP PID.

\begin{figure}[htbp]
	\centering
		\includegraphics[width=0.8\textwidth]{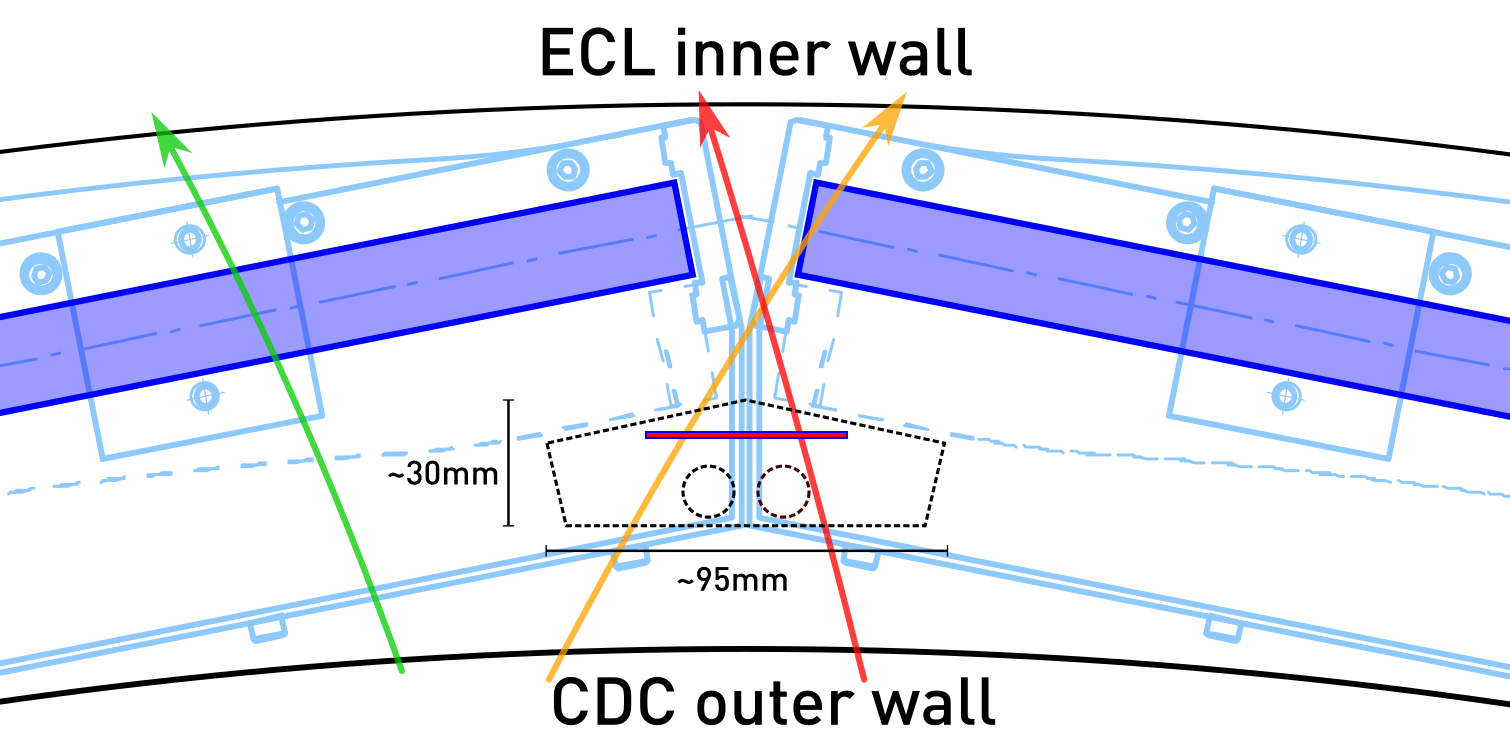}
	\caption[]{Conceptual sketch of supplemental TOP gap instrumentation module in the cross section of two adjacent TOP modules as seen from the backward side. The solid light blue lines show the outlines of the readout side of the TOP quartz bar box. The dashed light blue lines show the outline of the rest of the TOP quartz bar box. The TOP quartz cross section is shown in dark blue. The dashed black lines roughly show the dimensions of a possible STOPGAP module, with the two dashed circles indicating possible cooling lines or cabling channels and a layer of silicon sensors shown in pink.}
	\label{fig:schematic_view}
\end{figure}

A possible remedy is to install a supplemental time-of-flight detector that covers the non-instrumented areas between adjacent quartz bars. There is around \SI{45}{\mm} of radial free space between the the outer shell of the Belle\,II Central Drift Chamber (CDC) and the inner radius of the TOP module enclosures. Fig.\ref{fig:schematic_view} shows a sketch of the geometry around the gap between two TOP module enclosures and a possible STOPGAP module geometry.

\subsubsection{MC Performance Study}
To show the feasibility of a STOPGAP system, a Monte Carlo (MC) study based on the full Belle II simulation of $\Upsilon(4S) \to B\bar{B}$ events has been conducted. We directly compared the TOP performance with a full barrel time of flight system in the same geometry. This allows us to determine the design parameters of the sensors to be installed in the TOP gap regions.

\begin{figure}[htbp]
\begin{subfigure}[t]{0.48\textwidth}
  \centering
  \includegraphics[width=1.0\linewidth, trim=1cm 0cm 0cm 0cm ]{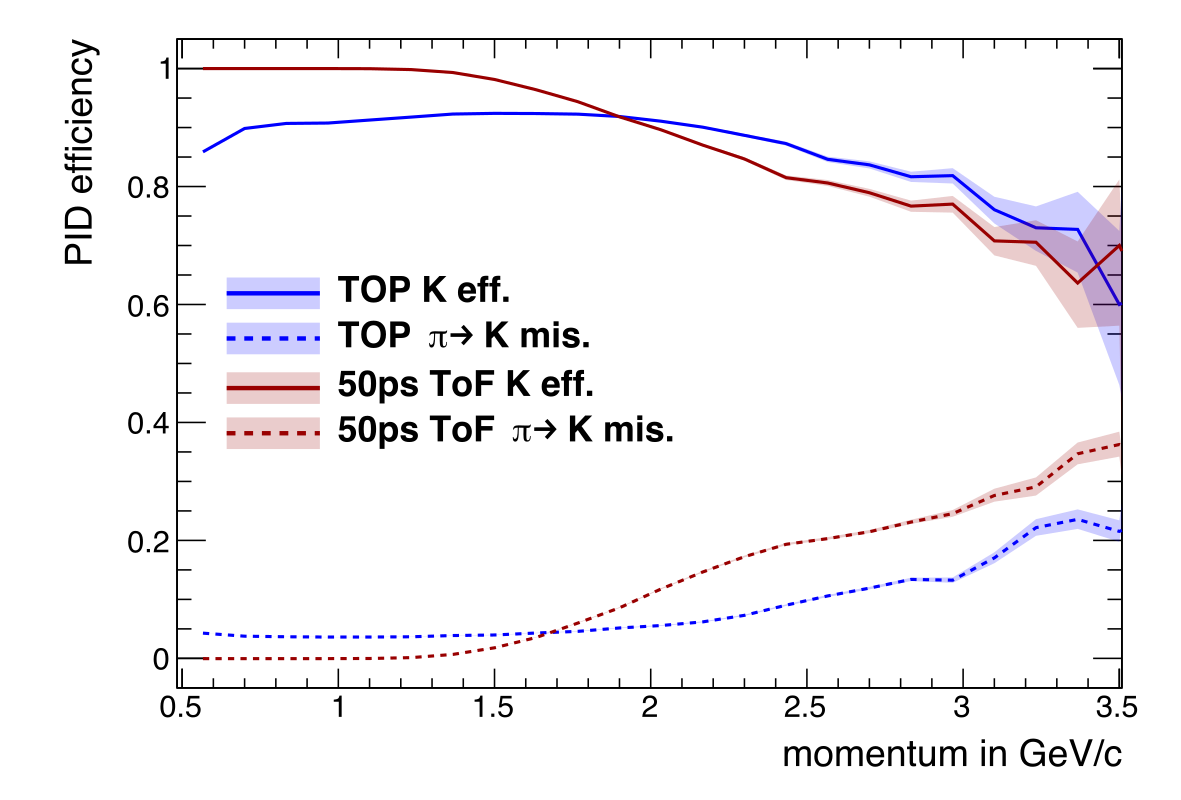}  
  \caption{Selection and mis-identification probabilities at $LL_{K}-LL{\pi}>0$ of a simulated STOPGAP PID system for charged kaons as a function of particle momentum, shown for a simulated sensor with \SI{50}{ps} MIP timing resolution.}
  \label{fig:tof_perf}
\end{subfigure}
\hfill
\begin{subfigure}[t]{.48\textwidth}
  \centering
  \includegraphics[width=1.0\linewidth, trim=1cm 0cm 0cm 0cm]{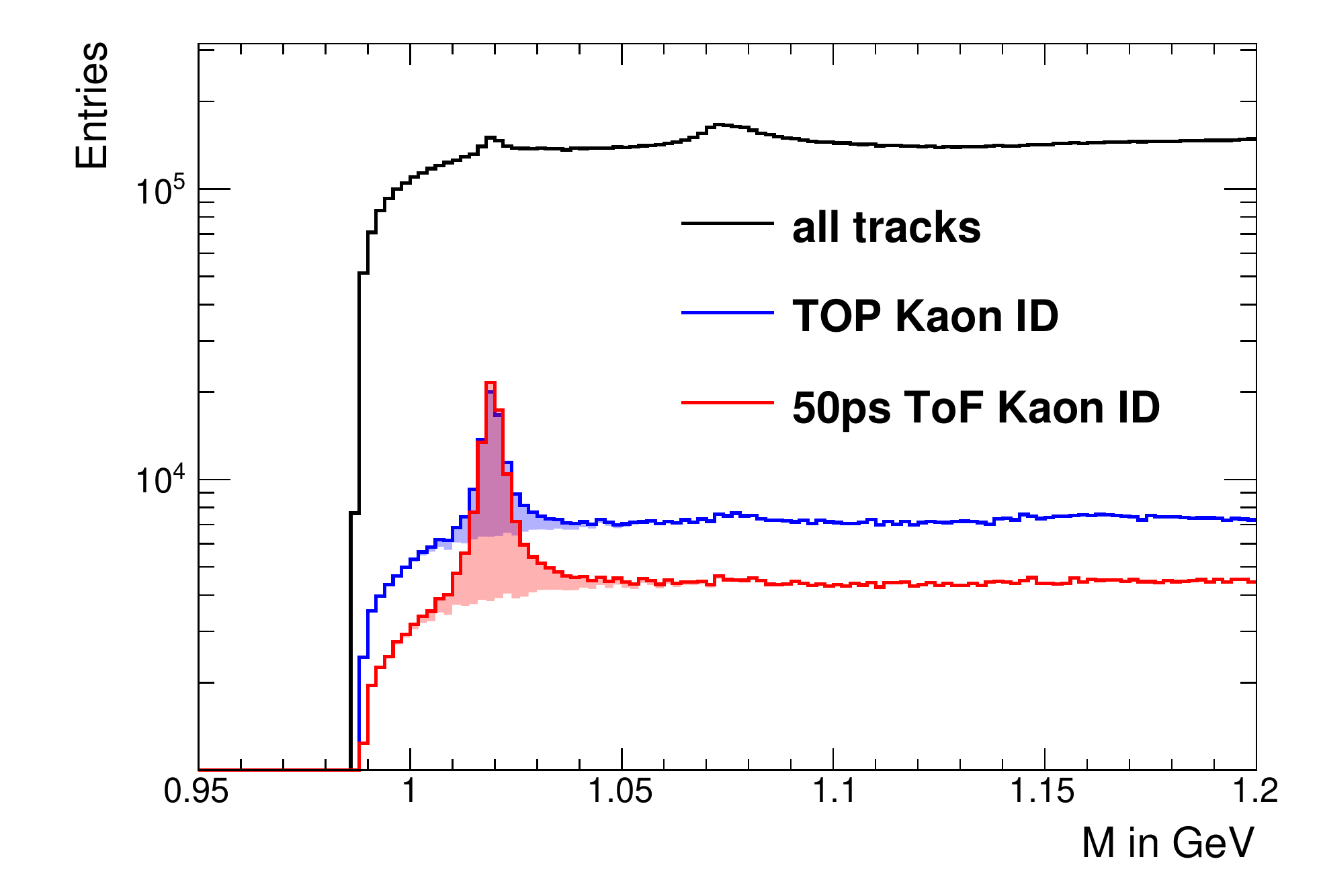}  
  \caption{Invariant mass distribution from charged kaons selected at $LL_{K}>LL{\pi}$ for both tracks, with both tracks in the TOP acceptance. The black line shows the unselected spectrum. The shaded areas indicates the fraction corresponding to true $\phi$ decays. The TOP system achieves a S/N of 0.37 ($\mu\pm2\sigma$), while the STOPGAP reconstruction achieves a S/N of 0.79 in the same invariant mass window.}
  \label{fig:phi_KK}
\end{subfigure}
\caption{Simulated particle identification performance of a time-of-flight system in the Belle\,II barrel region.}
\label{fig:stopgap_perf}
\end{figure}

To simulate the STOPGAP response several contributions to the total time resolution have been considered, both reducible and irreducible: sensor readout resolutions from \SIrange{20}{100}{\ps}, global clock distribution jitter (\SI{10}{\ps}), the SuperKEKB bunch overlap time (\SI{15}{\ps}), and track length uncertainties. The particle identification is then performed by evaluating a likelihood constructed from these time resolution components. The resulting selection and mis-identification fractions for charged pions and kaons are shown for a sensor resolution of \SI{50}{ps} in Fig.\ref{fig:tof_perf}. A STOPGAP system with a combined time resolution for sensor and readout of \SI{50}{\ps} would perform significantly better than the TOP system in the momentum range $p<\SI{2}{\GeV}$. A \SI{30}{\ps} sensor would outperform TOP in the whole momentum range. Fig.\ref{fig:phi_KK} shows the reconstructed invariant mass of $\phi\rightarrow K^+ K^-$ decays selected from the simulated dataset. A \SI{50}{\ps} STOPGAP system achieves more than twice better signal-to-noise ratio near the $\phi$ mass peak. The improved efficiency is entirely defined by the difference in kaon efficiency, which is only \SI{90}{\percent} for TOP, but driven by the MIP efficiency in STOPGAP.

\subsubsection{Requirements}
Depending on the details of the chosen STOPGAP geometry, \SIrange{1}{3}{\m\squared} of active area will be needed to fill the gaps in the TOP acceptance with a single sensor layer. In case a double-layer design is necessary, that area would double accordingly. For reference, a full replacement of the TOP system would require around \SI{18}{\m\squared} of timing sensitive sensor.

The particle identification performance of a time-of-flight system is fundamentally limited by the single MIP detection efficiency of the used sensor technology. A MIP efficiency as close as possible to \SI{100}{\percent} is thus a strong requirement for such a system. 

We studied the expected background hit rates and radiation doses for a STOPGAP system based on extrapolations of the expected beam background at full SuperKEKB luminosities at the radii of the currently installed Belle\,II vertex detector. We conclude that even with an additional safety factor of 25 and a conservative readout granularity and open time, the background occupancy of a STOPGAP system would not exceed $\SI{0.60+-0.23}{\percent}$. The levels of expected radiation exposure for a STOPGAP system are several orders of magnitude lower than those expected and observed in the (HL-)LHC environments. Radiation hardness is thus not expected to of significant importance in the selection of a suitable technology option.

Especially in comparison to the detector upgrades planned for the HL-LHC, the requirements on hit rates, radiation hardness and granularity are entirely negligible for a STOPGAP application. The driving factors of its performance are the single MIP timing resolution and a MIP efficiency of \SI{>99}{\percent}. As discussed above, STOPGAP modules are expected to yield useful particle identification capabilities for time resolutions \SI{<100}{ps}.

\subsubsection{Potential Sensor Technologies}
The existing options for sensors and readout technologies under active development and construction for the HL-LHC upgrades are similarly suitable for a STOPGAP system. Low Gain Avalanche diodes show very promising MIP timing resolutions down to \SI{20}{\ps}. However, currently available types are produced in expensive non-standard processes and require double layers for the required \SI{>99}{\percent} MIP efficiency, further increasing their price. LYSO+SiPM systems reach MIP timing resolutions down to \SI{30}{\ps}, but are fundamentally limited in their possible granularity to around \SI{1}{\cm\squared}. The necessary material budget is limited by the required thickness of the LYSO crystal of a few \si{mm} to yield enough scintillation photons for a timing measurement of the required precision.

A feasible sensor option for STOPGAP would be thin silicon sensors without avalanche amplification inside the sensors, which improve their signal-to-noise ratio by means of a lower noise and thus higher powered electronic amplification of the sensor output signal. This has been demonstrated in principle by the NA62 Gigatracker \cite{gigatracker}, which achieves \SI{130}{ps} time resolution on single MIP hits per layer with a \SI{200}{\um} thick silicon sensors without internal avalanche amplification. The natural next step is to integrate high-powered low noise preamplifiers and discriminators directly into a fast monolithic pixel sensor produced in a CMOS process, leveraging the cost benefits and scalability of industrial CMOS fabrication.

At least two groups are actively pursuing first steps towards MAPS with single MIP time resolutions of \SI{<100}{ps} \cite{geneva1, geneva2, cactus}. These projects show that it is indeed possible to achieve sub-\SI{100}{ps} timing with monolithic CMOS detectors, even though their prototypes are not yet feasible for practical use. The relatively tame requirements of a STOPGAP sensor on hit rates, radiation hardness and (in principle) readout granularity, at least compared to the requirements for HL-LHC applications, opens up the STOPGAP proposal as an ideal test bed for such emerging fast timing CMOS MAPS technologies.

While it is unlikely for any realistic fast MAPS sensor to reach the timing performance of (AC-)LGADs, it should be possible to get close to within a factor of two. The prospect of achieving almost comparable resolutions with significantly reduced cost is extremely appealing. Opening up the fast timing frontier for MAPS also enables future MAPS tracker projects to include as much time resolution into their designs as needed for the application. Establishing fast MAPS is thus of prime importance not only for HEP detectors, but for technologically related instrumentation as a whole. 

\subsubsection{Resource estimates and timeline}
The resource estimate for the whole project is divided into two parts: the development of the MAPS technology, and the actual construction of the detector modules.

\paragraph{R\&D for fast MAPS}
Making significant steps towards very fast CMOS sensors with timing resolution around \SI{50}{ps} is the most important fundamental building block of the STOPGAP proposal. The current efforts towards fast CMOS sensors use fabrication processes of \SI{130}{\nm} and \SI{150}{\nm} structure size. An immediate improvement in time resolution (or alternatively the power draw at a fixed time resolution) can be achieved moving the development to the next smaller node of \SI{65}{\nm}. At least two engineering submissions and extensive laboratory test bench and testbeam campaigns will be necessary to develop and validate a sensor layout that achieves the necessary time resolutions at an acceptable power budget. 

The intermediate goal of this proposed R\&D program is the construction of a small sized prototype detector with active area of a few \si{\cm\squared} in order to validate the sensor technology in a realistic environment. This prototype would be installed inside Belle\,II and operated synchronously with the other detectors while recording physics collisions. Demonstrating the time-of-flight resolution for single MIPs in the environment of an active HEP experiment will not only validate the sensor technology, but also uncover the practical issues with operating a sensor in realistic conditions compared to test beam or bench tests. The ultimate capstone of the program would be a working prototype of a fully integrated monolithic detector with integrated sparsified readout that demonstrates the required timing and hit rate capabilities at least in a test beam environment. 

Such a program seems unfeasible without significant funding for at least three to five years. Out of that time, we coarsely estimate around one to two years will be needed to finish the sensor design, at least one year is needed for the integrated timing frontend and readout logic, and an additional year will be needed for integration onto one common mixed signal chip. Depending on available person power and expertise, parts of this design and test effort can run in parallel. To acquire the necessary momentum, the strong support of an existing local work group with extensive expertise in CMOS sensor design and a profound interest in pushing the possibilities of fast MAPS to its limits is required at the very least for the first one to two years.

\paragraph{STOPGAP construction}
Ultimately, 16 STOPGAP modules with around \SIrange{1}{2}{\m\squared} of total active sensor area will have to be built and installed. Apart from the significant funding requirements for the sensor production itself, this will need detailed concepts for the mechanics, cooling, service routing and installation procedure for these modules. None of these tasks have even been considered so far. First steps towards a STOPGAP module design in the form of finite element simulations to estimate the required stiffness and eventual mechanical mock-ups could in principle start in parallel with the sensor R\&D program outlined above.

\subsubsection{Timing Layers at Lower Radii}\label{sec:stopgap_timing}
Several of the currently proposed and discussed upgrades of the Belle\,II vertexing system plan to increase the inner radius of the installed CDC system. The thin silicon sensors proposed for the vertexing upgrade are unlikely to provide enough $dE/dx$ discrimination for low momentum particle as the CDC does currently. Additionally, the CDC currently provides track triggering for transverse particle momenta $p_T$ down to around \SI{100}{\MeV}, which a full silicon inner tracking system might not be able to provide. 

We expect the principles and sensor technologies proposed for a STOPGAP system can be applied to full timing layers installed in an upgraded Belle\,II detector at lower radii with minor additional requirements. We thus explore here the potential of recovering the PID and track triggering capabilities for low $p_T$ particles with an analytical model of dedicated timing layers at \SI{250}{\mm} and/or \SI{450}{\mm} radius.

\begin{figure}[htbp]
\begin{subfigure}[t]{0.48\textwidth}
  \centering
  \includegraphics[width=1.0\linewidth, trim=1cm 0cm 2cm 1cm ]{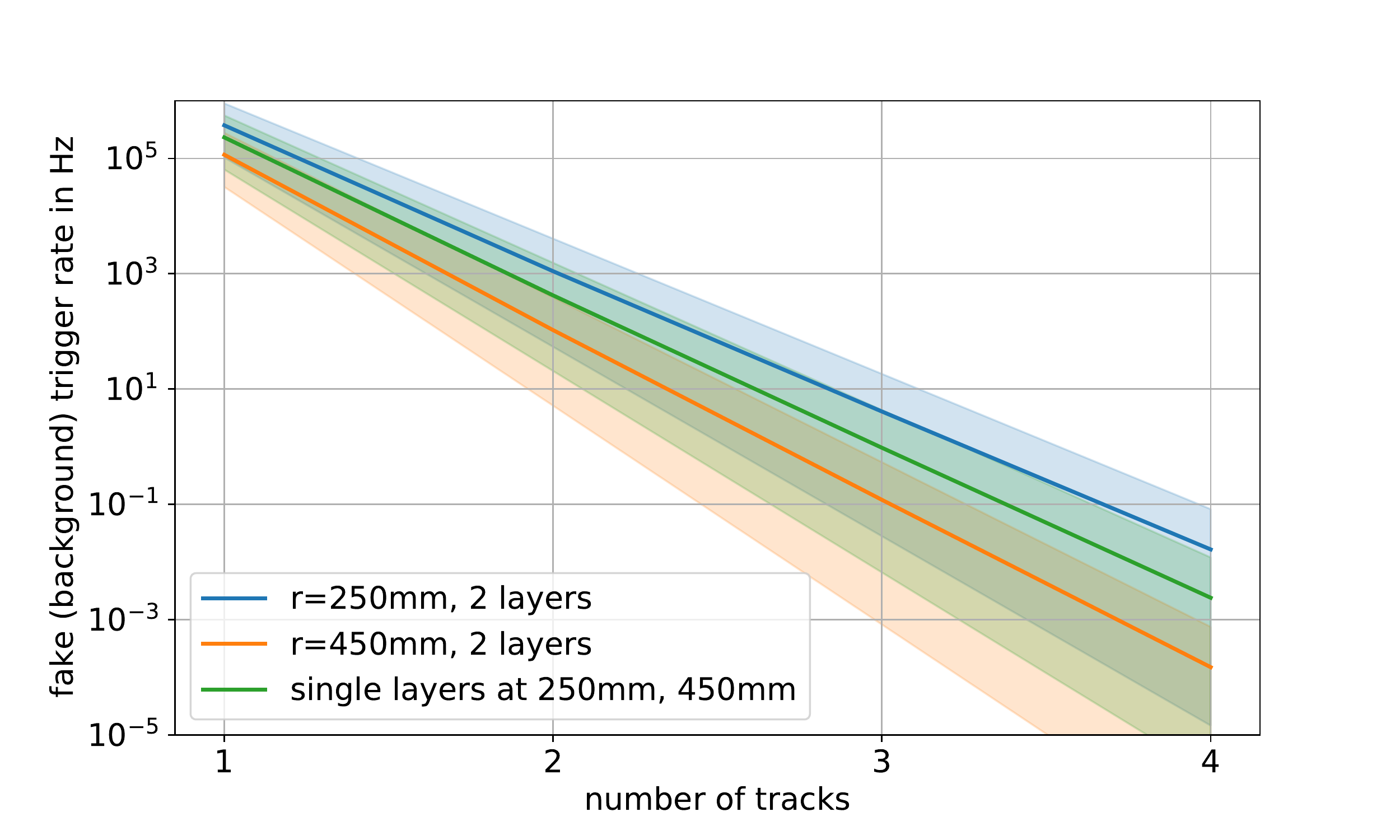}  
  \caption{Estimated background trigger rates of double layers of fast timing coincidence track trigger at \SI{250}{\mm} and \SI{450}{\mm} radii as well as for two single track trigger timing layers at \SI{250}{\mm} and \SI{450}{\mm} radius, based on the beam backgrounds expected at the full design luminosity of SuperKEKB.}
  \label{fig:timing_trig_twolayer}
\end{subfigure}
\hfill
\begin{subfigure}[t]{.48\textwidth}
  \centering
  \includegraphics[width=1.0\linewidth, trim=1cm 0cm 2cm 1cm]{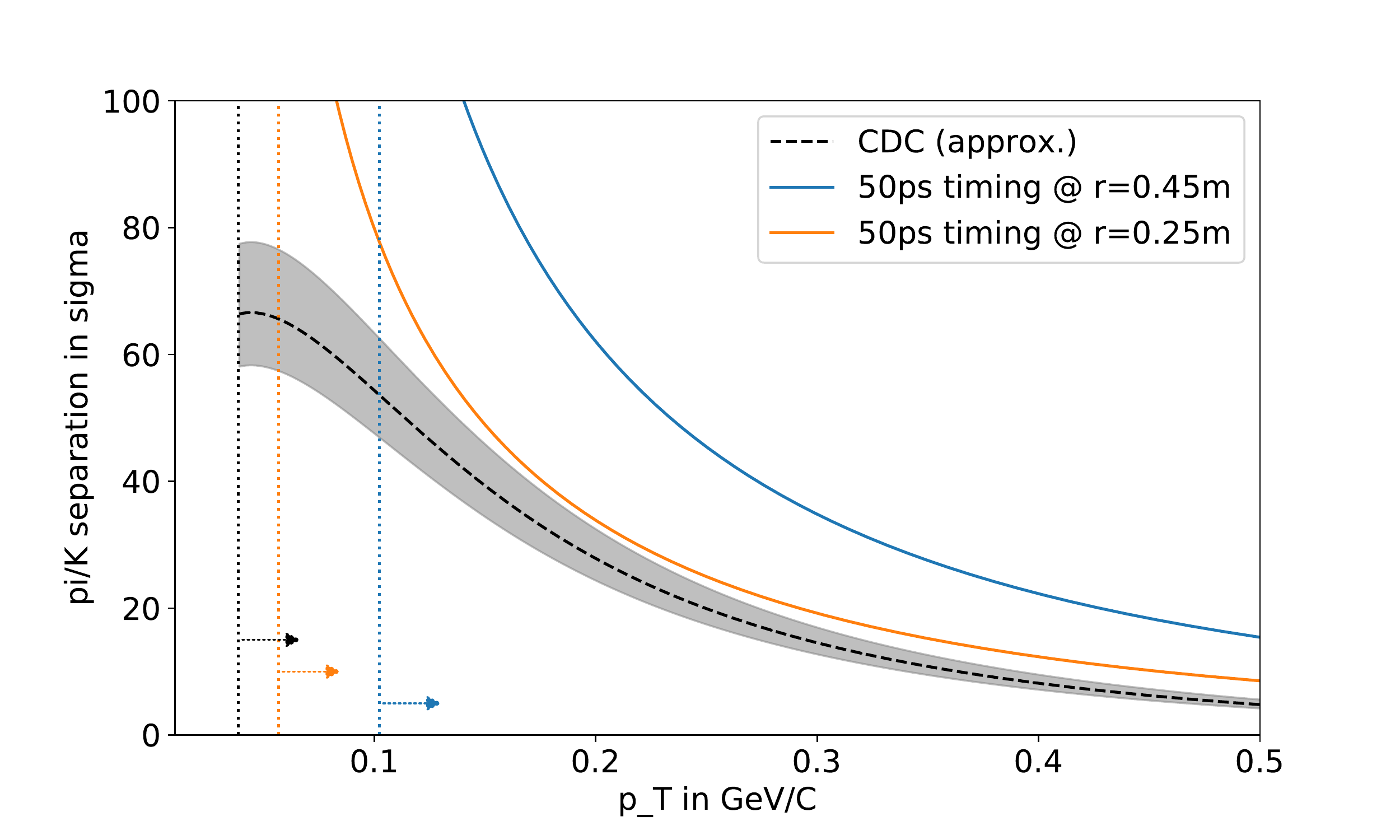}  
  \caption{Pion/kaon separation performance of a MIP timing layer compared to an estimated $dE/dx$ PID at different radii. The arrows indicate the lower $p_T$ cutoff for each system due to its radius.}
  \label{fig:tof_lowpt}
\end{subfigure}
\caption{Performance estimates for timing layers at lower radii.}
\label{fig:trackt}
\end{figure}

For pion/kaon separation purposes, the region of $p_T<\SI{500}{\MeV}$ is most relevant, as higher $p_T$ will reach the dedicated PID subdetectors TOP, ARICH and also generally provide a $dE/dx$ measurement from the remaining part of the CDC or its replacement. We calculate the time of flight differences between charged kaons and pions divided by the assumed time resolution of a timing layer, resulting in a measure of particle identification in standard deviations of separation shown in Fig.\ref{fig:tof_lowpt}. This indicates that a \SI{50}{\ps} MIP timing resolution layer at \SI{250}{\mm} radius yields consistently better pion/kaon separation than the CDC, while a \SI{450}{\mm} radius timing layer leads to at least double the separation power between pions and kaons over the CDC in the momentum region inspected here.

To estimate the  feasibility of timing layers to provide track triggers for Belle\,II, a simplified spatial and temporal coincidence trigger model is combined with extrapolations from the beam background estimates described before. The expected trigger rates from pure beam backgrounds are then calculated for double timing layers at \SI{250}{\mm} and \SI{450}{\mm}, as shown in Fig.\ref{fig:timing_trig_twolayer}. A configuration with two single timing layers at \SI{250}{\mm} and \SI{450}{\mm} radius, respectively, is estimated to yield background trigger rates right in between the double-layer options. Additionally a distanced two-layer setup would enable a coarse estimation of the Z-origin of a triggered track, enabling a selection of tracks from the IP, as well as a coarse estimation of the track momentum (or at least its charge) on the trigger level. Despite the large systematic uncertainties of this study, the resulting expected trigger rates show the fundamental feasibility of a timing track trigger system, at least for triggering events with two or more tracks in its acceptance. 

For timing layers at lower radii, the requirements on hit rate and radiation hardness are stronger than for the STOPGAP case, but still tame compared to HL-LHC developments. As part of the inner tracking detector, finer granularities and a much reduced material budget would be required compared to STOPGAP.

\printbibliography[heading=subbibliography]
\clearpage
\end{refsection}

\subsection{A TPC-based tracking system}
\begin{refsection}
\label{sec:TPC}
\editor{P.Lewis}
In the long term upgrade scenario, increased luminosities and beam-induced backgrounds may exceed the capabilities of drift chambers for charged particle tracking. We have evaluated a new tracking concept based on three elements: (1) replacing the CDC with a time projection chamber (TPC) with pixel readout~\cite{andreas}, (2) a new vertexing system based on the VTX with additional layers, and (3) new fast timing layers based on the STOPGAP technology to provide a trigger source and low-$p_T$ track PID (see Sec.~\ref{sec:stopgap_timing}). We present a conceptual design~\cite{andreas} for such a tracking system along with the results of preliminary simulation-based studies that validate its capabilities at $5\times$ the design luminosity of SuperKEKB. 

\begin{figure}
\begin{center}
	\includegraphics[width=12cm]{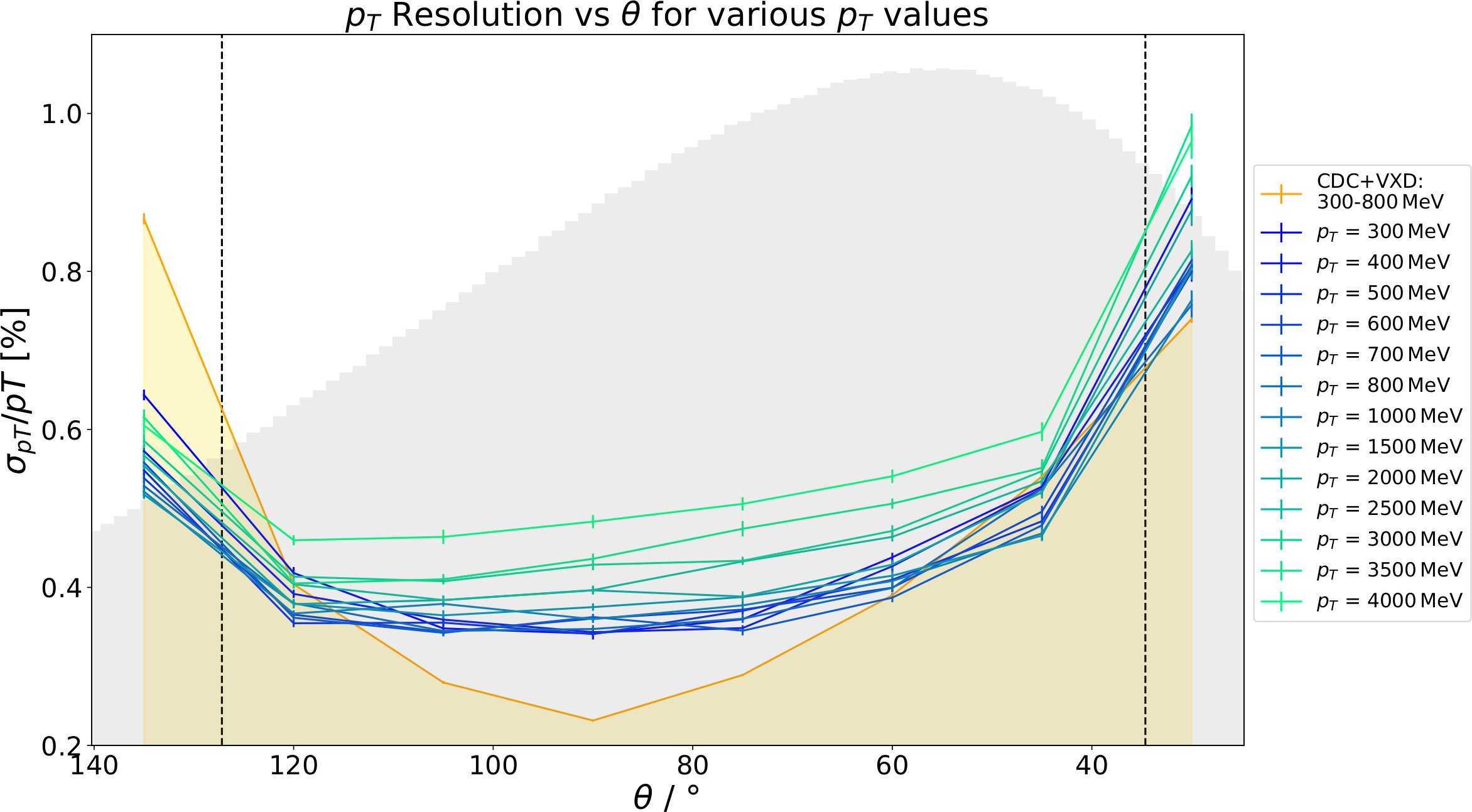}
	\caption{A comparison of the $p_T$ resolution for CDC+VXD compared to TPC+VTX vs. polar angle $\theta$. The gray histogram in the background indicates the distribution of tracks in $\Upsilon(4S)$ decays.}
	\label{fig:pTRes_PT}
	\includegraphics[width=10cm]{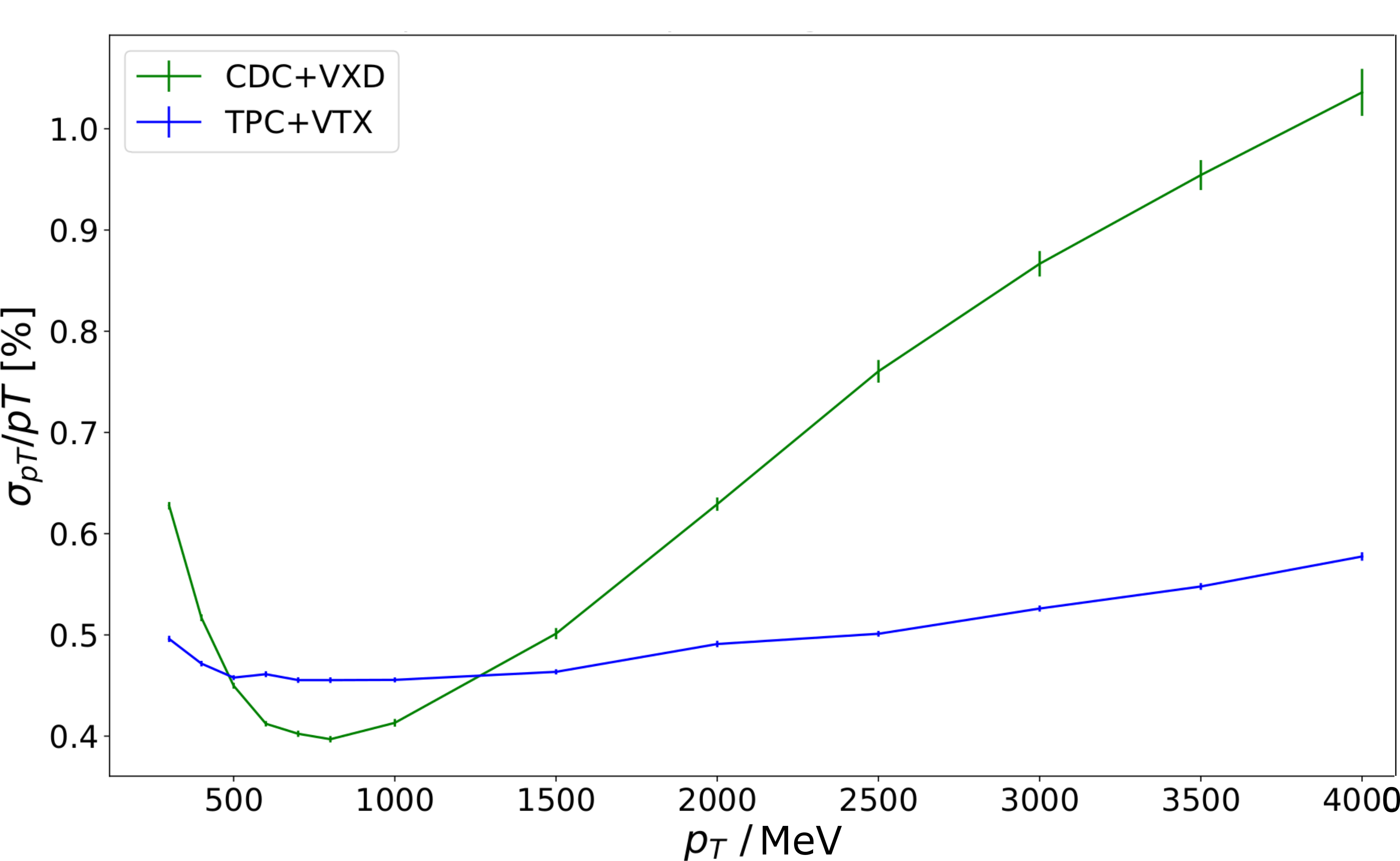}
	\caption{A comparison of the $p_T$ resolution for CDC+VXD compared to TPC+VTX vs. $p_T$.}
	\label{fig:pTRes_PTCOL}
\end{center}    
\end{figure}

\subsubsection{Design}
The basic TPC design consists of a single gas volume of length \SI{242}{cm} with high-resolution readout tiling the backward endcap. We do not use any form of ion gating and we assume that an external detector will provide the trigger, which will be used to isolate discrete event ``windows" in the continuous data. 

We assume that the TPC uses atmospheric pressure Ar-CF$_{4}$-iC$_{4}$H$_{10}$ (95-3-2), a mixture that has been studied for use in LCTPC~\cite{t2k_gas}. We use MAGBOLTZ~\cite{magboltz} to determine the drift field ($289$~V/cm) that minimizes the drift time of electrons ($v_D=\SI{7.89}{\centi\meter/\micro\second}$, leading to a maximum drift time of roughly $\SI{30}{\micro\second}$ for a maximum drift length of $\SI{242}{\centi\meter}$). From the same simulation we find that the longitudinal and transverse diffusion coefficients are $\sigma_L=\SI{200}{\micro\meter/\cm}^{0.5}$ and $\sigma_T=\SI{84}{\micro\meter/\cm}^{0.5}$.

For charge amplification and readout, we take inspiration from the GridPix system proposed for use in LCTPC~\cite{ild_gridpix}. GridPix consists of an array of $55\times \SI{55}{\micro\meter}$ pixels with a MICROMEGAS mesh mounted onto the surface. For our purposes, this technology presents a number of advantages: first, the small pixels and direct mapping between amplification cells and pixels constitute essentially a best-case resolution scenario. Second, in theory such a sensor can be operated in \textit{binary readout} mode in which each individual hit represents exactly one electron and consists only of the pixel ID and a time ID. This can dramatically reduce the data throughput, which we anticipate to be a significant technical challenge at ultrahigh luminosities and with continuous readout. Thirdly, without charge sharing the operational gain can be reduced; coupled with an intrinsically low ion backflow rate, distortions of the drift field due to backflowing ions should be minimized. Finally, it is very easy to implement such a detector in the digitization simulation. 

The timing layer concept includes both the $250$~mm and $450$~mm scenarios described in Sec.~\ref{sec:stopgap_timing}, corresponding to placement inside the VTX or between the VTX and the TPC, respectively. 

\subsubsection{Simulation}
We use the Belle II simulation framework with an expanded VTX along with a new TPC volume. We generate ionizations with GEANT4, then drift, diffuse, and overlay the primary electrons using gas parameters determined by MAGBOLTZ. After drift and diffusion, we bin into 3D voxels (nominally $55\times55\times\SI{55}{\micro\meter\cubed}$) to approximate amplification and digitization in a GridPix-like detector operating in binary readout mode. 

Finally, we center the time window on a single $\Upsilon(4S)$ event and discard hits from outside a $\SI{30}{\micro\second}$ window approximating the total drift time of a single event. 

\subsubsection{Studies}
We target our first studies at answering the primary difficulties of operating a tracking TPC at high luminosities. These difficulties include event pileup, background pileup, ion backflow, and continuous readout. We consider the proposal viable if the tracking performance, specifically $p_T$ resolution, is comparable to the CDC at $5\times$ the design luminosity of SuperKEKB. 

We find that event overlap is modest, with an average of nine background tracks (largely from Bhabha scattering) per triggered $\Upsilon(4S)$ event. These tracks are generally easy to identify offline via diffusion width and impact parameter. 

Beam-induced backgrounds cause a high rate of microcurlers that deposit large amounts of charge in the TPC, completely dominating the total ionization rate. We find, using $200\times 200\times\SI{200}{\micro\meter\cubed}$ voxels, an occupancy due to these background hits below $2\times10^{-4}$, suggesting no detector limitations due to occupancy. These background hits may present difficulties in data throughput; however, we find that microcurlers are easily identified and can in principle be largely rejected at the frontend, greatly reducing the required data throughput. 

Using an estimate of $1\%$ backflow at a gain of 2000, and using measurements of the mobilities of the primary ions in the T2K gas~\cite{t2k_gas}, we find 
a typical ion charge density 
comparable to other tracking TPCs. However, this estimate is certainly optimistic: injection backgrounds are not simulated and we expect them to be very high since continuous injection will presumably be used. Furthermore, the TPC ionization rate is dominated by microcurlers, most of which come from beam-induced low-energy photons. Therefore, the ion density depends strongly on our SuperKEKB beam background simulation that may not be suitable for the future upgrade scenario. We cannot rule out the possibility that high backflowing ion densities will substantially degrade the tracking performance.

For transverse tracks, the current simulation indicates better performance with the CDC with low-$p_T$ tracks (Fig.~\ref{fig:pTRes_PT}); however, this is largely do to multiple scattering in the VTX, timing layers, and simulated TPC walls. This simulated material budget is highly speculative. For tracks further from the transverse, the number of spatial hit points in the TPC increases, improving the resolution of the TPC compared to the CDC. In addition, the effective position measurement resolution of the TPC is determined by the diffusion and ionization statistics, and is significantly better than that of the CDC. Consequently, the TPC $p_T$ resolution degrades much slower than the CDC as $p_T$ increases (Fig.~\ref{fig:pTRes_PTCOL}).

Overall, we conclude that the extended VTX plus TPC tracking performance is comparable to the current VXD plus CDC, with some advantages and some disadvantages. However, the material budget of the inner detectors is responsible for the degradation of the $p_T$ resolution for low-$p_T$ tracks, not the TPC itself. Therefore, other upgrade scenarios involving an expansion of the vertexing detector will share the same challenges. 

We additionally conclude that the capabilities of the fast timing layers, particularly when installed at $r=250$~mm, more than compensate for the complete loss of triggering and partial loss of PID capabilities necessitated by the TPC design. In particular, the low-$p_T$ TOF PID capabilities of such a detector could, in principle, provide pion/kaon separation capabilities far superior to those currently achievable by the CDC (Fig.~\ref{fig:tof_lowpt}). 

\subsubsection{Conclusions}
Our preliminary studies indicate that a TPC, combined with a new fast timing layer and an expanded vertexing system, can provide comparable tracking performance to the Belle II tracking system at luminosities five times the design luminosity of SuperKEKB. However, these studies rely on a single set of assumptions about readout and detector technology, and more work is required, particularly to determine whether cheaper readout technologies can similarly achieve suitable performance, 
and to determine the quantity and effect of the backflowing ions. 
\printbibliography[heading=subbibliography]
\clearpage
\end{refsection}

\end{document}